\documentclass[lettersize,journal]{IEEEtran}
\usepackage{amsmath,amssymb}
\usepackage{psfrag}
\usepackage{epsfig}
\usepackage{cite}
\usepackage{ulem}
\usepackage{graphicx}
\usepackage{subfigure}
\usepackage{multirow}
\usepackage{threeparttable}
\usepackage{booktabs}
\usepackage{mathrsfs}
\usepackage{amsfonts}
\usepackage{amsmath}
\usepackage{psfrag}
\usepackage{graphicx}
\usepackage{amssymb}
\usepackage{subfigure}
\usepackage{algorithmic}
\usepackage{algorithm}
\usepackage{color}
\usepackage{overpic}
\usepackage{cases}
\usepackage{listings}
\usepackage{cite}
\usepackage{pifont}
\usepackage{amsmath,amssymb,amsfonts}
\usepackage{algorithmic}
\usepackage{graphicx}
\usepackage{textcomp}
\usepackage{amsmath,amssymb}
\usepackage{enumerate}
\usepackage{psfrag}
\usepackage{ragged2e}
\usepackage{url}
\usepackage{float}
\usepackage{epsfig}
\usepackage{color}
\usepackage{graphics}
\usepackage{tabularx}
\usepackage{booktabs}
\usepackage[T1]{fontenc}
\usepackage{subfigure}
\usepackage{fancybox}
\usepackage{multirow}
\usepackage{threeparttable}
\usepackage{booktabs}
\usepackage{mathrsfs}
\usepackage{bm}
\usepackage{algorithm}
\usepackage{algorithmic}
\usepackage{ulem}
\usepackage[labelfont=bf,format=plain,justification=raggedright,singlelinecheck=false]{caption}
\usepackage[colorlinks=true,linkcolor=black,urlcolor=black,citecolor=black]{hyperref}
\usepackage{algorithmic}
\usepackage{algorithm}
\usepackage{array}
\usepackage{textcomp}
\usepackage{url}
\usepackage{verbatim}
\usepackage{graphicx}
\usepackage{amsmath,amssymb}
\usepackage{psfrag}
\usepackage{epsfig}
\usepackage{cite}
\usepackage{ulem}
\usepackage{graphicx}
\usepackage{multirow}
\usepackage{threeparttable}
\usepackage{booktabs}
\usepackage{mathrsfs}
\usepackage{amsfonts}
\usepackage{amsmath}
\usepackage{psfrag}
\usepackage{graphicx}
\usepackage{amssymb}
\usepackage{algorithmic}
\usepackage{algorithm}
\usepackage{color}
\usepackage{overpic}
\usepackage{cases}
\usepackage{listings}
\usepackage {setspace}
\usepackage{cite}
\usepackage{pifont}
\usepackage{amsmath,amssymb,amsfonts}
\usepackage{algorithmic}
\usepackage{graphicx}
\usepackage{textcomp}
\usepackage{amsmath,amssymb}
\usepackage{enumerate}
\usepackage{psfrag}
\usepackage{ragged2e}
\usepackage{url}
\usepackage{epsfig}
\usepackage{color}
\usepackage{graphics}
\usepackage{tabularx}
\usepackage{booktabs}
\usepackage[T1]{fontenc}
\usepackage{fancybox}
\usepackage{multirow}
\usepackage{threeparttable}
\usepackage{booktabs}
\usepackage{mathrsfs}
\usepackage{bm}
\usepackage{algorithm}
\usepackage{algorithmic}
\usepackage{ulem}
\usepackage{subfigure}
\usepackage[labelfont=bf,format=plain,justification=raggedright,singlelinecheck=false]{caption}
\usepackage[colorlinks=true,linkcolor=black,urlcolor=black,citecolor=black]{hyperref}
\usepackage{nomencl}
\makenomenclature
\begin{document}

\title{A Predictive Cooperative Collision Avoidance for Multi-Robot Systems Using Control Barrier Function}
 \author{Xiaoxiao~Li,
             Zhirui~Sun,
             Hongpeng~Wang,
             Shuai~Li,~\IEEEmembership{Senior~Member,~IEEE,}
             Jiankun~Wang,~\IEEEmembership{Senior~Member,~IEEE}
\thanks{This work was supported in part by the Guangdong Provincial Key Laboratory of Novel Security Intelligence Technologies under Grant 2022B1212010005, in part by the Shenzhen Basic Research Project (Natural Science Foundation) under Grant JCYJ20210324132212030. (Corresponding author: Hongpeng Wang)}
\thanks{X. Li and H. Wang are with the Harbin Institute of Technology Shenzhen, P.R. China. (e-mails: lxx@stu.hit.edu.cn; wanghp@hit.edu.cn).}
\thanks{H. Wang is also with the Guangdong Provincial Key Laboratory of Novel Security Intelligence Technologies.}
\thanks{Z. Sun and J. Wang are with the Department of Electronic and Electrical Engineering, Southern University of Science and Technology, Shenzhen 518055, China (e-mail: sunzr2023@mail.sustech.edu.cn; wangjk@sustech.edu.cn).}
\thanks{S. Li is with the Faculty of Information Technology and Electrical Engineering, University of Oulu, 90570 Oulu, Finland, and also with the VTT-Technical Research Centre of Finland, 90590 Oulu, Finland (e-mail: shuai.li@oulu.fi).}}




\maketitle

\begin{abstract}
Control barrier function (CBF)-based methods provide the minimum modification necessary to formally guarantee safety in the context of quadratic programming, and strict safety guarantee for safety-critical systems. However, most CBF-related derivatives myopically focus on present safety at each time step, a reasoning over a look-ahead horizon is exactly missing.
In this paper, a predictive safety matrix is constructed. We then consolidate the safety condition based on the proposed safety matrix’s smallest eigenvalue. A predefined deconfliction strategy of motion paths is embedded into the trajectory tracking module to manage robots’ deadlock conflicts, which computes the deadlock escape velocity with the minimum attitude angle.
Comparison results show that the introduction of the predictive term is robust for measurement uncertainty and is immune to oscillations. The proposed deadlock avoidance method avoids a large detours,  without obvious stagnation.
\end{abstract}

%

\section{Introduction}
\IEEEPARstart{T}{he} pivot of a multi-robot system is how to design a control system that gives attention to both accuracy and efficiency while ensuring that each robot does not collide with other robots/obstacles during task execution. However, the local controllers deployed on robots are usually a combination of a nominal controller responsible for completion of a primary objective and a collision avoidance (CA) controllers for ensuring safety. As the number of robots increases, the CA controller are dominant, and as a result, robots would not make progress toward the primary objective.
Compared to velocity obstacle (VO) series with enlarged conservative bounding volumes, and learning-based methods that lack provably safety assurance, control barrier function (CBF)-based methods provide the minimum modification necessary to formally guarantee safety in the context of quadratic programming (QP), and strict safety guarantee for safety-critical systems, \emph{e.g},  \cite{SBC1,SBC2,SBC3} which extend the CBF to the multi-robot system by incorporating all pairwise collision-free constraints into an admissible control space. However, CBF-related derivatives myopically focus on present safety at each time step, a reasoning over a look-ahead horizon is exactly missing.
This allows a trajectory to get too close to the unsafe set, may crashing into other robots or obstacles because of the inaccurate sensing from the onboard sensors on spatial targets or response delay from the actuation system, \emph{etc}.

This drawback can be mitigated by the design of a robust safety controller, such as PrSBC \cite{NIPs2020}. Although the PrSBC produces a more permissive set, the deployment of the chance constraint enforces a more active CA behavior for robots, more likely to cause deadlocks or large evasive actions. Another line follows the idea in \cite{IJRR2014} that the measurement uncertainty is compensated by enlarging the safety threshold. This provides an additional buffer zone for robot, however, the magnitude is limited by the environment, and a large magnitude causes the CA path oscillations.
Therefore, more predictive states of the robots should be incorporated in the safety constraints.
A recent approach for making CBFs consider the future state evolution is PCBFs \cite{breeden2022predictive} and ff-CBF \cite{black2023future}. However, PCBFs do not guarantee input constraint satisfaction.
Unlike PCBFs, ff-CBF does not require numerical integration of the system trajectories forward in time, however, it takes on an increased computational load in exchange for applicability to more general predicted trajectories.
It should be noted that both PCBFs and ff-CBF methods, deployed on a small number of robot systems, achieve collision-free and deadlock deconfliction through acceleration/deceleration actions in a fixed path.
They cannot guarantee that the resulting motion will be deadlock-free and that all robots will always eventually reach their destination in a large-scale robot system.

Deadlock results from symmetry in initial conditions and goals locations and that heterogeneity in the controller parameters is not sufficient \cite{IJRR2023}. In \cite{RAL2019,IJRR2023}, degree of evoluation of the system state is continuously compared against small thresholds. Once these thresholds are satisfied, the system is declared deadlock in the sense that deadlocks would be detected after only they happen, leading a slow-reaction system.
In the context of deadlock deconfliction, two ideas are usually applied: one is the speed adjustment strategy for the systems where robots have unchanged paths as previously mentioned \cite{breeden2022predictive} and \cite{black2023future}. In \cite{SMC2017}, the hold-and-wait action is inserted to coordinate robots' behaviors.
Another one concentrates on the change of robots’ trajectories. A commonly-used strategy is inspired by the right-hand rules.
For instance, an artificial perturbation to the right-hand side of each robot is introduced in \cite{TRO2018}. This however can lead to limited extension in practice as the magnitude of such perturbations is hard to determine. Similar clockwise maneuver is adopted in \cite{TAChe2024} which shows that the tangent motion can resolve deadlocks.
Regrettably, this method is easy to lead to the Zeno phenomenon.
\cite{grover2021deadlock,IJRR2023} analyze the condition of deadlocks based on the Karush-Kuhn-Tucker (KKT) condition and characterize the deadlock as a force-equilibrium on robots. Deadlock is avoided by rotating the robots around each other to swap positions, however, which can be theoretically shown for the case with no more than three robots. As pointed out in \cite{grover2021deadlock,IJRR2023}, deadlock analysis complexity increases exponentially with the number of robots, algorithms that can avoid deadlock in general cases still await.

In this paper, a predictive safety matrix is constructed in a vector format. We then consolidate the safety condition based on the proposed safety matrix’s smallest eigenvalue. Our work is related to \cite{shahriari2021novel}, the difference is that we trimmed the differential calculations related to the smallest eigenvalue, this liberates from redundant computational amount and is immune to oscillations, without the need for designing specialized control laws.
For the compensation on measurement uncertainty, we continue to use the principle from \cite{IJRR2014}. What is better than \cite{IJRR2014} is that our method is independent of the environment.
This work is an extension of our previous works \cite{TCYB2022Li,TITs2024}, where we focus on the CBF's safety guarantee over a look-ahead horizon and the rationality of deconfliction direction. As pointed out in \cite{RAL2019}, fixed-side avoidance is not suitable for all situations. However, compared to \cite{SBC3,RAL2019}, a predefined rules-based deconfliction strategy of motion paths is embedded into the trajectory tracking (TT) module to manage robots’ deadlock conflicts, which computes the deadlock escape velocity with the minimum attitude angle.
The main contributions of this paper can be summarized as:
\begin{itemize}
\item[1)] A predictive safety matrix is constructed. We then consolidate the safety condition based on the proposed safety matrix’s smallest eigenvalue.
\item[2)] A predefined rules-based deconfliction strategy of motion paths is embedded into the TT, which computes the deadlock escape velocity with the minimum attitude angle.
\end{itemize}
%
%
\section{Related Work}\label{sec.2a}
In this paper, we investigate multi-robot motion coordination problem in the context of the wheeled mobile robot. Therefore, we review some current popular CA methods on the wheeled mobile robot, then give a brief survey on CBFs.
\vspace{-10pt}
\subsection{CA Methods: Review}
Reactive cooperative CA methods mainly have VO-based \cite{TRO2018,RVO,ORCA,VRORCA,MCCA}, field-based \cite{TAChe2024, SAPF} and CBF-based, \emph{etc}.

VO guides the robot to avoid collisions by defining a conical collision zone in the velocity space and selecting speeds outside the collision zone. Reciprocal velocity obstacle (RVO) and optimal reciprocal collision avoidance (ORCA) are the most pivotal branches within the realm of VO, serving as the foundational pillars upon which subsequent derivatives are developed. ORCA gives attention to both CA efficiency and trajectory smoothness while avoiding the reciprocal dance in RVO, by trimming VO region and introducing a half-plane constraints of the allowable velocity for the pairwise robots, and selecting the velocity closest to the preferred velocity through a linear programming algorithm. However, the assumption of reciprocal CA responsibility induces deadlock in crowed environments. The variable responsibility ORCA \cite{VRORCA} allows robots to take on different CA responsibilities, however, this method still allocates zero speed to robots in dense environments. MCCA method determines priority for each robot to solve the deadlock by introducing the concept of 'masked velocity' \cite{MCCA}. Despite many improvements, the enlargement principle of the VO still makes it remain conservative.
In \cite{IJRR2014}, the CA module would be activated once when the other robot enters the sensing range of a robot. Although \cite{ICSL2020,ACC2020,zhang2021motion} reduce its conservatism by establishing the collision sector, they continue to use the conic concept from the VO in danger region.
\cite{SAPF} proposes SAPF method that can adapt to a narrow corridor, by modifying the repulsive potential field in the traditional artificial potential field (APF) method as the combination of repulsive potential and vortex potential. Although SAPF is shown to be superior to dynamic window method and APF method, there are speed jumps.
Moreover, \cite{IROS} indicates that the CBF is better than the APF in term of the minimal invasion of CA.
\subsection{CBFs: Review}
The CBF is first conceptualized in \cite{wieland2007constructive}, which is use to map an inequality constraint defined over system states onto a constraint on the control input. Aaron D. Ames \emph{et al.} carry forward the ensuing development and application of the CBF. \cite{ames2014control} establishes the unification of control Lyapunov functions and CBF  through a QP for simultaneous achievement of multi-objectives. Proceed from different control requirements and system characteristics, the CBF is refined as the zeroing CBF (ZCBF) and reciprocal CBF in \cite{ames2016control}. By incorporating all pairwise safety constraints into a permissible control space, this concept is extended to the multi-robot system in \cite{borrmann2015control} and \cite{wang2016safety}, and sublimed in \cite{SBC3}.
\cite{SBC3} laies the foundation for all future research on CBF-based multi-robot motion coordination. Conic CBF is proposed in \cite{ibuki2020distributed} where collision-free motion coordination on a sphere is investigated. Subsequent research tends to focus on the design of robust safety controllers.
\cite{NIPs2020} proposes PrSBC for both measurement uncertainty and motion uncertainty. \cite{alan2023control} is concerned with the design of a robust controller disturbed by the model uncertainty through the combination of the CBF and the notion of Input-to-State Safety. \cite{xiao2021adaptive} proposes adaptive CBF that can accommodate time-varying control bounds and noise in the system dynamics. We just to name a few here. These proposals are aimed at addressing the bottlenecks and challenges in ensure security using CBF. To compensate for the lack of formal guarantees in learning-based methods, CBFs has been used for safety learning \cite{cheng2019end,peng2023design, emam2022safe}. This is left for our future work.
Among existing works, only a small portion of the work, such as \cite{breeden2022predictive,black2023future}, casts light to CBF over a forward-looking horizon.

\section{PRELIMINARIES}\label{sec.2}
In this part, we review the basic concept of the CBF, model the measurement uncertainty, and give the predictive safety condition.
\vspace{-10pt}
\subsection{Control Barrier Function}
The basic idea of CBF-based safety control method is to define a safe set $\mathcal{C}$, \emph{i.e.}, the system states $\mathbf{x}$ having no collisions, and then use the CBF to formally guarantee the forward invariance of the safe set, \emph{i.e.}, if the initial state $\mathbf{x}(0)\in \mathcal{C}$ , then $\mathbf{x}(t)\in \mathcal{C}$ for all $t\geq0$.
Specifically, the safe set is defined as the superlevel set of a continuously differentiable function $h$:
\begin{equation}
\mathcal{C}=\left\{\mathbf{x} \in \mathbb{R} \mid h(\mathbf{x}) \geq 0\right\}
\end{equation}
we say $h$ is a CBF \cite{history} if $\partial h/\partial \mathbf{x} \neq0$ for all $\mathbf{x}\in\partial \mathcal{C}$ and there exists an extended class-$\mathscr{K}$ function $\psi(\cdot)$ such that $h$ satisfies
\begin{equation}\label{eqn.bf}
\dot{h}(\mathbf{x}) \geq-\psi(h(\mathbf{x}))
\end{equation}
The extended class--$\mathscr{K}$ function $\psi(\cdot)$ regulates the rate of the system states converge to the boundary of $\mathcal{C}$. \emph{E.g}, $\psi(h(\mathbf{x})$ is chosen as $\kappa h^3$ \cite{SBC3}. Different choices of $\psi(\cdot)$ lead to different behaviors near the boundary. In this paper, we need to make the error of the trajectory tracking (TT) task converge to zero. Therefore, ZCBF $\psi(h(\mathbf{x}))=\kappa h$ is used. Given a ZCBF,
the admissible control space $\mathcal{S}(\mathbf{x})$ for a robot can be defined as
\begin{equation}
\mathcal{S}(\mathbf{x})=\left\{\mathbf{\mu} \in \mathbb{R} \mid \dot h(\mathbf{x})+ \kappa h\geq 0\right\}
\end{equation}
where $\mu$ is the control input.
\begin{figure}
  \centering
  \includegraphics[scale=0.35]{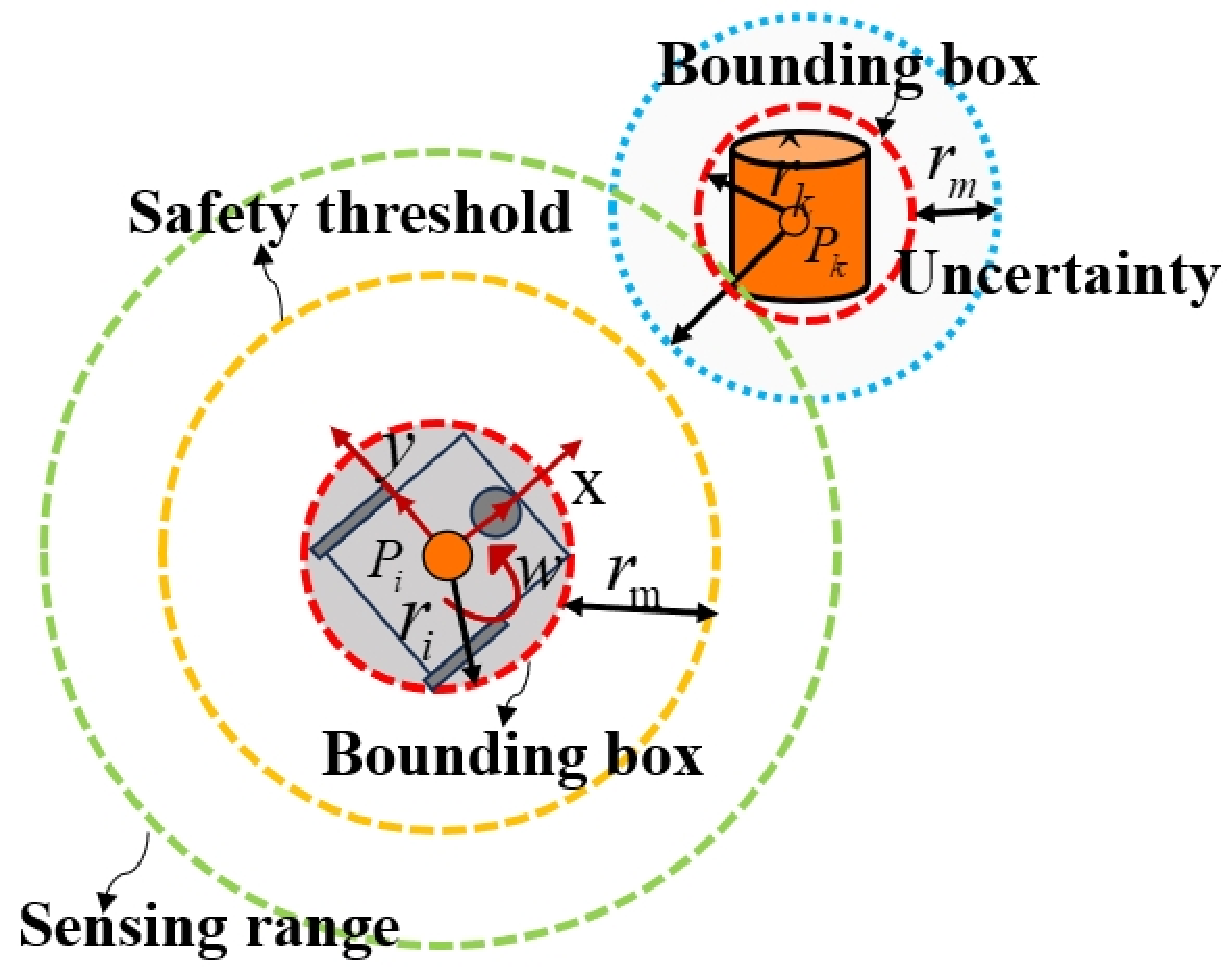}
  \caption{\justifying The measurement error or sensing uncertainty is assumed to be bounded by constant $r_m$.}\label{uncer}
\end{figure}
\vspace{-10pt}
\subsection{Measurement Uncertainty Model}
In this paper, we restrict our analysis to circular robots and obstacles. We model all robots and obstacles by the smallest enclosing disk of radius $r_i$ and $\hat{r}_k$, respectively, $i=1,2,\cdots,N$, $k=1,2,\cdots,n$. Therefore, the area occupied by the $i$-th robot at position $\mathbf{p}_i$ can be described as $\mathscr{A}_i\left(\mathbf{p}_i\right)=D\left(\mathbf{p}_i, r_i\right) \subset \mathbb{R}^2$.  Similarly, the area occupied by the $k$-th obstacle is denoted by $\mathscr{A}_k\left(\mathbf{p}_k\right)=D\left(\mathbf{p}_k, \hat{r}_k\right) \subset \mathbb{R}^2$.
The safety condition is usually formulated as $||\mathbf{p}_{ij}||\geq r_i+r_j$ and $||\mathbf{p}_{ik}||\geq r_i+\hat{r}_k$ in a shared environment with $N$ robots and $n$ obstacles, $j=1,2,\cdots,N, j\neq i, j\in \mathscr{N}_i$, $\mathscr{N}_i$ denotes the neighbour of the robot $i$.

However, the effectiveness of such safety condition depends on perfect environmental perception. Robots generally rely on onboard sensors to perceive surrounding environmental information, and disturbance from external noise and environmental conditions may sometimes cause inaccurate measurement data. This is a feasible way against measurement uncertainty by increasing radius of the enclosing disk by $r_m$, as shown in Fig. \ref{uncer}, $r_m$ is the radius of the safety buffer zone set for the robot. This implies that the robot’s position estimate lies within the circular area with radius $r_i+r_m$ and centered at $\mathbf{p}_i$ of its true coordinate.
This however can lead to limited extension as the magnitude $r_m$ is hard to determine when facing heterogeneous robots/obstacles. More importantly, a large magnitude causes the CA path oscillations.
In this paper, the measurement error or sensing uncertainty is also assumed to be bounded by $r_m$. We aims at designing a predictive safety condition for robot to guarantee smoothness of the CA's trajectory while providing an additional buffer zone.
\begin{figure}[htb]
  \centering
  \hspace{12pt}
  \includegraphics[scale=0.175]{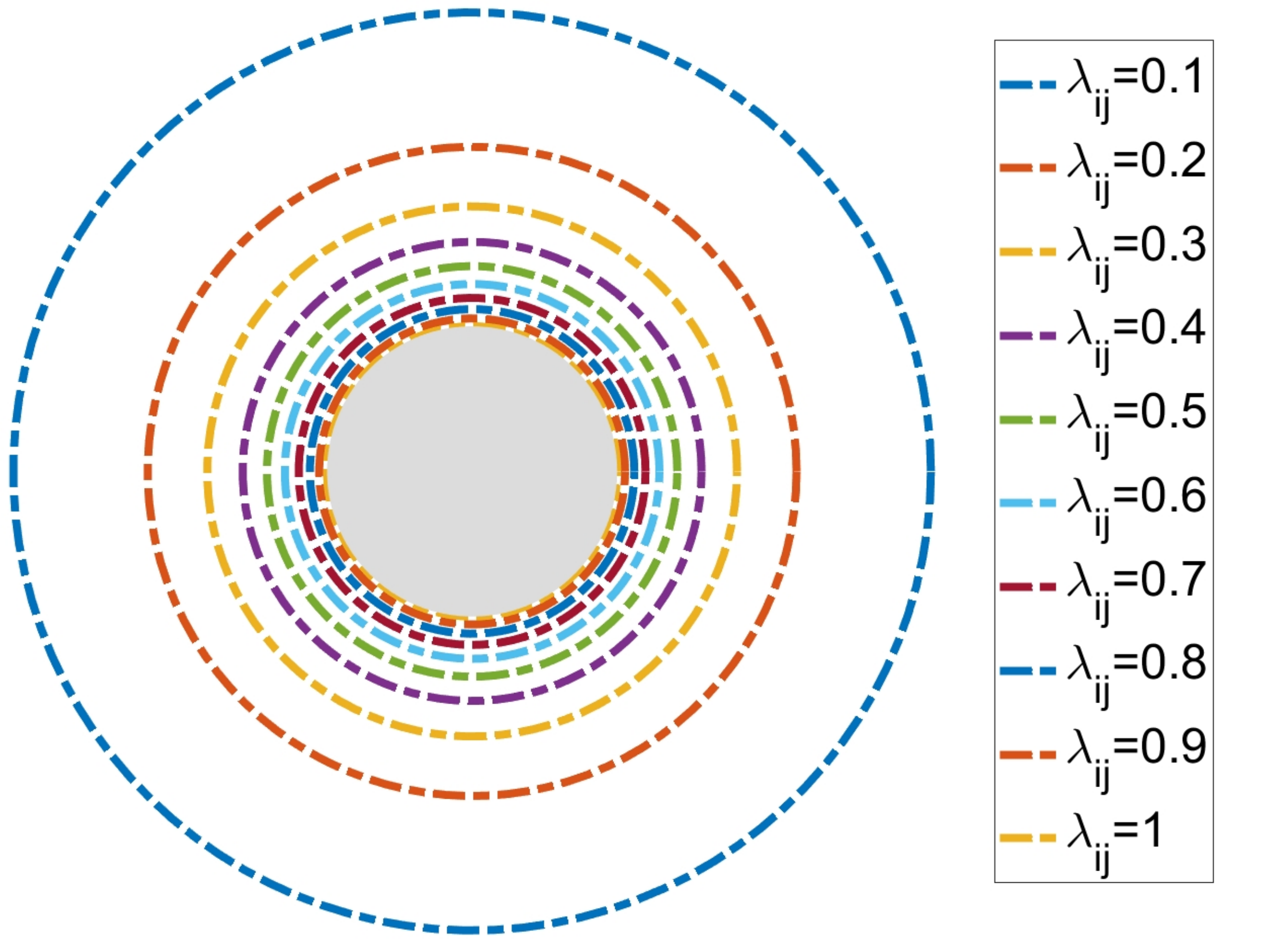}
  \caption{\justifying Influence of different $\lambda_{i j}$ on the admissible control space $\mathcal{S}(\mathbf{x})$. When $\lambda_{ij}=1$, the predictive term is not considered in the safe constraint, and the circle surrounded by it, \emph{i.e}, the gray region, is a dangerous area that a robot is prohibited from entering. Following it, it is observed that the smaller the $\lambda_{i j}$, the greater the shrunken $\mathcal{S}(\mathbf{x})$.}\label{fig0}
\end{figure}
\vspace{-10pt}
\subsection{Predictive Safety Condition}
The motion for the $i$-th robot with the generalized coordinates $\mathcal{P}_i =[x_i,y_i,\theta_i]^\text{T}$ is described as
\begin{equation}\label{eqn.k}
\left[\begin{array}{c}
\dot x_i \vspace{1ex}\\
\dot y_i \vspace{1ex}\\
\dot \theta_i \vspace{1ex}
\end{array}\right]=\left[\begin{array}{cc}
\cos\theta_i&0\\
\sin\theta_i&0\\
0&1
\end{array}\right]
\left[\begin{array}{c}
v_i \vspace{1ex}\\
w_i
\end{array}\right]
\end{equation}
where $\mathbf{p}_i=[x_i, y_i]^\text{T}$ and $\theta_i$ are the position and orientation of the $i$-th robot, respectively.  $v_i$ and $w_i$ are the translational and angular velocities of the $i$-th robot, respectively. $[\dot x_i, \dot y_i, \dot \theta_i]^\text{T}$ is the time derivative of $[x_i, y_i, \theta_i]^\text{T}$.

Inspired by \cite{shahriari2021novel}, assume that the displacements of the pairwise robots $i, j$ after a short period of time ($\tau$) are $\delta_i$ and $\delta_j$, the predictive safety condition can be quantified as the following lemma.

{\it{Lemma 1}}: The predictive safety condition between the pairwise robots $i, j$ can be expressed as
\begin{equation}\label{eqn.1}
\lambda_{i j}|| \mathbf{p}_{i j}||^2 \geq \rho_{i j}^2
\end{equation}
where the safety threshold $\rho_{i j}=r_i+r_j$, the relative position $\mathbf{p}_{ij}=\mathbf{p}_{i}-\mathbf{p}_{j}$.
$\lambda_{i j}$ denotes the prediction window, $|| \mathbf{p}_{i j}||$ is the Euclidean distance.

\text{Proof:}
Assumes that the robot continue to walk with the same velocity and orientation as the current timestep. Therefore, the positions of the robots $i, j$ after $\tau$ are:
\begin{equation}\label{eqn.3}
\begin{split}
&\mathbf{p}_i^{\prime}=\mathbf{p}_i+\left[\cos(\theta_i)~ \sin(\theta_i)\right]^{\mathrm{T}} \delta_i\\
&\mathbf{p}_j^{\prime}=\mathbf{p}_j+\left[\cos(\theta_j)~ \sin(\theta_j)\right]^{\mathrm{T}} \delta_j
\end{split}
\end{equation}
To ensure the collision-free, we need to enable
$||\mathbf{p}_i^{\prime}-\mathbf{p}_j^{\prime}|| \geq \rho_{ij}$. We have
\begin{equation}
\begin{aligned}
& \left(x_i-x_j+\delta_i \cos \theta_i-\delta_j \cos \theta_j\right)^2 \\
& +\left(y_i-y_j+\delta_i \sin \theta_i-\delta_j \sin \theta_j\right)^2 \geq \rho_{i j}^2
\end{aligned}
\end{equation}
Further, we can obtain
\begin{equation}\label{eqn.4}
\begin{aligned}
& \left(-|| \mathbf{p}_{i j}|| \cos \beta_{i j}+\delta_i \cos \theta_i-\delta_j \cos \theta_j\right)^2 \\
& \quad+\left(-|| \mathbf{p}_{i j}||\sin \beta_{i j}+\delta_i \sin \theta_i-\delta_j \sin \theta_j\right)^2 \geq \rho_{i j}^2
\end{aligned}
\end{equation}
where $\beta_{ij}$ is the bearing angle,  $\beta_{ij}=\tan^{-1}(\mathbf{p}_i-\mathbf{p}_j)$. Eq. (\ref{eqn.4}) is equivalent to
\begin{equation}\label{eqn.5}
\begin{aligned}
& || \mathbf{p}_{i j}||^2+\delta_i^2+\delta_j^2-2 || \mathbf{p}_{i j}|| \delta_i \cos \left(\beta_{i j}-\theta_i\right) \\
& \quad-2 \delta_{i} \delta_j \cos \left(\theta_i-\theta_j\right)+2 || \mathbf{p}_{i j}|| \delta_j \cos \left(\beta_{i j}-\theta_j\right) \geq \rho_{i j}^2
\end{aligned}
\end{equation}
Letting $\boldsymbol{D}_{i j}=[\delta_{i},\delta_{j},|| \mathbf{p}_{i j}||]^\mathrm{T}$, Eq. (\ref{eqn.4}) is rewritten as
\begin{equation}\label{eqn.6}
\boldsymbol{D}_{i j}^\mathrm{T}\boldsymbol{S}_{i j}\boldsymbol{D}_{i j}\geq \rho_{i j}^2
\end{equation}
where
\begin{equation}
\boldsymbol{S}_{i j}=\left[\begin{array}{ccc}
1 & -\cos \left(\theta_{ij}\right) & -\cos \left(\varrho_i\right) \\
-\cos \left(\theta_{ij}\right) & 1 & \cos \left(\varrho_j\right) \\
-\cos \left(\varrho_i\right) & \cos \left(\varrho_j\right) & 1
\end{array}\right]
\end{equation}
where $\theta_{ij}=\theta_i-\theta_j$, $\varrho_i=\beta_{i j}-\theta_i$, $\varrho_j=\beta_{i j}-\theta_j$. For a symmetric matrix $\boldsymbol{S}_{i j}$, it is known that
\begin{equation}\label{eqn.7}
\boldsymbol{D}_{i j}^\mathrm{T} \boldsymbol{S}_{i j} \boldsymbol{D}_{i j} \geq \lambda_{i j}\left\|\boldsymbol{D}_{i j}\right\|^2
\end{equation}
where $\lambda_{ij}$ is the smallest eigenvalue of $\boldsymbol{S}_{i j}$. Due to
$\left\|\boldsymbol{D}_{i j}\right\|^2=\delta_{i}^2+\delta_{j}^2+|| \mathbf{p}_{i j}||^2 \geq || \mathbf{p}_{i j}||^2$, we have
\begin{equation}\label{eqn.8}
\lambda_{i j}\left\|\boldsymbol{D}_{i j}\right\|^2 \geq \lambda_{i j} || \mathbf{p}_{i j}||^2
\end{equation}
if $\lambda_{i j}\geq 0$. Therefore, if $\lambda_{i j} || \mathbf{p}_{i j}||^2\geq \rho_{ij}^2$, Eq. (\ref{eqn.6}) holds.
$\hfill\blacksquare$

In this subsection, a predictive safety matrix is first constructed. We then consolidate the safety condition based on the proposed safety matrix’s eigenvalue. Unlike \cite{shahriari2021novel}, we trimmed the differential calculations related to $\lambda_{ij}$, $\lambda_{ij}$ is set as a constant with a value range of $(0,1]$. The advantage of doing so is immune to oscillations and the CBF-based safety control method can be applied directly without the need for designing specialized control laws.
 Fig. \ref{fig0} shows influence of different $\lambda_{i j}$ on the admissible control space $\mathcal{S}(\mathbf{x})$.
When $\lambda_{i j}=1$, the predictive term is not considered in the safe constraint, and the circle surrounded by it is a dangerous area that a robot is prohibited from entering. Following Fig. \ref{fig0}, it is observed that the smaller the $\lambda_{i j}$, the greater the shrunken $\mathcal{S}(\mathbf{x})$. Therefore, in this paper, $\lambda_{ij}=0.5$ is chosen to make a trade-off.
\section{Method}
Based on the feedback linearization technology \cite{TCYB2022Li}, Eq. (\ref{eqn.k}) is further written as
 \begin{equation}\label{eqn.kinematic}
\left[\begin{array}{c}
\dot{\mathcal{P}}_i \vspace{1ex}\\
\dot{\mathcal{V}}_i
\end{array}\right]=\left[\begin{array}{c}
\mathbf{\Gamma}_i\vspace{1ex}\\
\dot{\mathbf{\Gamma}}_i
\end{array}\right]
\mathbf{u}_i +\left[\begin{array}{c}
0 \vspace{1ex}\\
\mathbf{\Gamma}_i
\end{array}\right] \dot{\mathbf{u}}_i
\end{equation}
for the robot $i$, $i= {1, 2, ..., N }$, where $\mathbf{\Gamma}_i=[\mathbf{A}_i,\mathbf{\Lambda}_i]^\mathrm{T}$, $\mathbf{u}_i$ and $\mathbf{\dot u}_i$ are Jacobian matrix, wheel velocity and the control input of the robot $i$, respectively.
$\dot{\mathcal{V}}_i=[\mathbf{\dot p}_i,\dot \theta_i]^\text{T}$, and
$\mathbf{\dot p}_i=\mathbf{A}_i\mathbf{u}_i$ are the velocity of the robot $i$.
Assume that the robot $i$ has a circular collision region centred at $\mathbf{p}_i$ with the radius $r_i$. The pairwise robots $i,j$ would be collision-free if $\lambda_{ij}||\mathbf{p}_{ij}||^2\geq \rho_{ij}^2$ based on Eq. (\ref{eqn.1}), where $\rho_{ij}=r_i+r_j$, and $\mathbf{p}_{j}$, $r_j$ are the position and safety radius of the robot $j$, respectively. $N$ denotes the number of robots in a team. For a differential-driven robot,
\begin{equation}
\mathbf{A}_i=\left[\begin{array}{ll}
\frac{r_i\cos \theta_i}{2}+\frac{r_id_i \sin \theta_i}{L_i} & \frac{r_i \cos \theta}{2}-\frac{r_i d_i \sin \theta_i}{L_i} \\
\frac{r_i\sin \theta_i}{2}-\frac{r_i d_i \cos \theta_i}{L_i} & \frac{r_i\sin \theta}{2}+\frac{r_i d_i \cos \theta_i}{L_i}\\
\end{array}\right]
\end{equation}
$\mathbf{\Lambda}_i=[-r_i/L_i,r_i/L_i]^\text{T}$, where the explanation about $r_i, L_i, d_i$ is given in \cite{TCYB2022Li}.
\vspace{-10pt}
\subsection{Predictive Collision Avoidance}
{\it{Lemma 2}}: The predictive collision-free motion constraint between the neighboring pairwise robots $i, j$ is
\begin{equation}\label{eqn.9}
\begin{aligned}
-2\lambda_{ij}\mathbf{p}_{i j}^\mathrm{T}C\mathbf{A}_{\text{blk}} \mathbf {\dot u}\leq
&2\lambda_{ij}(||\mathbf{v}_{i j}||^2+\mathbf{p}_{i j}^\mathrm{T}C\mathbf{\dot A}_{\text{blk}} \mathbf{u}\\
&+(\kappa_{ij_1}+\kappa_{ij_2})\mathbf{p}_{i j}^\mathrm{T}\mathbf{v}_{i j})\\
&+\kappa_{ij_1}\kappa_{ij_2}(\lambda_{ij}||\mathbf{p}_{i j}||^2-\rho_{ij}^2)\\
 &\forall i,j\in N, ~i \neq j, ~j\in \mathscr{N}_i
\end{aligned}
\end{equation}
where $C=[0,\cdots,\overbrace{I}^{\text{robot}~ i},\cdots,\overbrace{-I}^{\text{robot}~j},\cdots,0]$, $\mathbf{A}_{\text{blk}}$ is a diagonal matrix which diagonally places $[\mathbf{A}_1,\cdots,\mathbf{A}_i,\cdots,$
$\mathbf{A}_j,\cdots,\mathbf{A}_N]$, $\mathbf{u}=[\mathbf{u}_1,\cdots$,
$\mathbf{u}_i,\cdots,\mathbf{u}_j,\cdots,\mathbf{u}_N]^\text{T}$,
the control parameter $\kappa_{ij_1}, \kappa_{ij_2}>0$.

\text{Proof:}
By a double integrator, the pairwise safe set $\mathcal{C}_{ij}$ for the robots $i, j$ is defined as:
\begin{equation}\label{eqn.10}
h_{ij}\geq 0,~~
h_{ij}=2\lambda_{ij}\mathbf{p}_{i j}^\mathrm{T}\mathbf{v}_{i j}+\kappa_{ij_1}(\lambda_{ij}\left\|{\mathbf{p}}_{ij}\right\|^2-\rho_{ij}^2)
\end{equation}
Based on the ZCBF, we have
\begin{equation}\label{eqn.11}
\dot h_{ij}+\kappa_{ij_2}h_{ij}\geq 0
\end{equation}
The extended class-$\mathscr{K}$ function $\kappa_{ij_2}h_{ij}$ is used to regulate the rate of $h_{ij}\geq 0$.
Due to
\begin{equation}
\dot h_{ij}=2\lambda_{ij}(||\mathbf{v}_{ij}||^2+\mathbf{p}_{ij}^\mathrm{T}\mathbf{u}_{ij}+\kappa_{ij_1}\mathbf{p}_{ij}^\mathrm{T}\mathbf{v}_{ij})
\end{equation}
substituted into Eq. (\ref{eqn.11}), the collision-free constraint Eq. (\ref{eqn.9}) is obtained.
$\hfill\blacksquare$

{\it{Theorem 1}}: Given the pairwise safe set defined by Eq. (\ref{eqn.10}),
the extended class-$\mathscr{K}$ function $\kappa_{ij_2}h_{ij}$, and the initial state $h_{ij}(0)=(\mathbf{p}(0), \mathbf{v}(0))\in h_{ij}$,
$\mathbf{p}(0)=[\mathbf{p}_i(0),\mathbf{p}_j(0)]^\text{T}$, $\mathbf{v}(0)=[\mathbf{v}_i(0),\mathbf{v}_j(0)]^\text{T}$,
 the pairwise safe set $h_{ij}\geq 0$ is forward invariant.

\text{Proof:}
Given
\begin{equation}
h_{ij}=2\lambda_{ij}\mathbf{p}_{i j}^\mathrm{T}\mathbf{v}_{i j}+\kappa_{ij_1}(\lambda_{ij}\left\|{\mathbf{p}}_{ij}\right\|^2-\rho_{ij}^2)
\end{equation}
\begin{equation}
\dot h_{ij}\geq -\kappa_{ij_2}h_{ij}
\end{equation}
 Construct $S_{ij}=h_{ij}(0)e^{-\kappa_{ij_2}t}$ with $S_{ij}(0)=h_{ij}(0)$, where $h_{ij}(0)$ denotes the initial value of $h_{ij}$. The time derivative of $S_{ij}$ is $\dot S_{ij}=-\kappa_{ij_2}h_{ij}(0)e^{-\kappa_{ij_2}t}=-\kappa_{ij_2}S_{ij}$.
 Based on the comparison principle, $h_{ij}\geq S_{ij}$ for all $t\geq 0$. Therefore,
 \begin{equation}
 h_{ij}\geq h_{ij}(0)e^{-\kappa_{ij_2}t}
\end{equation}
Considering $h_{ij}(0)\geq 0$, and $h_{ij}(0)=0$ only if $h_{ij}(0)=0$, therefore
 \begin{equation}
 h_{ij}\geq 0~~\text{for}~~\text{all}~~t\geq 0
\end{equation}
The forward invariance of the safety set $\mathcal{C}_{ij}$ is guaranteed.
$\hfill\blacksquare$

Notice that for a multi-robot system, the safe set can be defined as the intersection of all possible safe set. Therefore, if all possible
pairwise safety constraints are satisfied and their initial states are within
the safe set, the multi-robot system is guaranteed to be safe.
\begin{figure}
  \centering
  \includegraphics[width=3.25in]{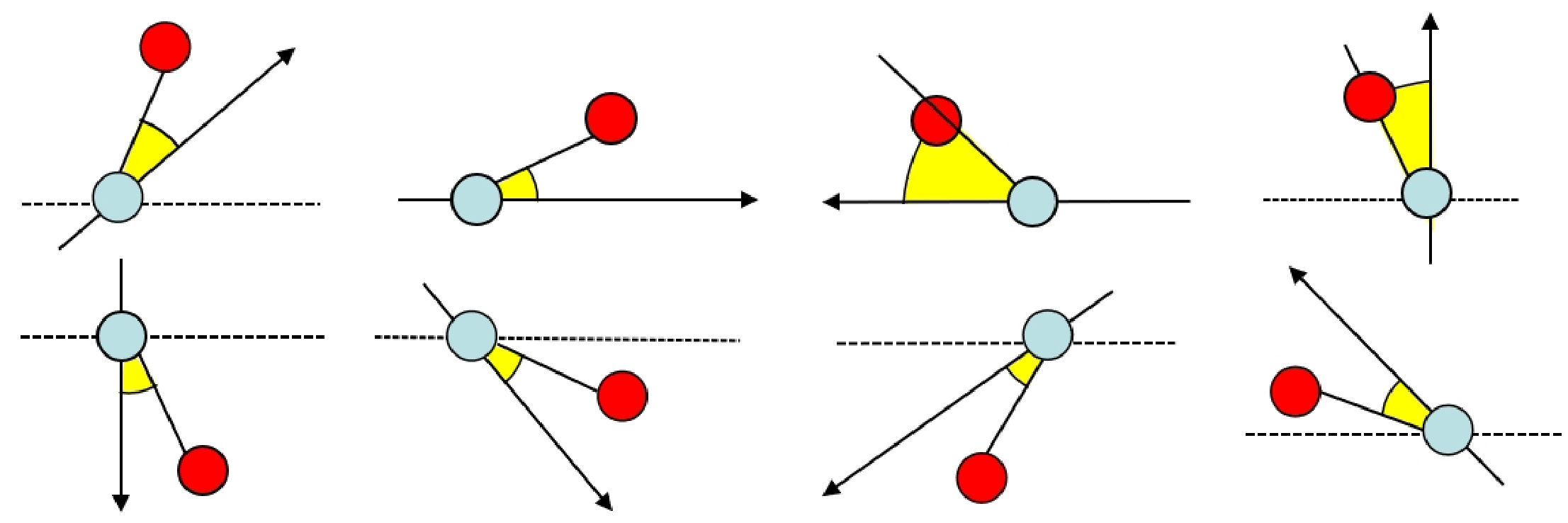}
  \caption{\justifying Considering the possible eight move direction for a WMR which is colored by cyan, we can find that the red obstacle is either located in left (positive) side or right (negative) side of the WMR's movement direction in local coordination frame.}\label{g}
\end{figure}
\vspace{-8pt}
\subsection{Deadlock Detection and Deconfliction Direction Decision}
\vspace{-2pt}
In our previous work \cite{TITs2024}, the deadlock escape strategy attached as an auxiliary velocity vector is embedded into the TT module, which has been proven that the introduction of it has no influence of  the global convergence of TT error and it only changes the evolution mode of error function.
However, the escape direction is determined manually. Considering the possible eight move direction for a robot which is colored by cyan, see Fig. \ref{g}, we can find that the red obstacle is either located in left (positive) side or right (negative) side of the robot's movement direction in local coordination frame. The robot should turn with a small attitude for CA. Therefore,
inspired by it, the deconfliction direction is established based on the obstacles’ bearing angles.

Specifically, by a double integrator, TT strategy in this paper can be described as:
\begin{equation}\label{eqn.tt}
\mathcal{\ddot P}_{i}-\mathcal{\ddot P}_{d_i}+\kappa_{i_3}(\mathcal{\dot P}_{i}-\mathcal{\dot P}_{d_i})=-\kappa_{i_4}
(\mathcal{\dot P}_{i}-\mathcal{\dot P}_{d_i}+\kappa_{i_3}\xi_i),
\end{equation}
where $\xi_i=\mathcal{P}_{i}-\mathcal{P}_{d_i}$ denotes the error between the reference posture and the actual posture of the robot $i$, and $\kappa_{i_3},\kappa_{i_4}$ is used to regulate the tracking accuracy of $\mathcal{P}_{i}\rightarrow\mathcal{P}_{d_i}$.
$\mathcal{P}_{d_i}, \mathcal{\dot P}_{d_i}$ and $\mathcal{\ddot P}_{d_i}$ denote the desired posture, velocity and acceleration of the robot $i$.
To break the deadlock, an auxiliary velocity vector $-\zeta_i\mathscr{Q}_i(\mathcal{\dot P}_{i}-\mathcal{\dot P}_{d_i}+\kappa_{i_3}\xi_i)$ is added to Eq. (\ref{eqn.tt}).
Combined with Eq. (\ref{eqn.kinematic}), we have
\begin{equation}\label{eqn.position}
\begin{aligned}
\mathbf{\Gamma}_{i}\mathbf{\dot u}_{i}&=
\mathbf{\hat{a}}_{d_i}-\mathbf{\dot \Gamma}_{i}\mathbf{u}_{i}-(\kappa_{i_3}+\kappa_{i_4})(\mathbf{\Gamma}_{i}\mathbf{u}_{i}-\mathcal{V}_{d_i})\\
&-\kappa_{i_3}\kappa_{i_4}\xi_i-\zeta_i\mathscr{Q}_i(\mathbf{\Gamma}_{i}\mathbf{u}_{i}-\mathcal{V}_{d_i}+\kappa_{i_3}\xi_i)
\end{aligned}
\end{equation}
where $\mathbf{\hat{a}}_{d_i}=[\mathbf{a}_{d_i}, \ddot \theta_d]^\text{T}$, $\mathbf{a}_{d_i}$ is the preferred acceleration of the wheel.
The matrix $\mathscr{Q}_i$ is defined as
\begin{equation}\label{eqn.14a}
\begin{split}
&\mathscr{Q}_i=\begin{bmatrix}
\cos(g_iq_i)&-\sin(g_iq_i)&0\\
\sin(g_iq_i)&\cos(g_iq_i)&0\\
0&0&0
\end{bmatrix}\in \mathbb{R}^{3\times 3}
\end{split}
\end{equation}
$q_i\in[0^{\circ},180^{\circ}]$, where the sign of $g_i$ is used to determine the deadlock escape direction,
\begin{equation}
\left\{
\begin{aligned}
&g_{i}=-1~~  \text{if}~\operatorname{sgn}\left(\mathrm{min}(\theta_{{r_i}{o}})\right)\geq 0~~\\
&g_{i}=1~~~~\text{others}
\end{aligned}
\right.
\end{equation}
$\theta_{{r_i}{o}}=[\theta_{{r_i}{o_1}},\cdots,\theta_{{r_i}{o_j}}]$, $j\in \mathscr{N}_i$.
$\theta_{{r_i}{o_j}}$ denotes the bearing angle between the $i$-robot and its neighboring $j$-th obstacle,
\begin{equation}
\theta_{{r_i}{o_j}}=\operatorname{atan} 2\left(-y_{{r_i}{o_j}},-x_{{r_i}{o_j}}\right) .
\end{equation}
where $y_{{r_i}{o_j}}=y_{i}-y_{o_j}$, $x_{{r_i}{o_j}}=x_{i}-x_{o_j}$, $x_{o_j}$ and $y_{o_j}$ are the $x$-axis and $y$-axis coordinates of the $j$-th obstacle in the local coordinate frame, respectively.
$\zeta_i$ is used to regulate the intensity deviating from the desired trajectory,
\begin{equation}\label{eqn.15}
\left\{
\begin{aligned}
&\zeta_i>0~~~~  \text{if}~\eta_{\in i}>0~~\\
&\zeta_i=0~~~~\text{others}
\end{aligned}
\right.
\end{equation}
usually valued as 2 if $\eta_{\in i}>0$. In our approach, the pairwise robots avoid collisions by deviating from their respective perferred trajectories, and the deadlock escape strategy is also based on the same principle. Therefore, in this paper, we set the deadlock detection condition to be activated when the CA strategy of the robots is triggered, \emph{i.e}, $\eta_{\in i}>0$, where $\eta$ is Lagrange multiplier. This looking-ahead decision-making condition  reduces stagnation compared to previous methods \cite{SBC3,TAChe2024,RAL2019}.

{\it{Theorem 2}}: Trajectory tracking error of the robot $i$ that subjects to Eq. (\ref{eqn.position}) globally converges to zero when $t\rightarrow\infty$.

\text{Proof:}
Define
\begin{equation}\label{eqn.a}
\begin{aligned}
&e_i=\mathcal{\dot P}_{i}-\mathcal{\dot P}_{d_i}+\kappa_{i_3}(\mathcal{P}_{i}-\mathcal{P}_{d_i})\\
&\dot e_i=-\kappa_{i_4}e_i-\zeta_i\mathscr{Q}_ie_i
\end{aligned}
\end{equation}
Construct a Lyapunov function $V_{i_1}=e_i^\text{T}e_i/2$. Obviously, $V_{i_1}\geq0$, and $V_{i_1}=0$ only when $e_i=0$.
Taking the time derivation of $V_{i_1}$ and substituting by (\ref{eqn.a}) yields
\begin{equation}
\dot V_{i_1}=-(\kappa_{i_4}+\zeta_i\mathscr{Q}_i)e_i^2
\end{equation}
To ensure convergence, we need to ensure $\kappa_{i_4}+\zeta_i\mathscr{Q}_i>0$, \emph{i.e}, $\kappa_{i_4}>-\zeta_i\mathscr{Q}_i$. Due to $\zeta_i\geq0$,
\begin{equation}
\mathscr{Q}_i=\left[\begin{matrix}
\sin(g_iq_i)&-\cos(g_iq_i)\\
\cos(g_iq_i)&\sin(g_iq_i)
\end{matrix}\right],
\end{equation}
$g_iq_i\in[-180^{\circ},180^{\circ}]$, $e_i$ converges to zero when $t\rightarrow \infty$, provided that $\kappa_{i_4}>\zeta_i$. Define
\begin{equation}\label{eqn.a2}
\begin{aligned}
&\xi_i=\mathcal{P}_{i}-\mathcal{P}_{d_i}\\
&\dot \xi_i=-\kappa_{i_3}\xi_i
\end{aligned}
\end{equation}
Construct a Lyapunov function $V_{i_2}=\xi_i^\text{T}\xi_i/2$. Obviously, $V_{i_2}\geq0$, and $V_{i_2}=0$ only when $\xi_i=0$.
Taking the time derivation of $V_{i_2}$ and substituting by (\ref{eqn.a2}) yields
\begin{equation}
\dot V_{i_2}=-\kappa_{i_3}\xi_i^2
\end{equation}
Because $\kappa_{i_3}>0$, $\dot V_{i_2}\leq0$, and $\dot V_{i_2}=0$ only when $\xi_i=0$. Using LaSalle’s invariant principle, we can conclude that $\xi_i$ globally converges to zero as when $t\rightarrow \infty$.
$\hfill\blacksquare$

For a multi-robot system, globally convergence of every robot's TT error is guaranteed by $\kappa_{i_4}>\max{(\zeta)}$, $\zeta=[\zeta_1,\cdots,\zeta_i,\cdots,\zeta_N]^\text{T}$.
In this paper, we set same $\zeta_i$, $\kappa_{i_4}$ and $\kappa_{i_3}$ for every robot. Therefore, for the multi-robot system, $\mathcal{\dot P}-\mathcal{\dot P}_d$ converges to zero when $t\rightarrow \infty$, provided that $\kappa_{i_4}>\zeta_i$,
$\mathcal{\dot P}=[\mathcal{\dot P}_1,\cdots,\mathcal{\dot P}_i,\cdots,\mathcal{\dot P}_N]^\text{T}$,
$\mathcal{\dot P}_d=[\mathcal{\dot P}_{d_1},\cdots,\mathcal{\dot P}_{d_i},\cdots,\mathcal{\dot P}_{d_N}]^\text{T}$.
\vspace{-10pt}
\subsection{QP Controller}
 The right side of Eq. (\ref{eqn.position}) and Eq. (\ref{eqn.9}) are abbreviated as $\text{dzr}_i$ and $B_{\text{right}_{ij}}$, respectively, and the left side of Eq. (\ref{eqn.9}) is abbreviated as$B_{ij}$, we formulate the balance of TT and CA as a minimization of a QP:
\begin{subequations}\label{eqn.qp}
\begin{align}
&\mathop{\text{min}}\limits_{\mathbf{\dot u_i}}~~~(\mathbf{\Gamma}_i\mathbf{\dot u}_i-\text{dzr}_i)^\text{T}(\mathbf{\Gamma}_i\mathbf{\dot u}_i-\text{dzr}_i)\label{equ.16a}\\
&~\text{s.t.}~~~~~~~~~\mathbf{\dot u_i}^{-}\le \mathbf{\dot u_i}\le \mathbf{\dot u_i}^{+}~~~~~~~~\label{equ.16c}\\
&~~~~~~~~~~~B_{ij}\mathbf{\dot u}\leq B_{\text{right}_{ij}}
, ~\forall i,j\in N, ~i \neq j, ~j\in \mathscr{N}_i\label{equ.16d}
\end{align}
\end{subequations}
Eq. (\ref{eqn.qp}) is solved using the Lagrange multipliers method. The details consult \cite{TCYB2022Li,TITs2024}, which is omitted due to similarity.

\begin{figure}[htb]
  \centering
  \setlength{\abovecaptionskip}{-0.02cm}   
  \includegraphics[scale=0.35]{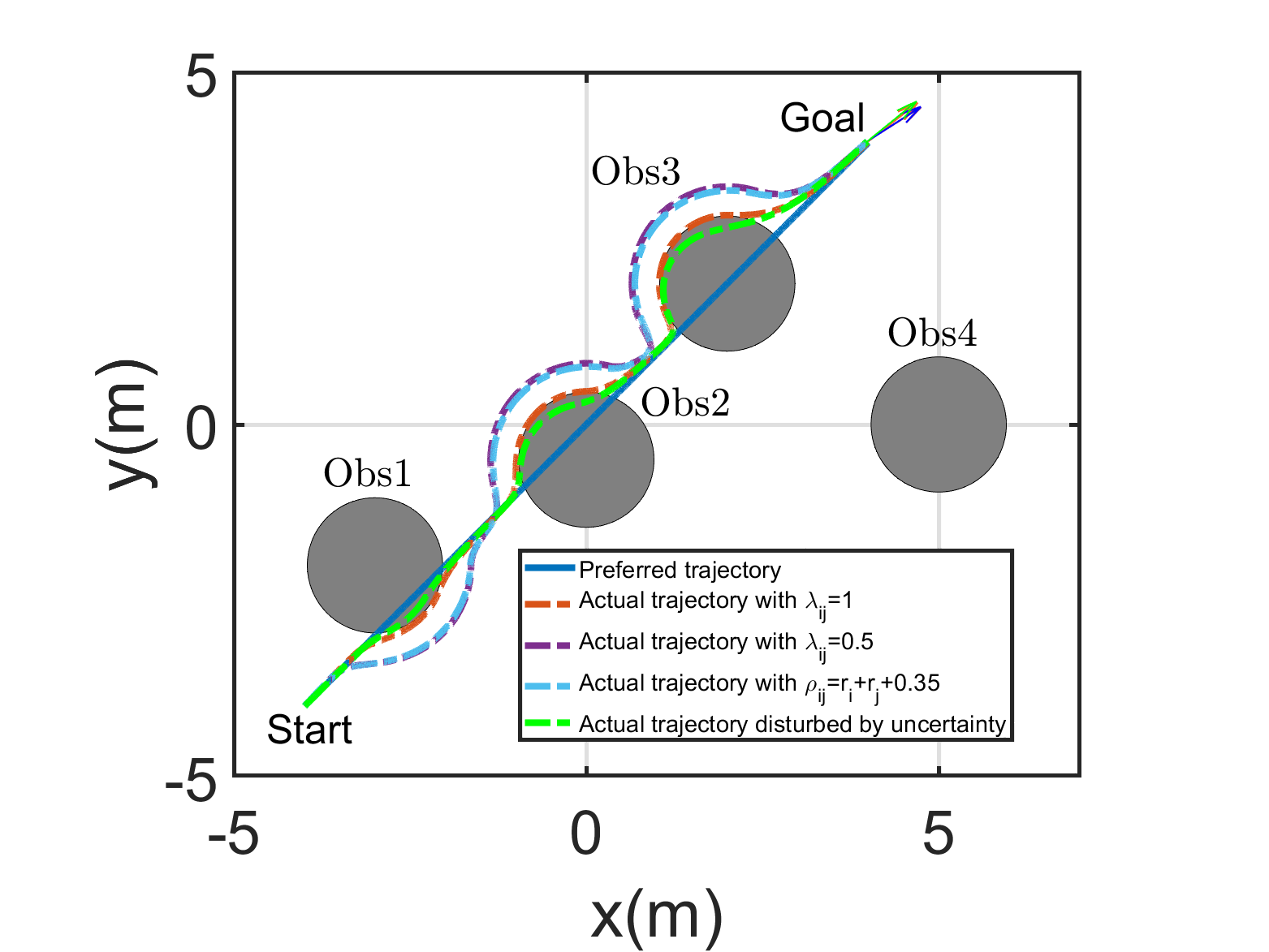}
  \caption{\justifying Example I: SOATT result of single robot with multiple static obstacles under our method.}\label{h}
\end{figure}

\begin{figure}
  \centering
  \setlength{\abovecaptionskip}{-0.005cm}   
  \hspace{20pt}
  \includegraphics[scale=0.24]{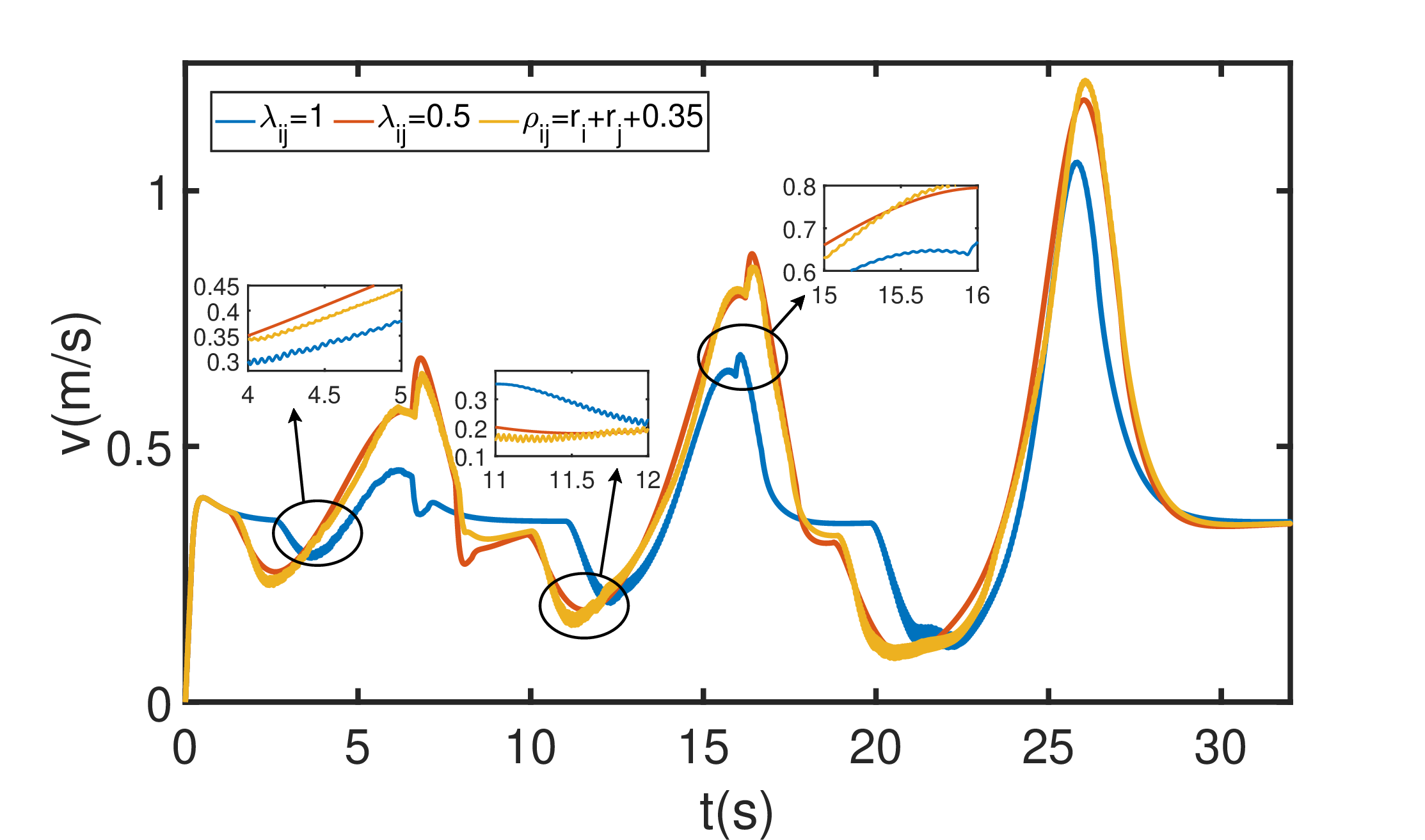}
  \caption{Velocity profiles corresponding to Fig. \ref{h}.}\label{h_v}
\end{figure}

 \begin{figure*}
  \centering
\setlength{\abovecaptionskip}{-0.1cm}   
\begin{minipage}[]{7.5in}
\subfigcapskip=-4pt
\hspace{-25pt}
\subfigure[]{\includegraphics[width=2.1in]{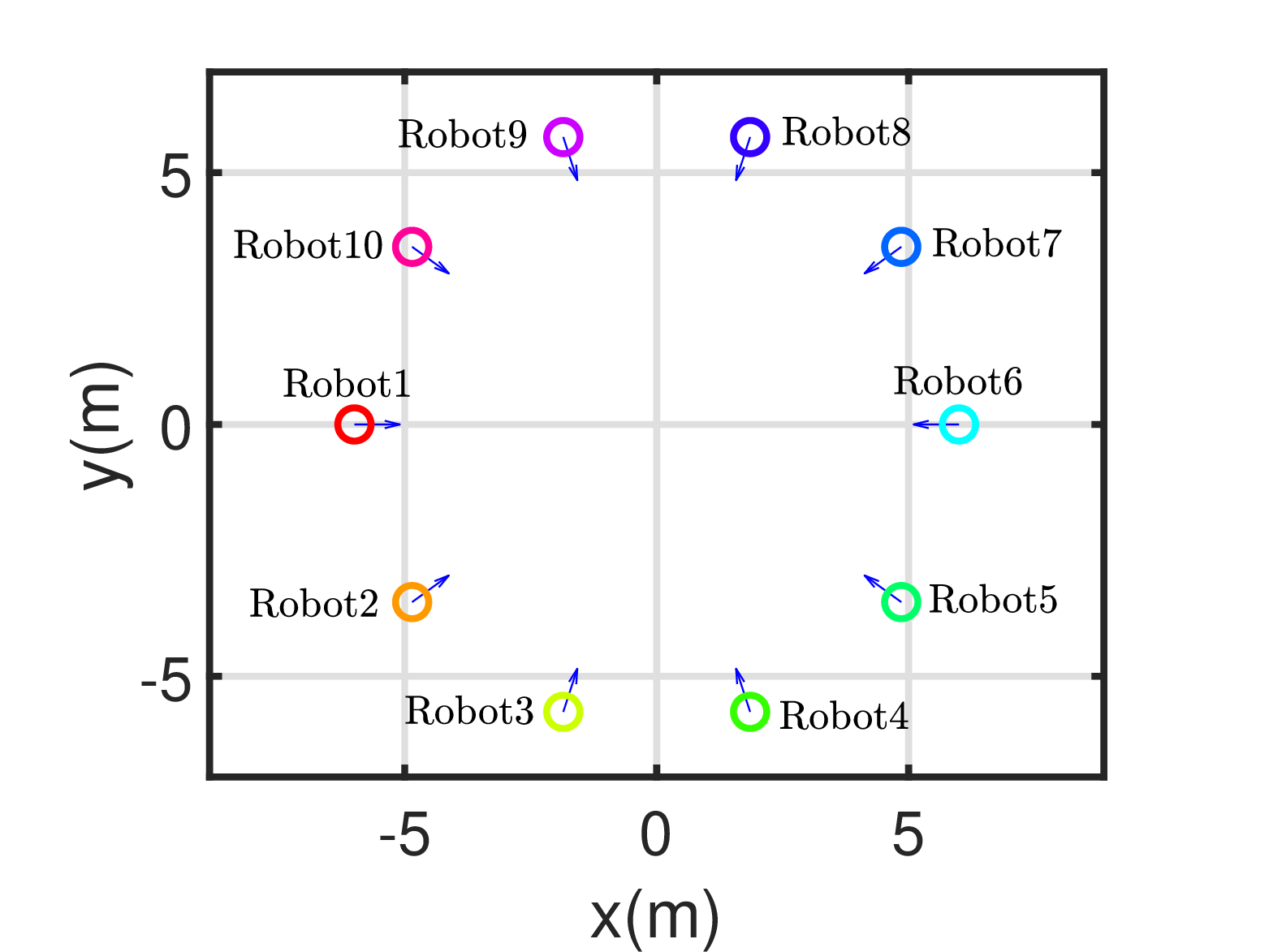}}
\hspace{-25pt}
\subfigure[]{\includegraphics[width=2.1in]{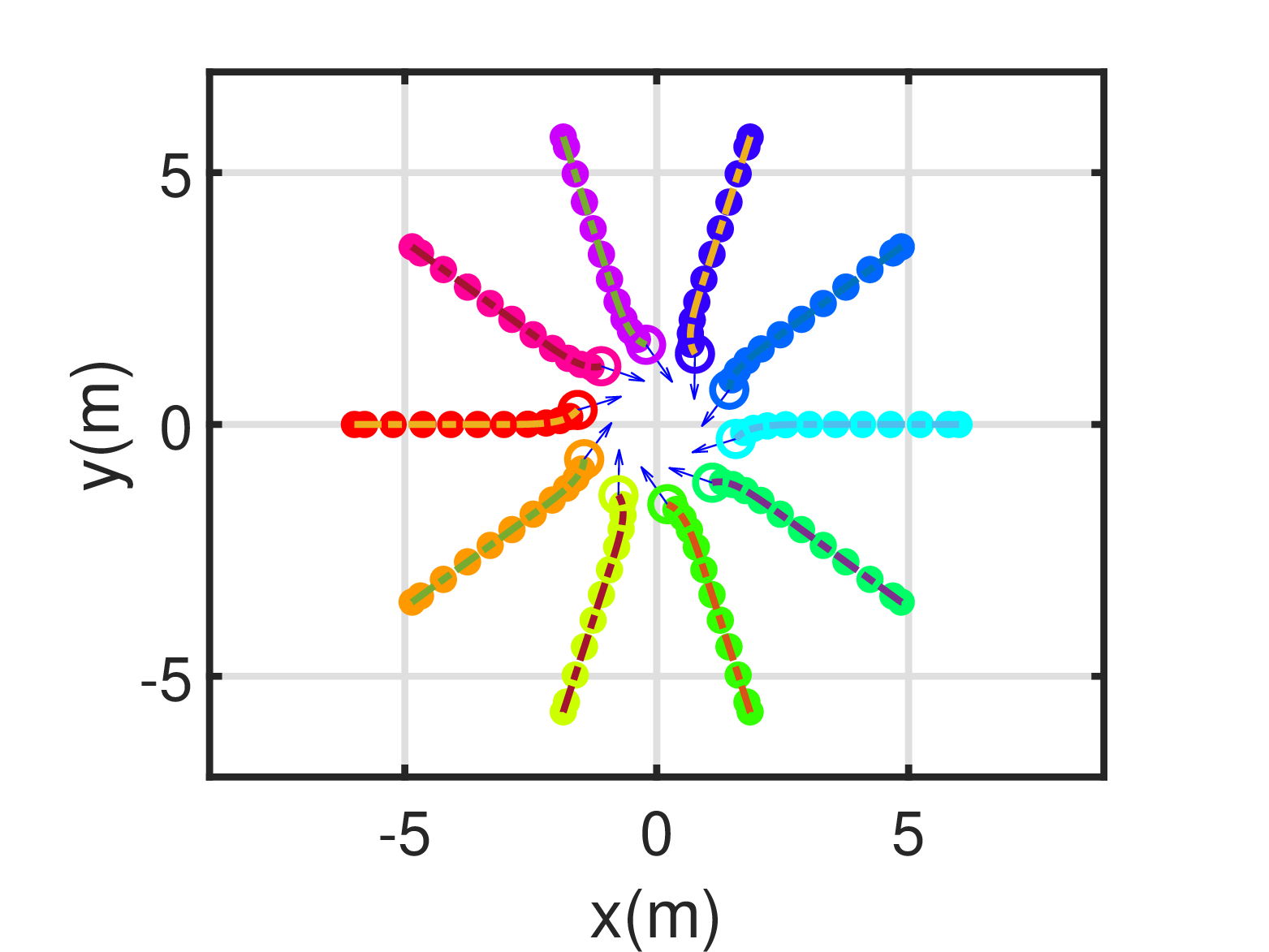}}
\hspace{-25pt}
\subfigure[]{\includegraphics[width=2.1in]{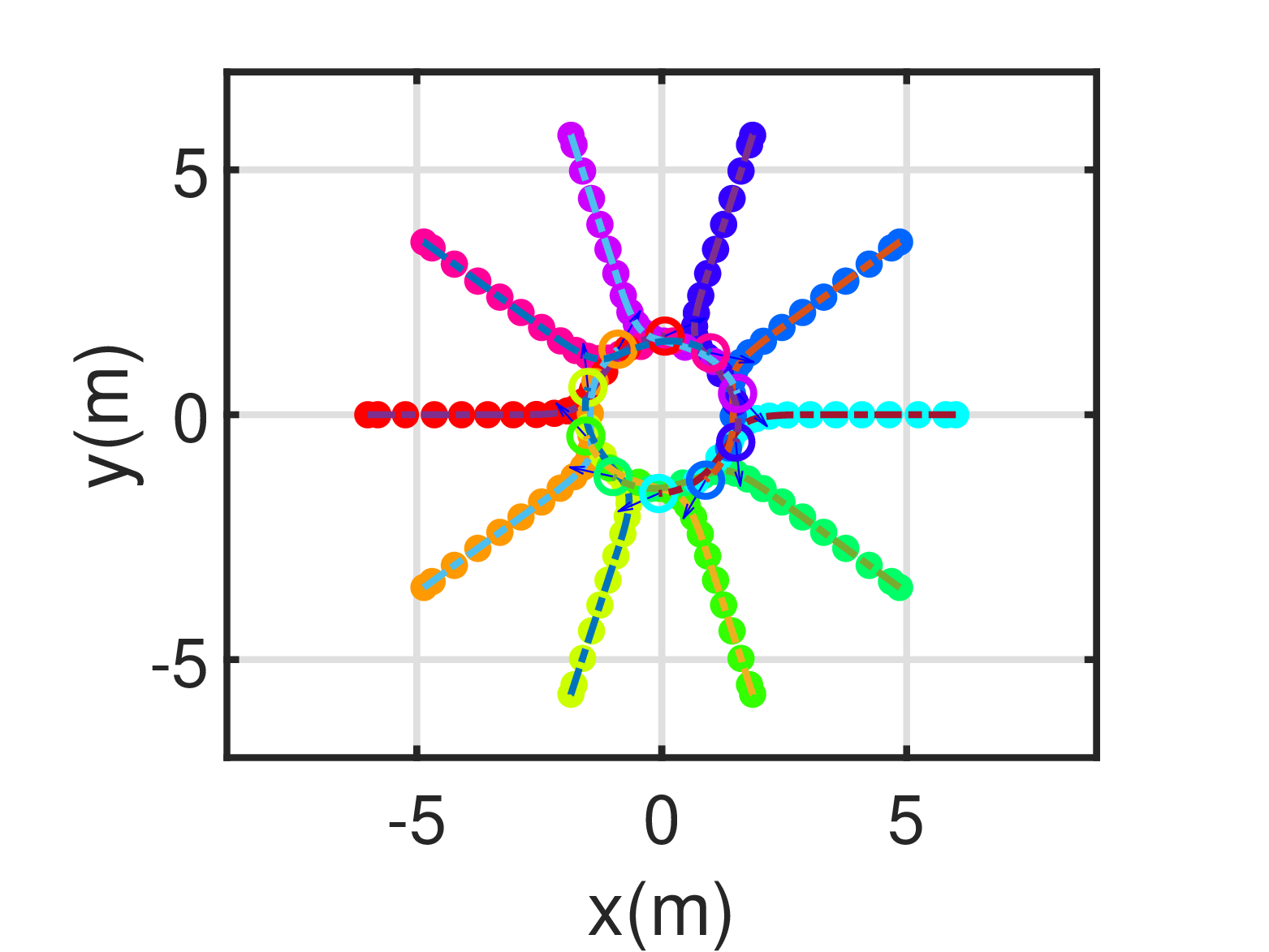}}
\hspace{-25pt}
\subfigure[]{\includegraphics[width=2.1in]{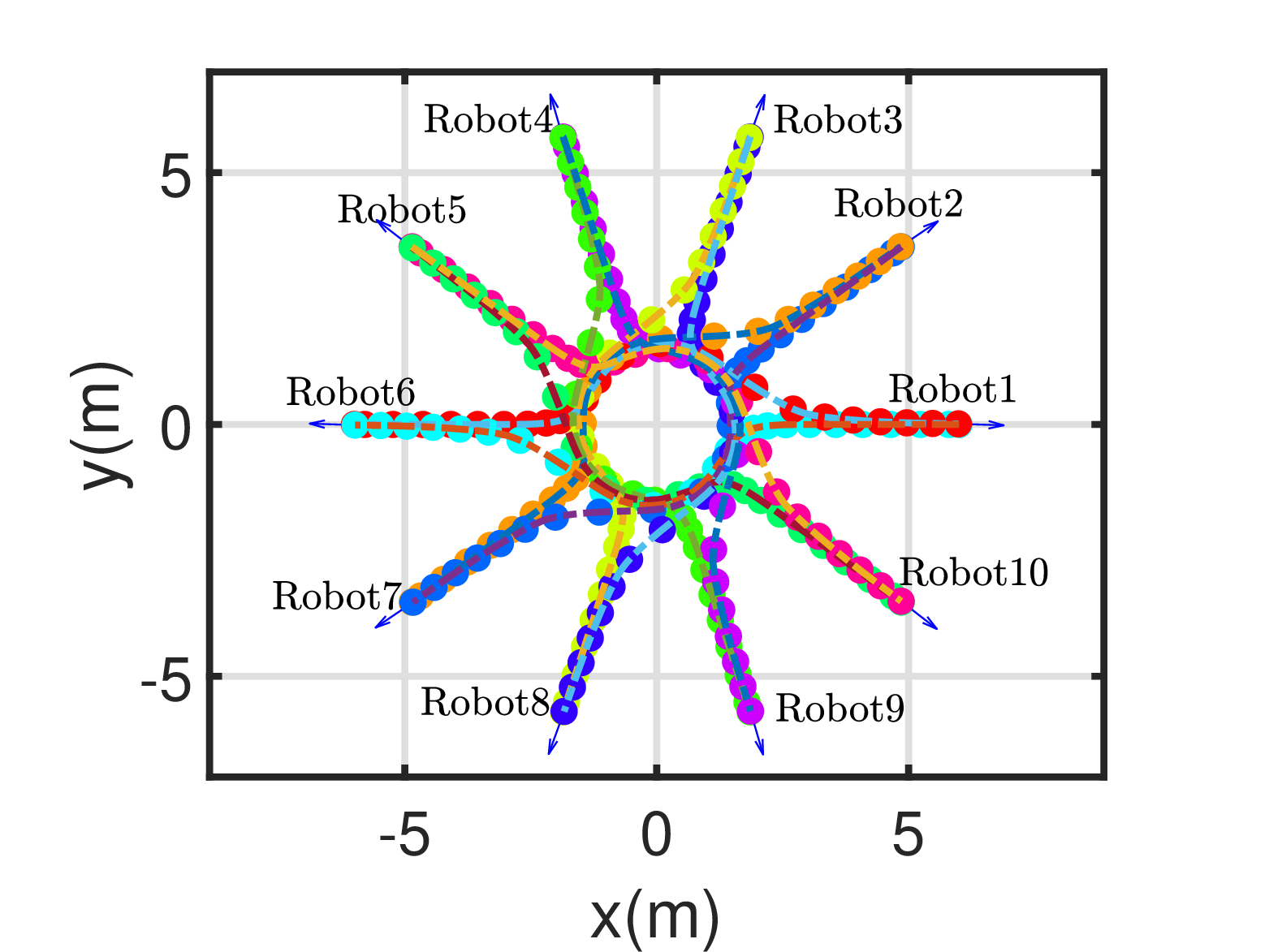}}
\end{minipage}
 \caption{\justifying Example II: SOATT snapshot of ten robots are trying
to move through the center of a circle to antipodal positions at different times, where $\lambda_{ij}=0.5$. (a) $t=0$. (b) $t=6s$. (c) $t=9s$. (d) $t=12s$.}\label{figm}
 \end{figure*}
 \begin{figure}
  \centering
  \vspace{-10pt}
\setlength{\abovecaptionskip}{-0.1cm}   
\begin{minipage}[]{5.5in}
\subfigcapskip=-4pt
\hspace{-25pt}
\subfigure[]{\includegraphics[width=2.1in]{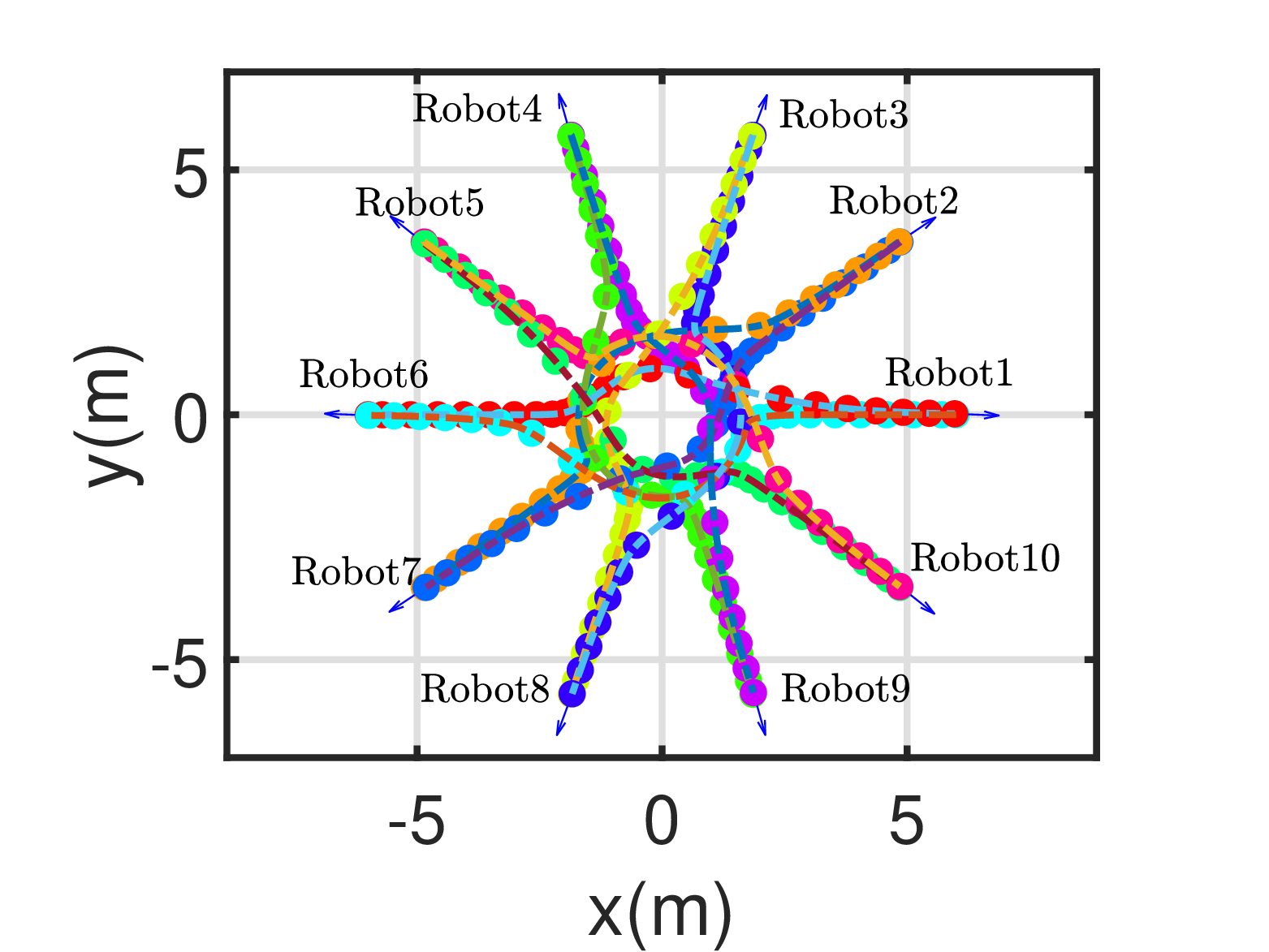}}
\hspace{-25pt}
\subfigure[]{\includegraphics[width=2.1in]{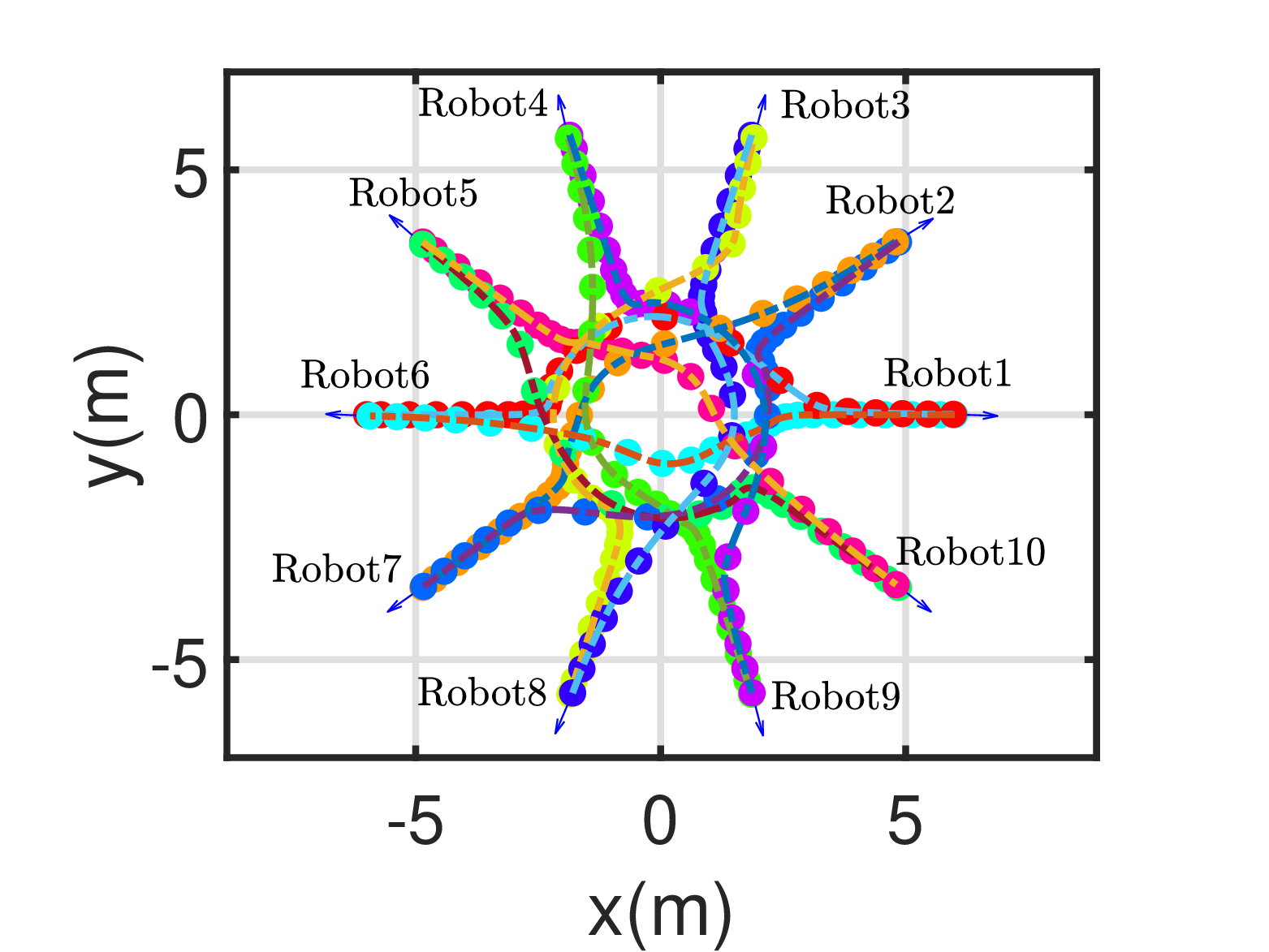}}
\end{minipage}
 \caption{\justifying Example II: An illustration of the trajectories generated by (a) $\lambda_{ij}=1$ and (b) $\rho_{ij}=r_i+r_j+0.35$, respectively.  }\label{figmm}
 \end{figure}

 \begin{figure}
  \centering
  \vspace{-5pt}
\setlength{\abovecaptionskip}{-0.1cm}   
\begin{minipage}[]{5.5in}
\subfigcapskip=-4pt
\subfigure[]{\includegraphics[width=1.23in]{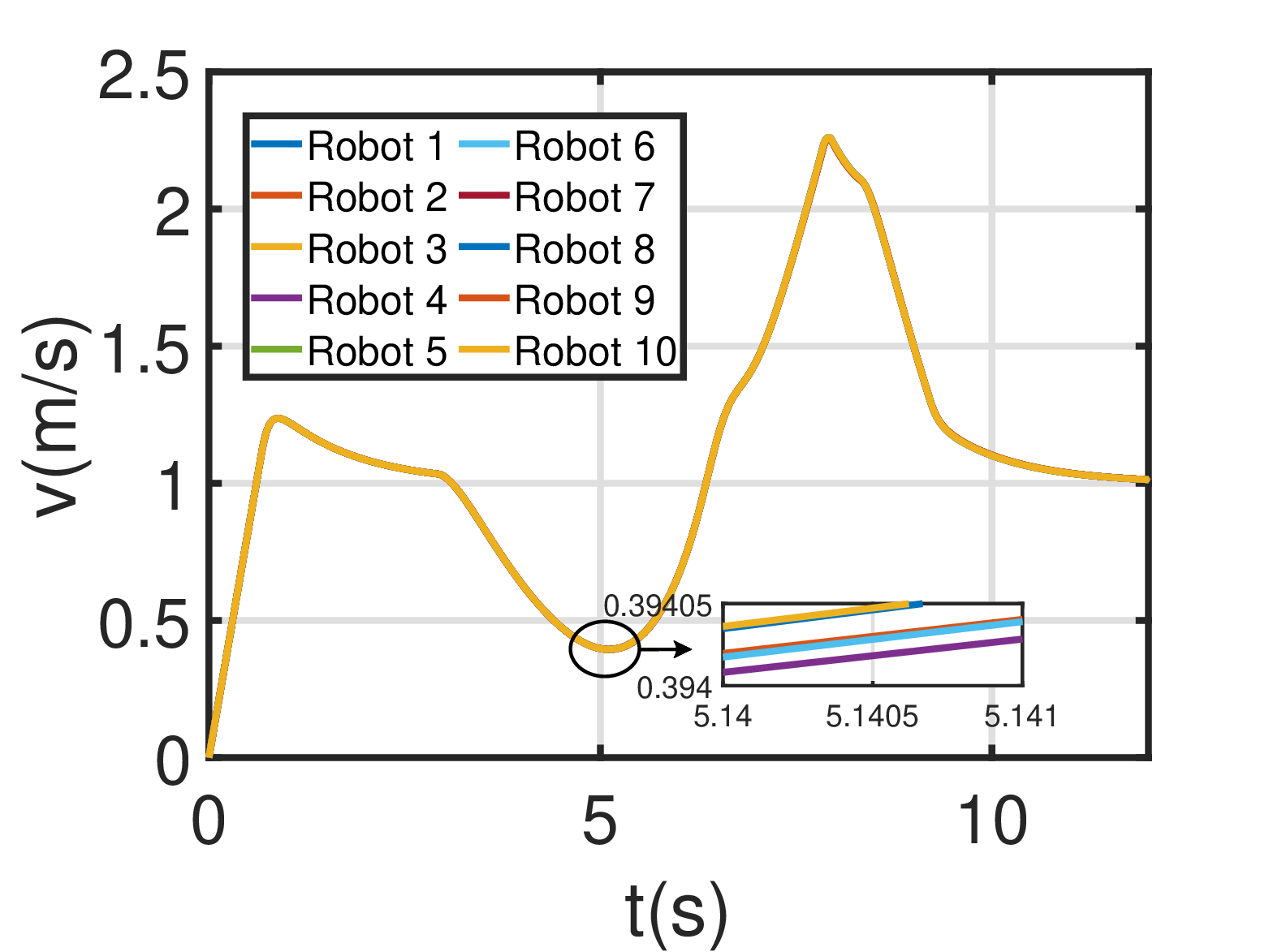}}
\hspace{-12pt}
\subfigure[]{\includegraphics[width=1.23in]{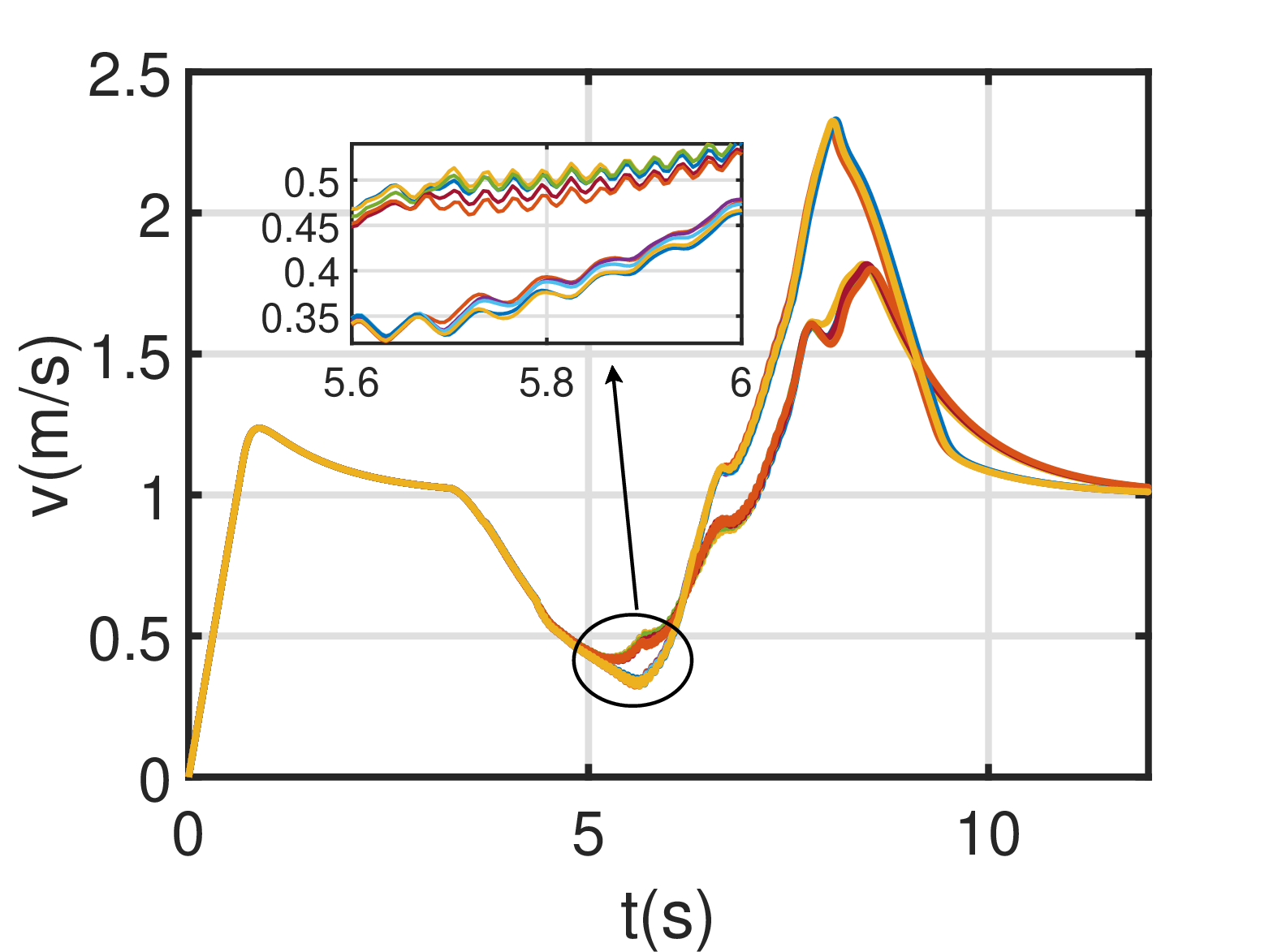}}
\hspace{-12pt}
\subfigure[]{\includegraphics[width=1.23in]{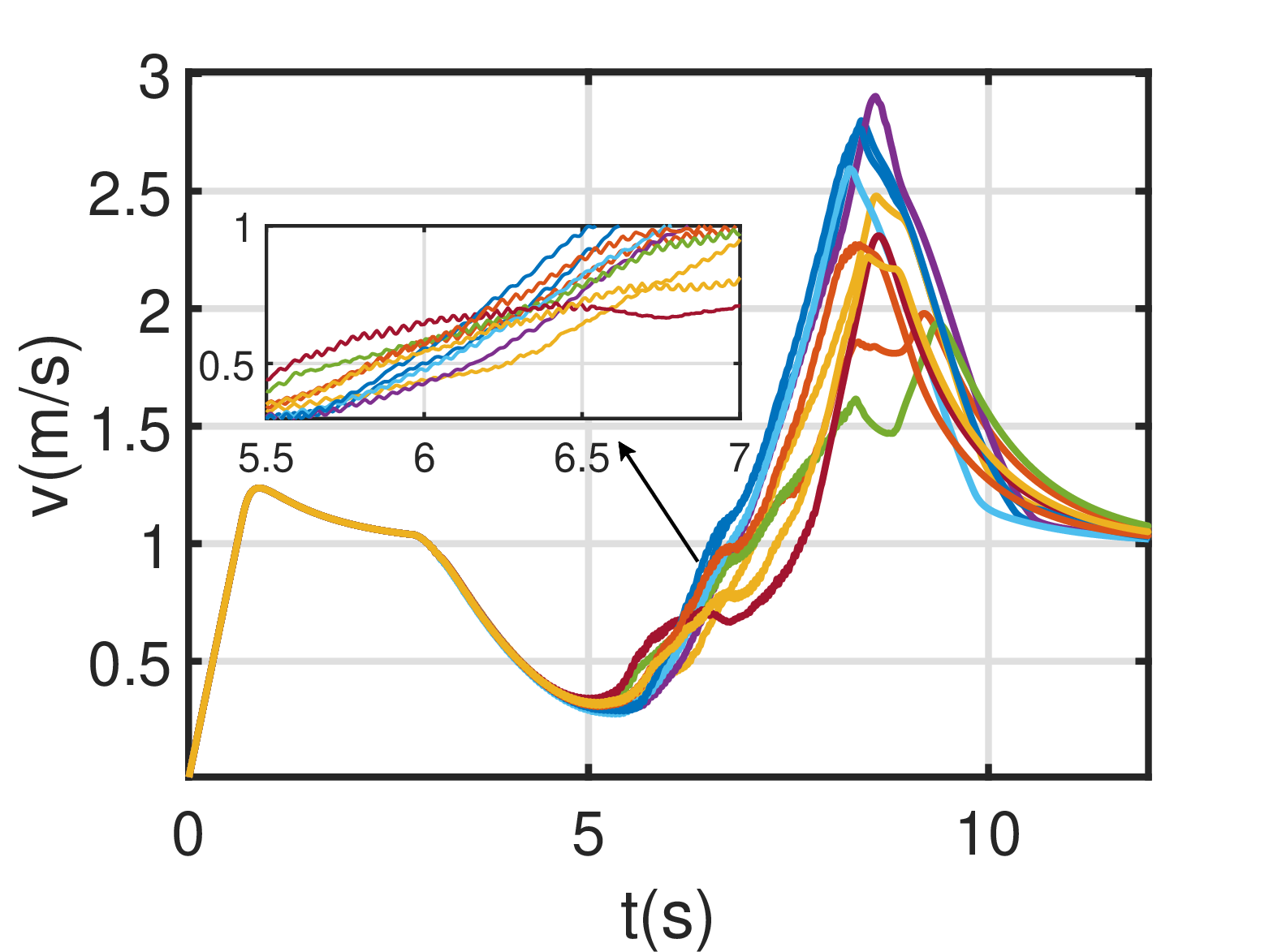}}
\end{minipage}
 \caption{\justifying Velocity profiles corresponding to Fig. \ref{figm} and Fig. \ref{figmm}. (a): $\lambda_{ij}=0.5$. (b): $\lambda_{ij}=1$. (c): $\rho_{ij}=r_i+r_j+0.35$.}\label{figs5}
 \end{figure}

 \begin{figure}
  \centering
  \vspace{-10pt}
\setlength{\abovecaptionskip}{-0.1cm}   
\begin{minipage}[]{5.5in}
\subfigcapskip=-4pt
\hspace{-25pt}
\subfigure[]{\includegraphics[width=2.35in]{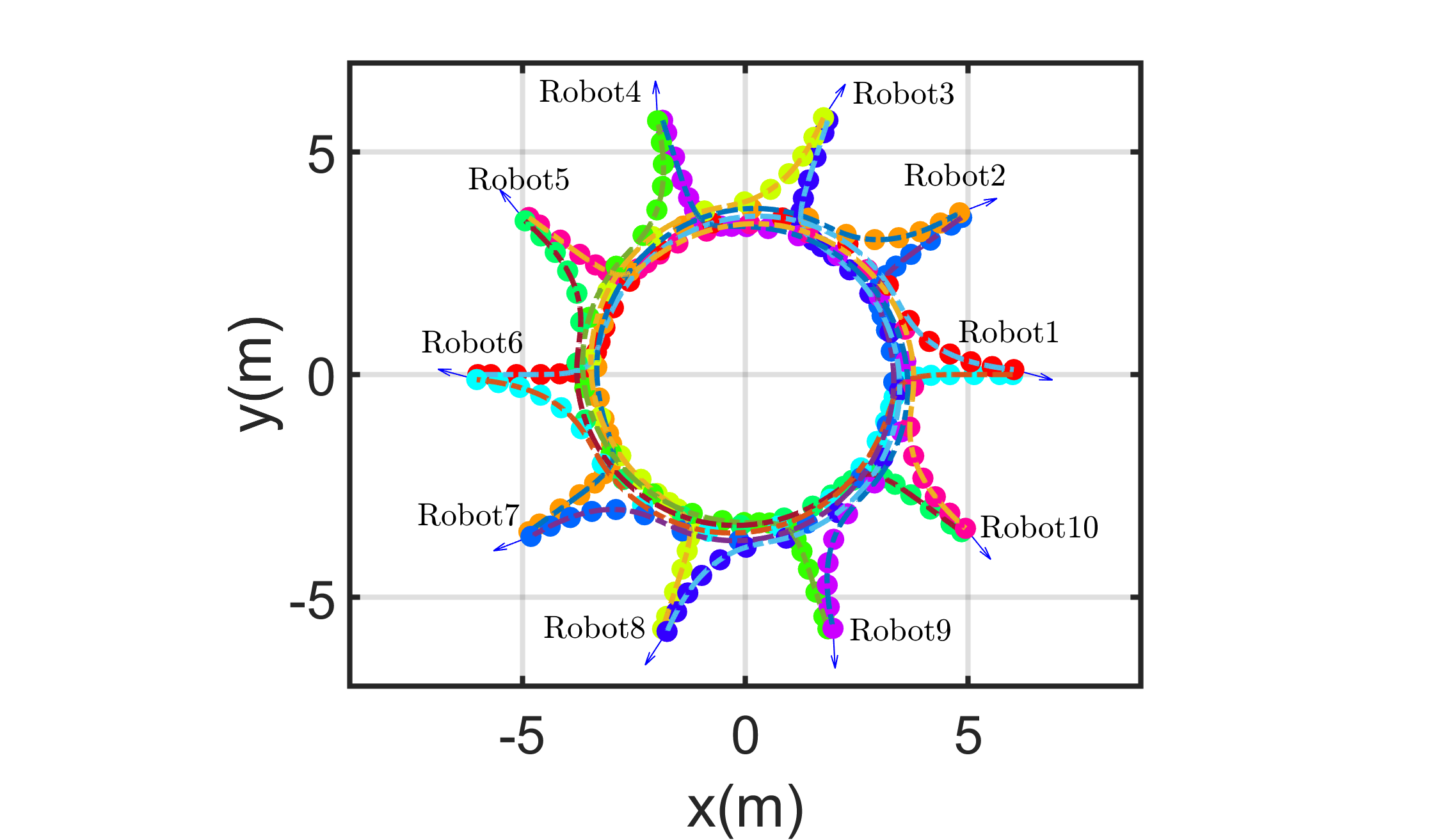}}
\hspace{-30pt}
\subfigure[]{\includegraphics[width=1.85in]{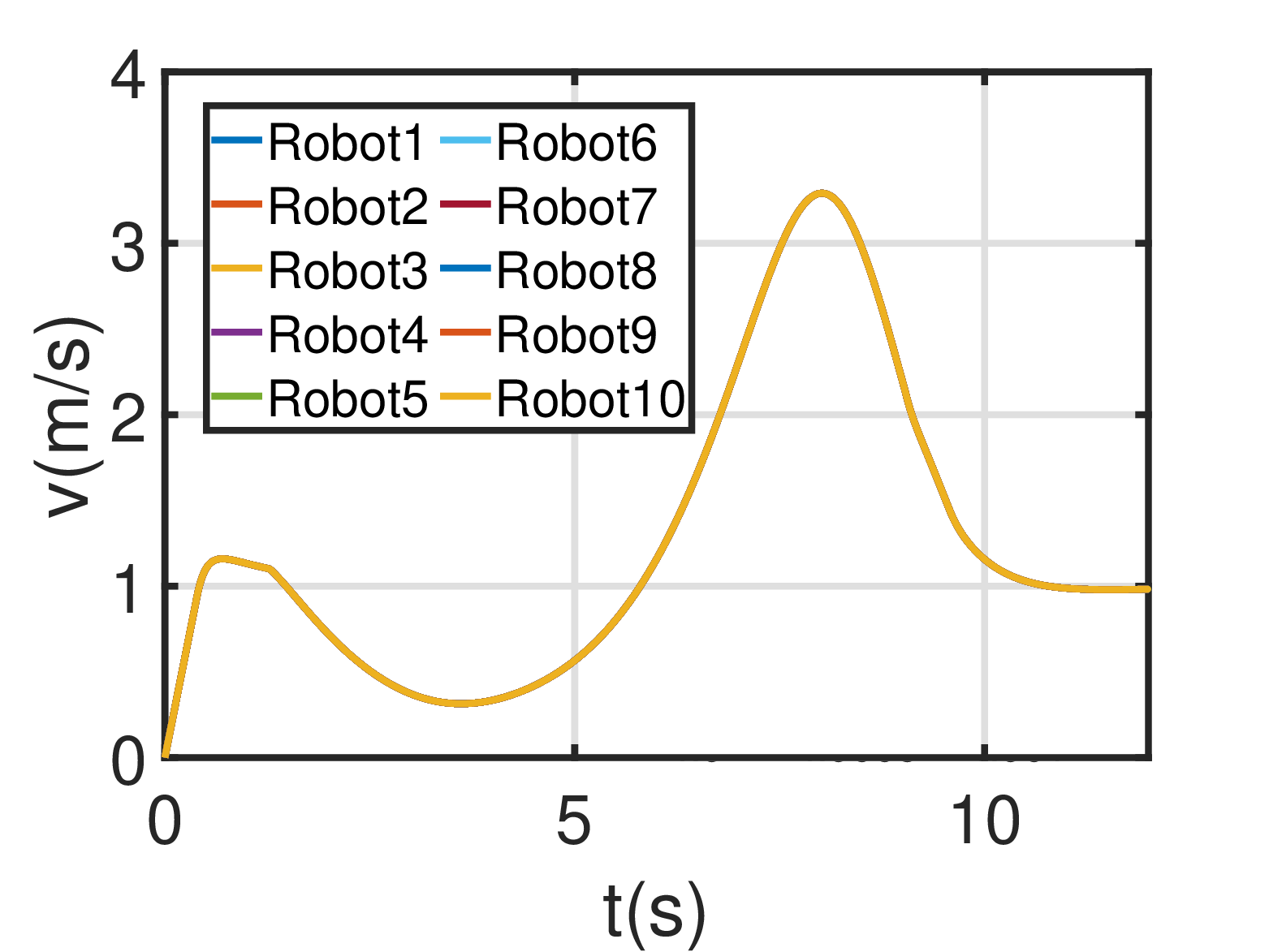}}
\end{minipage}
\begin{minipage}[]{5.5in}
\vspace{-10pt}
\subfigcapskip=-4pt
\hspace{-25pt}
\subfigure[]{\includegraphics[width=2.35in]{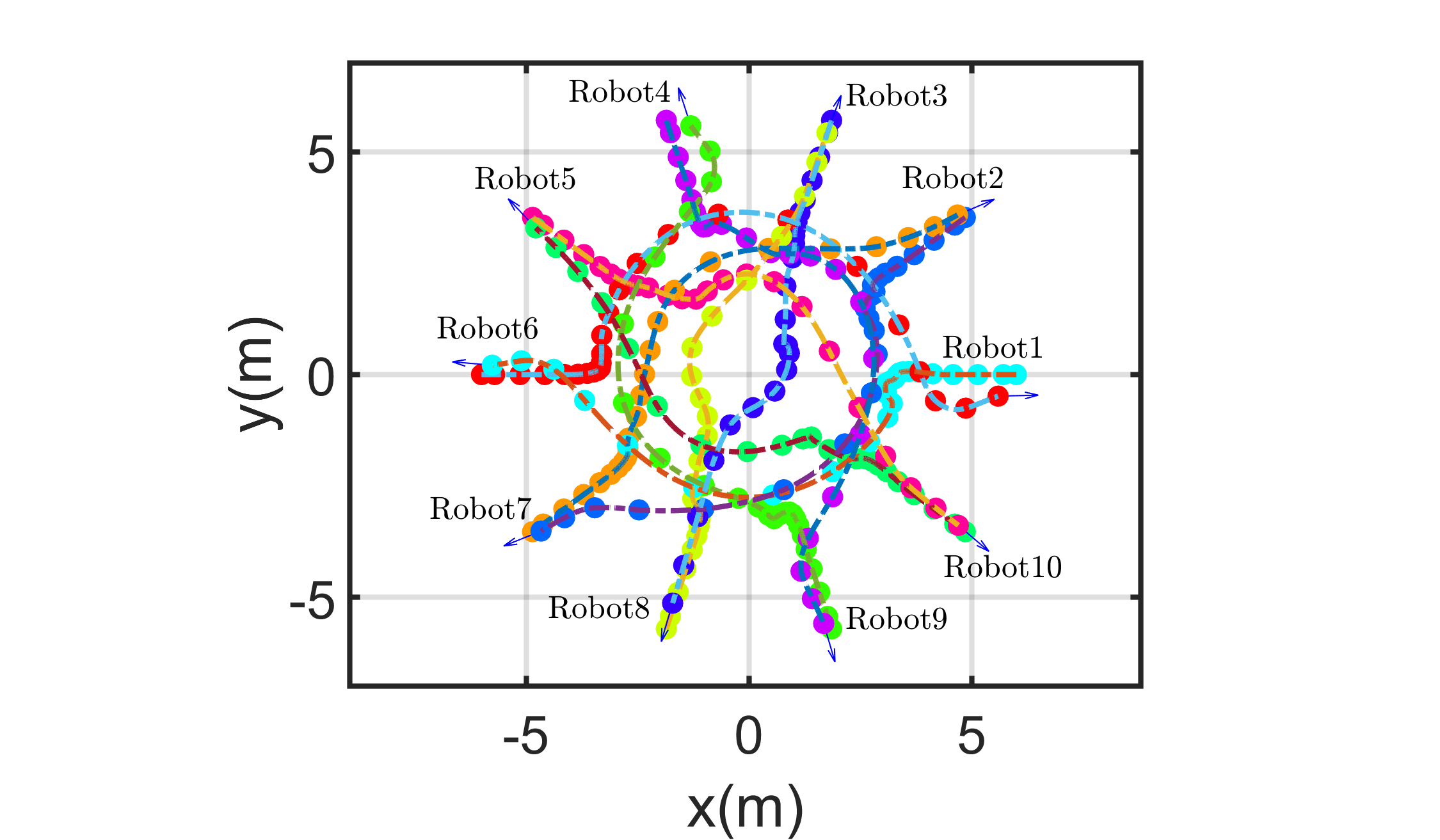}}
\hspace{-30pt}
\subfigure[]{\includegraphics[width=1.85in]{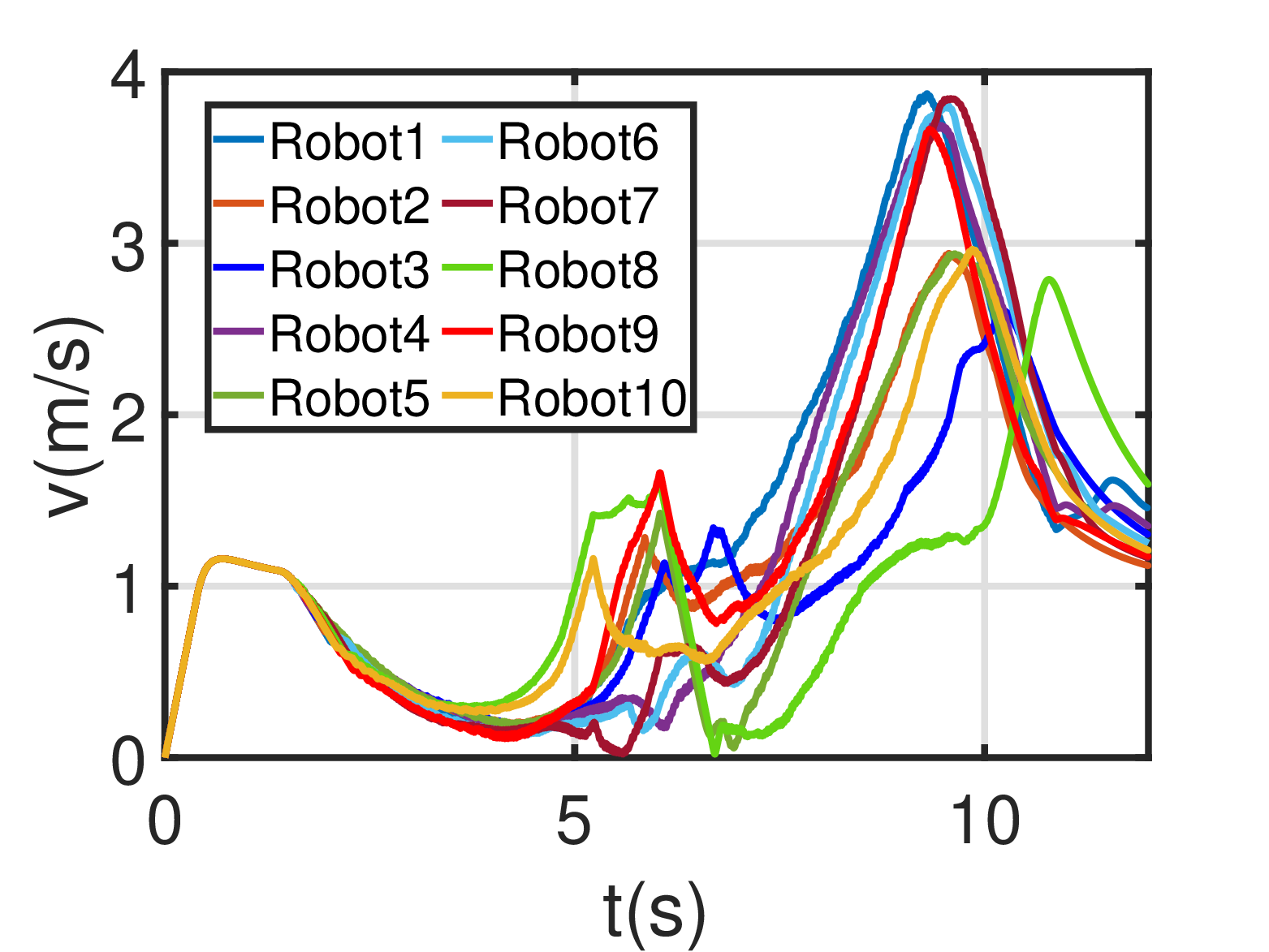}}
\end{minipage}
 \caption{\justifying Trajectory illustration and velocity profiles corresponding to Fig. \ref{figm}(a) when (a)-(b): $\lambda_{ij}=0.2$. (c)-(d): $\rho_{ij}=r_i+r_j+1$.}\label{figs5a}
 \end{figure}

\begin{figure}
  \centering
  \vspace{-5pt}
\setlength{\abovecaptionskip}{-0.1cm}   
\begin{minipage}[]{6.5in}
\subfigcapskip=-4pt
\hspace{-15pt}
\subfigure[]{\includegraphics[width=2.25in]{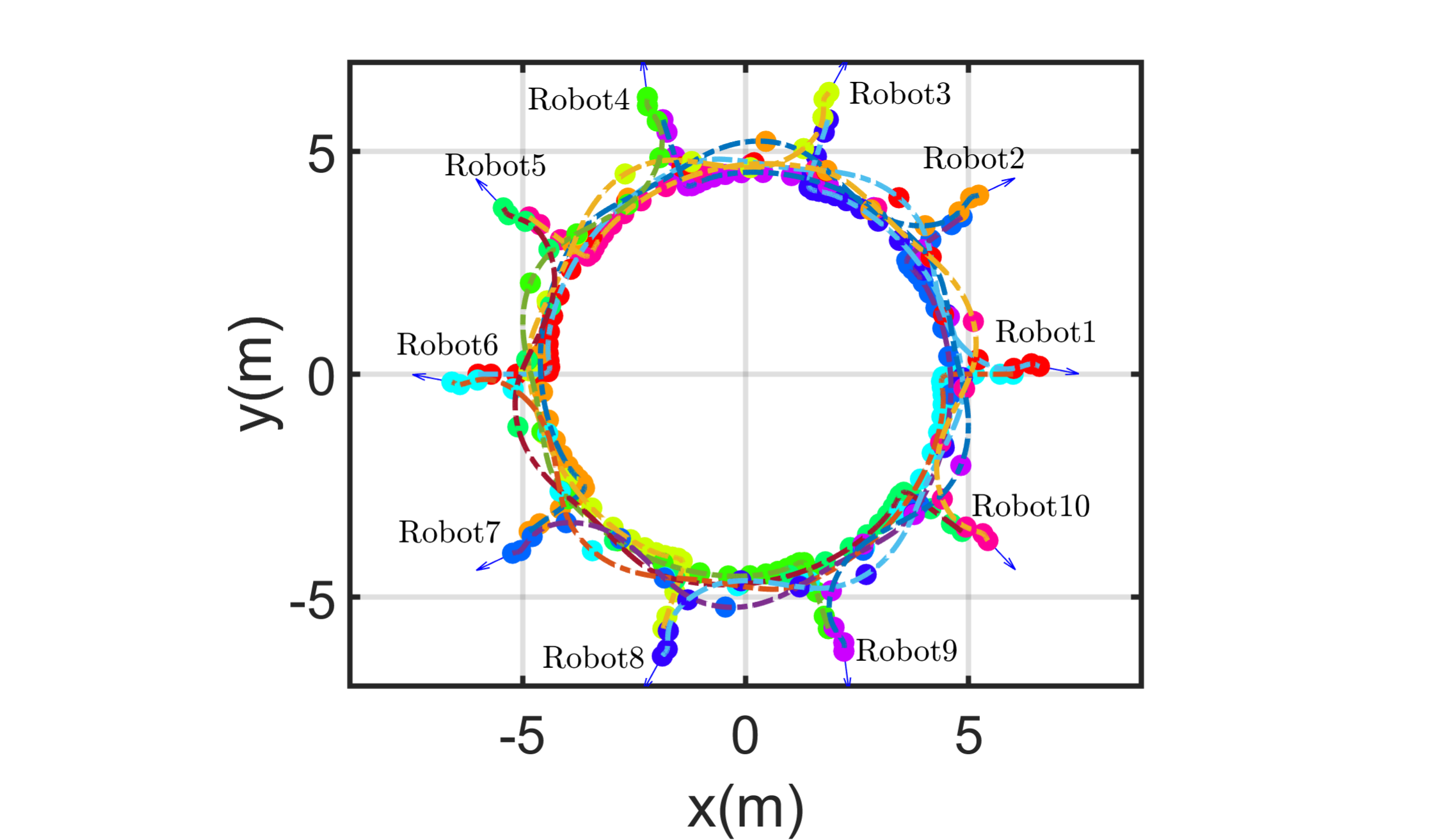}}
\hspace{-30pt}
\subfigure[]{\includegraphics[width=1.75in]{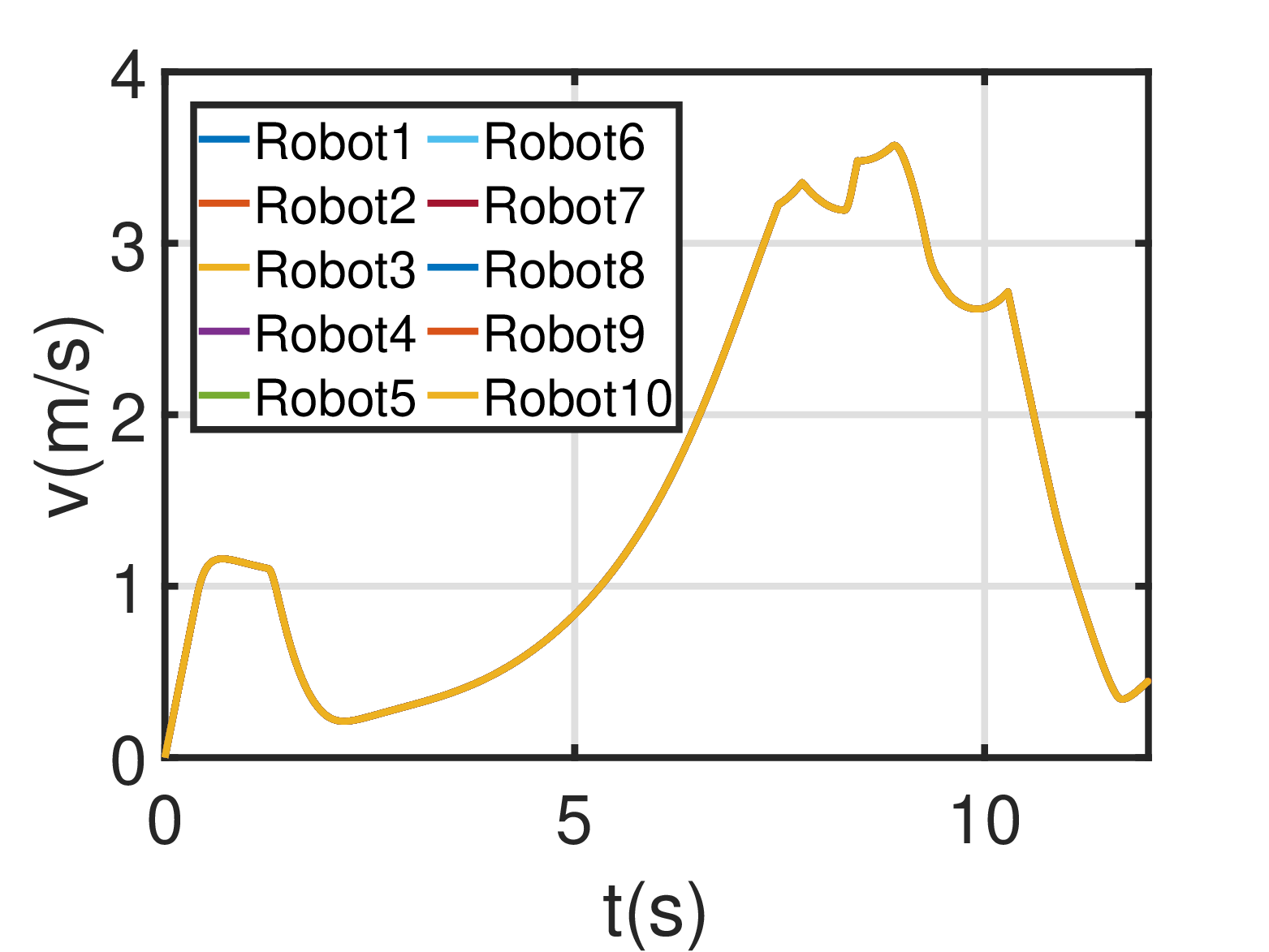}}
\end{minipage}
\begin{minipage}[]{6.5in}
\subfigcapskip=-4pt
\hspace{-15pt}
\subfigure[]{\includegraphics[width=2.25in]{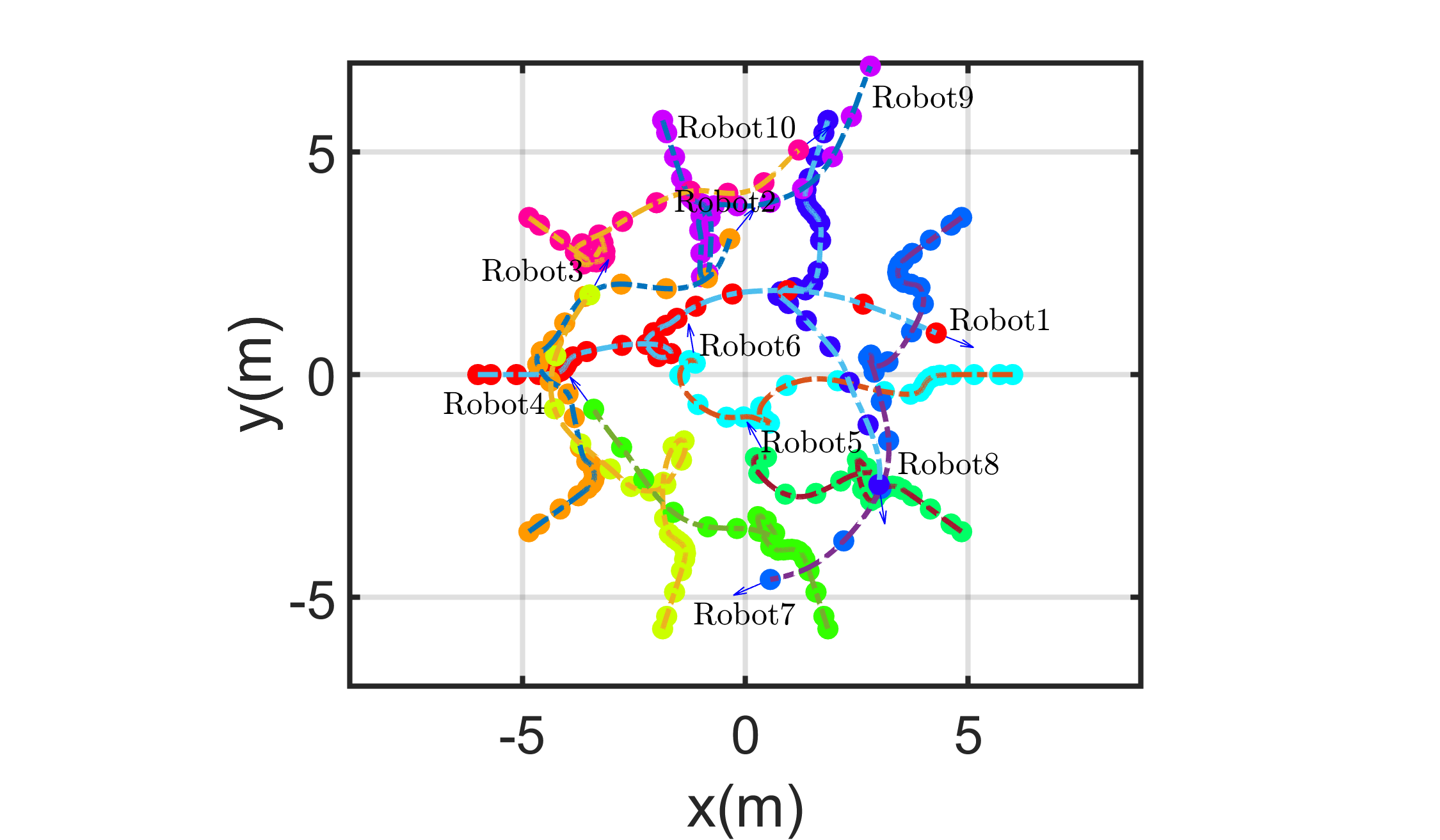}}
\hspace{-30pt}
\subfigure[]{\includegraphics[width=1.75in]{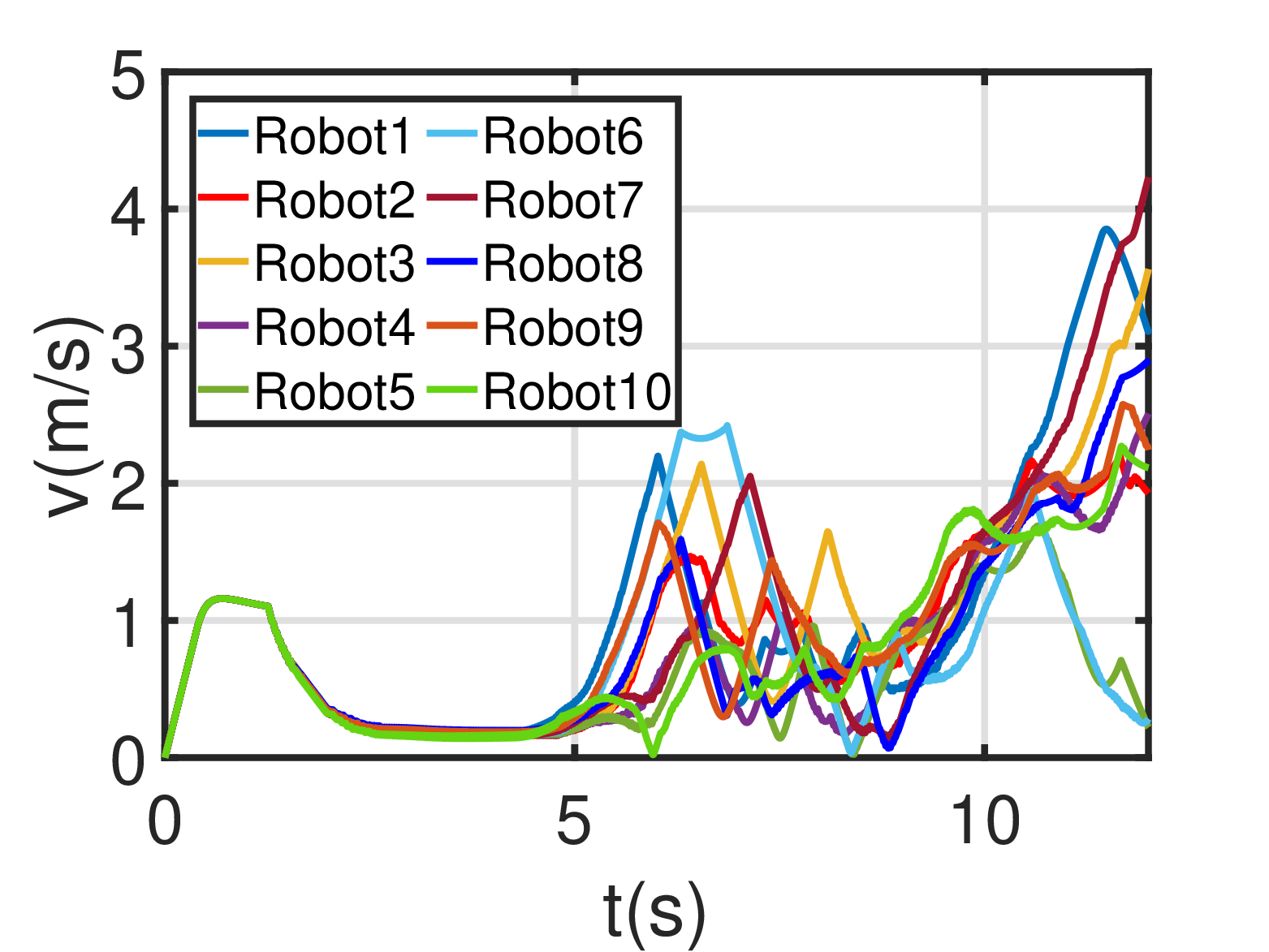}}
\end{minipage}
 \caption{\justifying Trajectory illustration and velocity profiles corresponding to Fig. \ref{figm}(a) when (a)-(b): $\lambda_{ij}=0.1$. (c)-(d): $\rho_{ij}=r_i+r_j+1.5$.}\label{figs5b}
\end{figure}

\section{Results}
In this part, we show effectiveness and superiority of the predictive cooperative CA method by three examples where $\kappa_{ij_1}=\kappa_{i_3}=1$, $\kappa_{ij_2}=\kappa_{i_4}=8$, $q_i=68^\circ$. The same parameters are used for every robot, which is to validate efficiency of the deadlock escape because of its ease of trapping into deadlock.

Simultaneous obstacle avoidance and trajectory tracking (SOATT) result of single robot with multiple static obstacles, and the corresponding velocity profiles are shown in Fig. \ref{h} and Fig. \ref{h_v}, respectively. In this example, the robot is expected to move $8\mathrm{m}$ at a constant speed from its initial position within $16\mathrm{s}$. When the predictive term is not considered, \emph{i.e}, $\lambda_{ij}=1$, the robot moves along to the boundary of the obstacle closely. This is achieved based on the perfect sensing information. The CA would fail when disturbed by measurement errors, as shown by the green line in Fig. \ref{h} where the positioning error is set to $0.15\mathrm{m}$, the robot crashes into obstacles.
Moreover, it could be observed from Fig. \ref{h_v} that oscillation happens in CA when $\lambda_{ij}=1$. Although SOATT achieved by increasing the safety threshold $\rho_{ij}$ by $0.35\mathrm{m}$ seems to be indistinguishable with comparison that one achieved by the predicted term that is considered in the safety constraint, the predictive term generates a smooth velocity profile. To further illustrate our advantages, a circle scenario is considered in Example II, where ten robots are trying to move through the center of a circle with radius of $6\mathrm{m}$ to antipodal positions within $12\mathrm{s}$. Fig. \ref{figm} and Fig. \ref{figmm} give an illustration of the trajectories generated by $\lambda_{ij}=0.5$, $\lambda_{ij}=1$ and $\rho_{ij}=r_i+r_j+0.35$, respectively. The corresponding vehicle velocity profiles are shown in Fig. \ref{figs5}.
Following Fig. \ref{figs5}, the predictive term $\lambda_{ij}=0.5$ still ensures the smoothness of the vehicle velocity compare to a way of increasing $\rho_{ij}$ in \cite{IJRR2014}. Next, we considered an extreme case under Fig. \ref{figm}(a), the robots navigate away from each other at larger distances. Trajectory illustration and velocity profiles are shown in Fig. \ref{figs5a} and Fig. \ref{figs5b}.
When $\rho_{ij}=r_i+r_j+1$, the robot fails in tracking the preferred trajectory and avoiding the neighboring robots.
Combined Fig. \ref{h}-Fig. \ref{figs5b}, we know that the proposal of the predictive cooperative CA method is to evade risks in advance at the cost of shrinking feasible space and has a certain tolerance for measurement uncertainty. Compared with the method of increasing safety thresholds, it relatively ensures the smoothness of the CA trajectory.

\begin{figure*}
  \centering
\setlength{\abovecaptionskip}{-0.1cm}   
\begin{minipage}[]{8.5in}
\subfigcapskip=-4pt
\hspace{-10pt}
\subfigure[]{\includegraphics[width=1.62in]{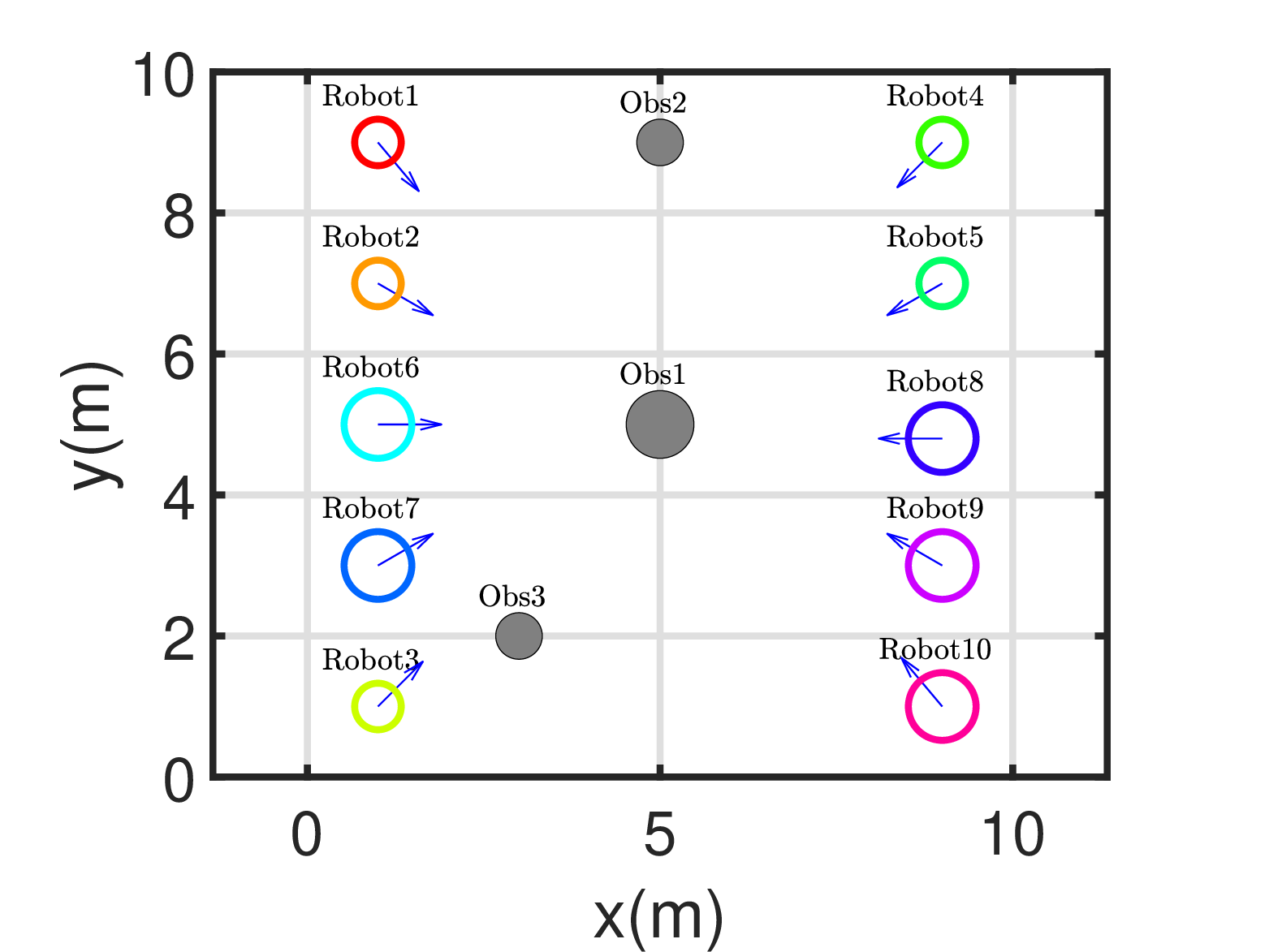}}
\hspace{-20pt}
\subfigure[]{\includegraphics[width=1.62in]{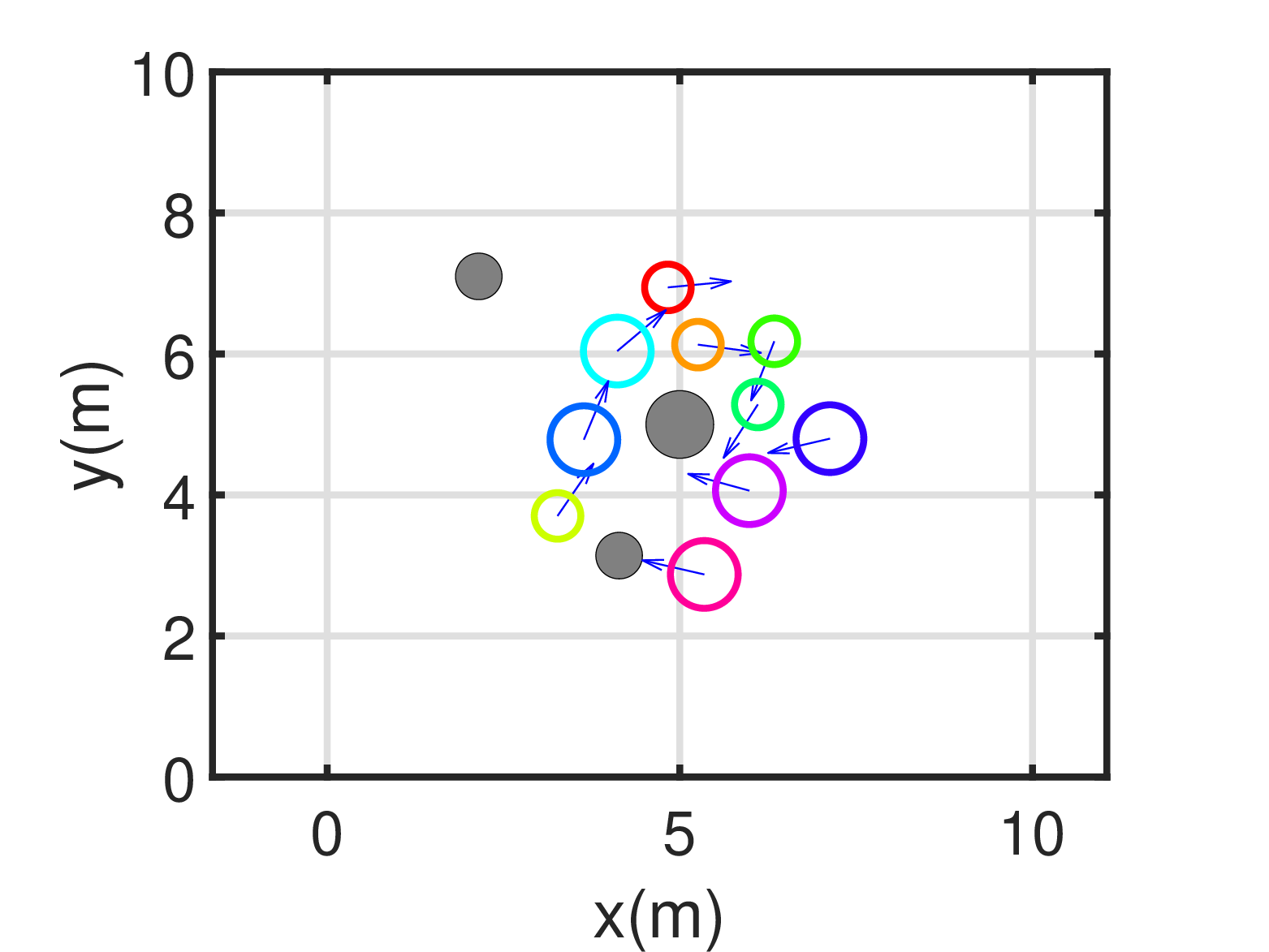}}
\hspace{-20pt}
\subfigure[]{\includegraphics[width=1.62in]{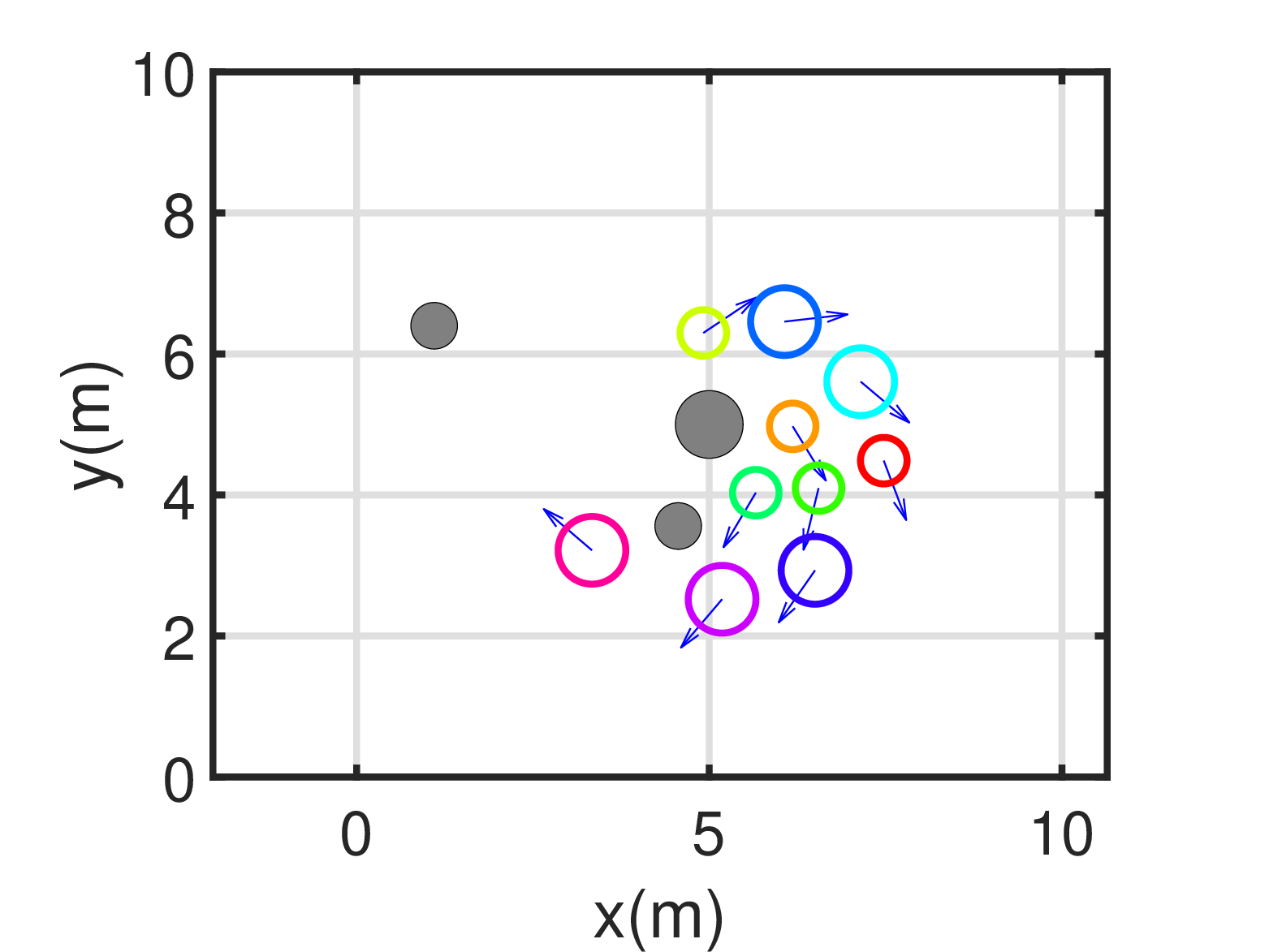}}
\hspace{-20pt}
\subfigure[]{\includegraphics[width=1.62in]{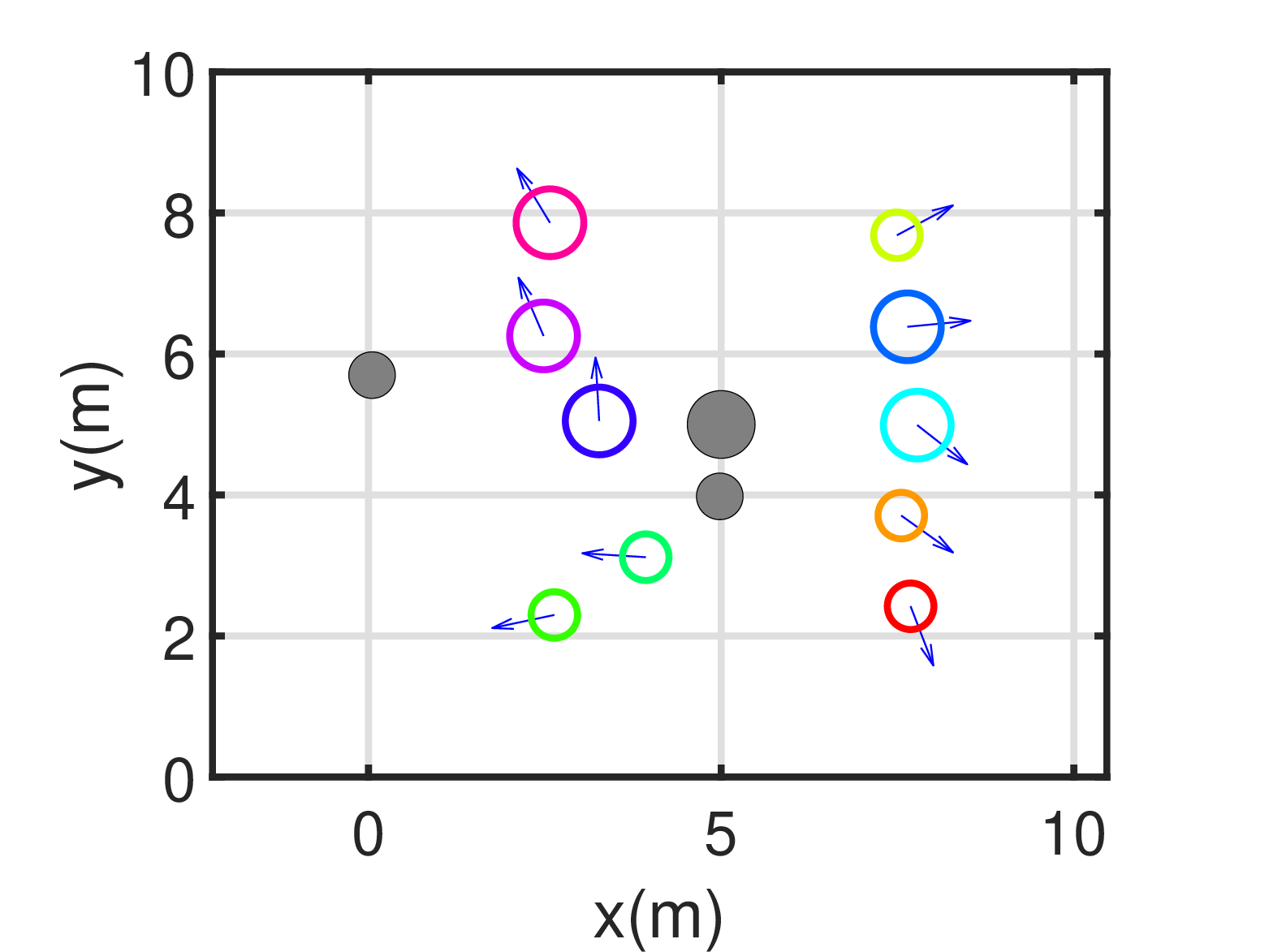}}
\hspace{-20pt}
\subfigure[]{\includegraphics[width=1.62in]{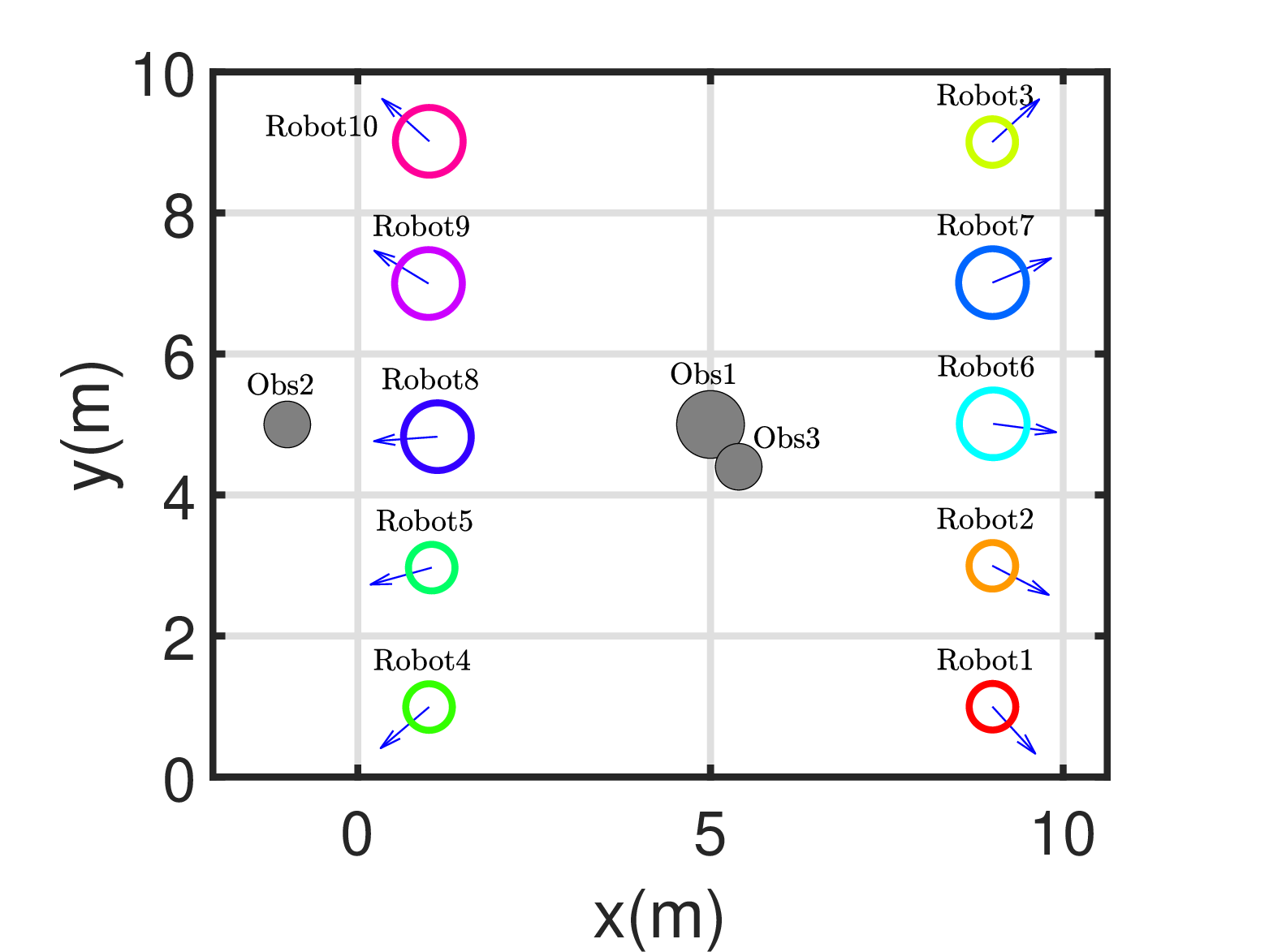}}
\end{minipage}
\begin{minipage}[]{8.5in}
\vspace{-10pt}
\subfigcapskip=-4pt
\hspace{-10pt}
\subfigure[]{\includegraphics[width=1.62in]{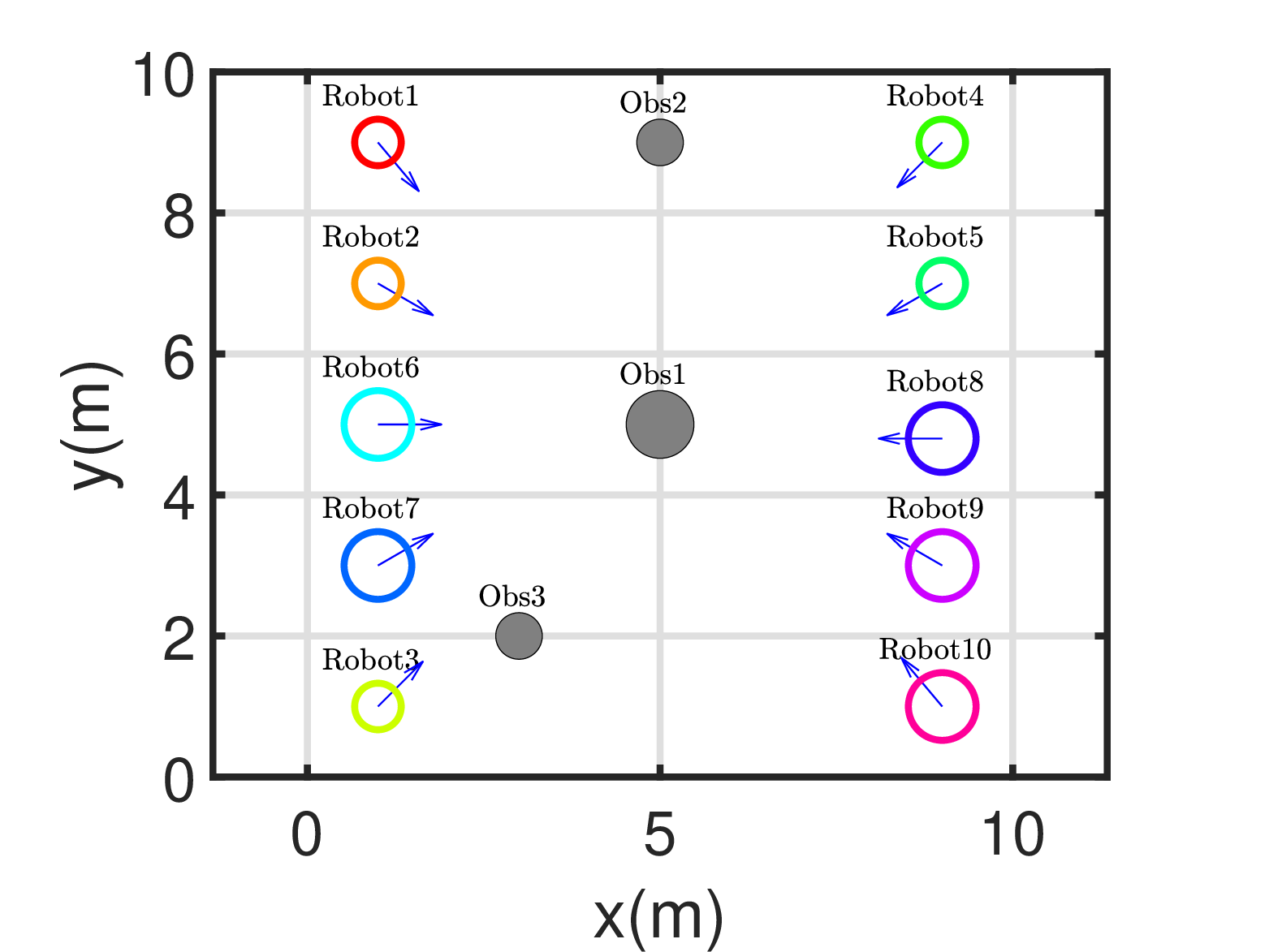}}
\hspace{-20pt}
\subfigure[]{\includegraphics[width=1.62in]{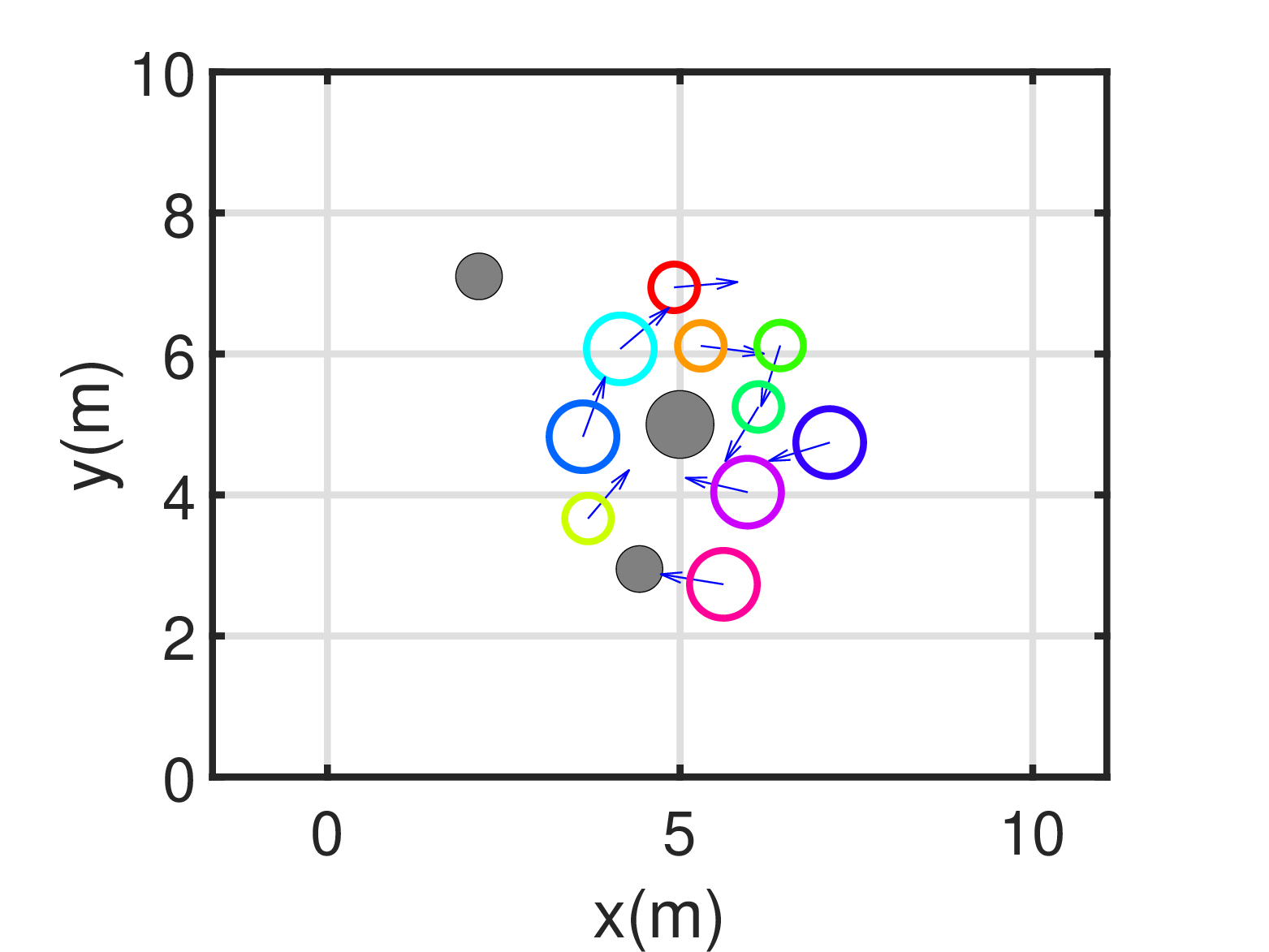}}
\hspace{-20pt}
\subfigure[]{\includegraphics[width=1.62in]{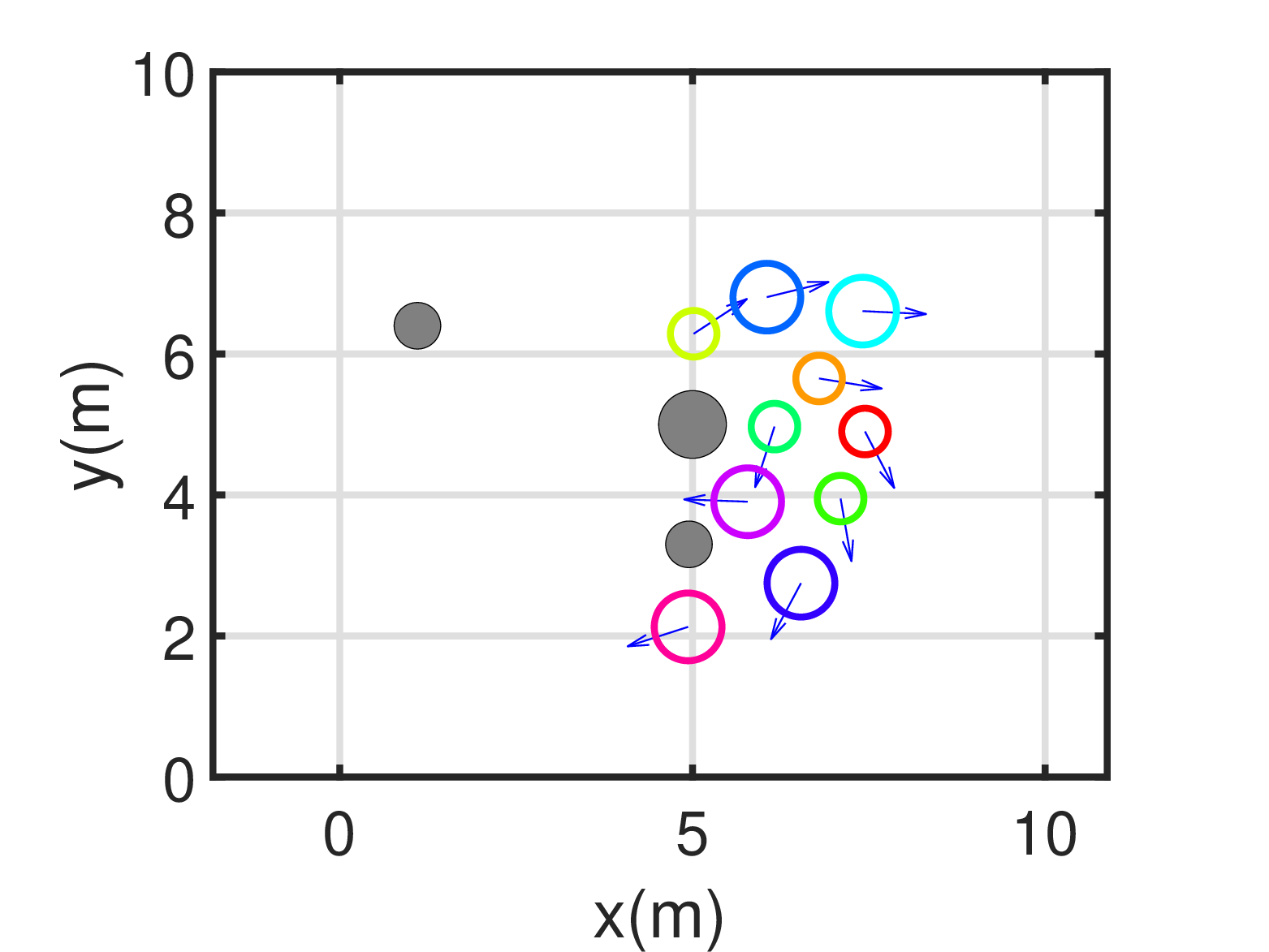}}
\hspace{-20pt}
\subfigure[]{\includegraphics[width=1.62in]{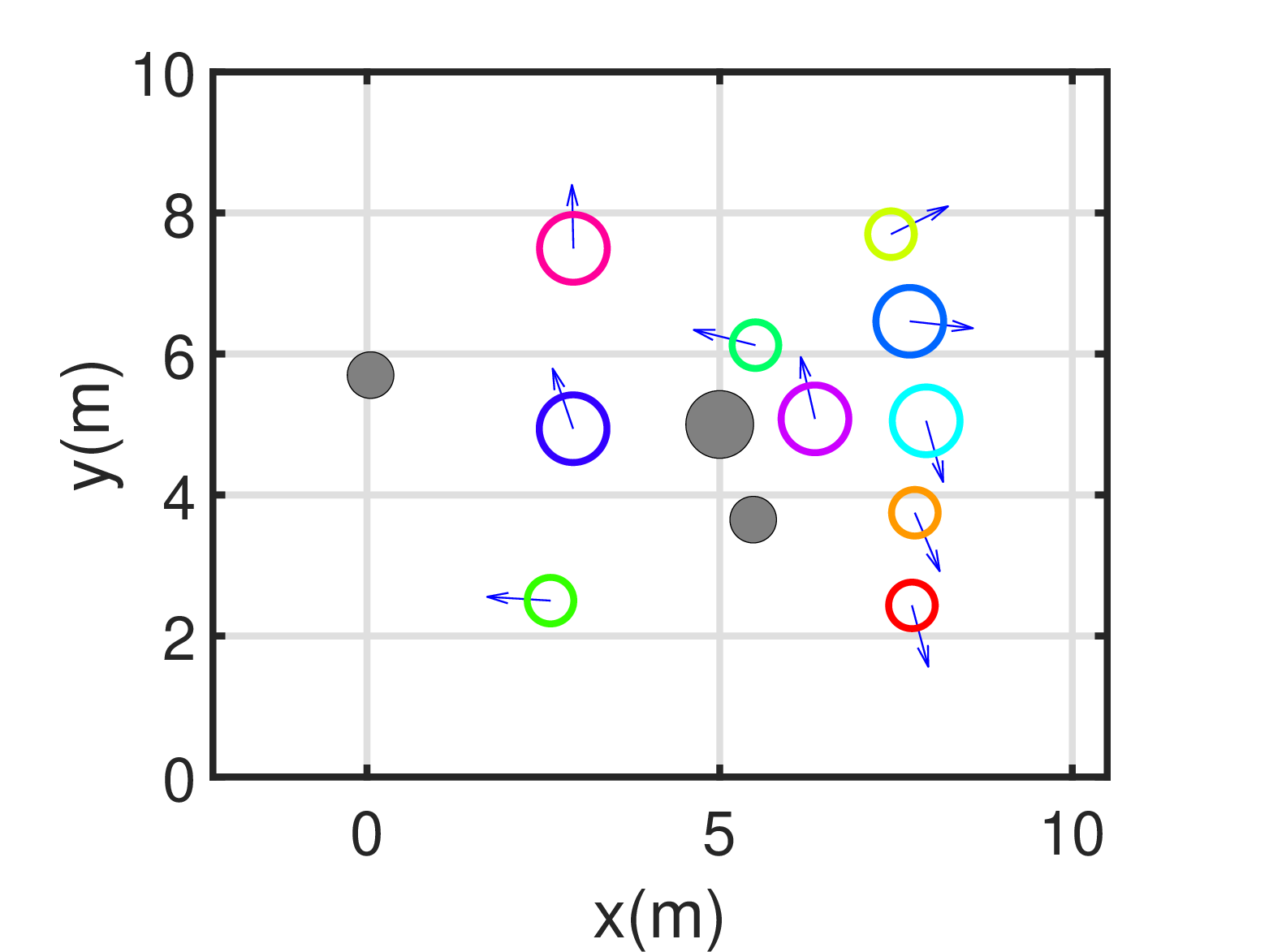}}
\hspace{-20pt}
\subfigure[]{\includegraphics[width=1.62in]{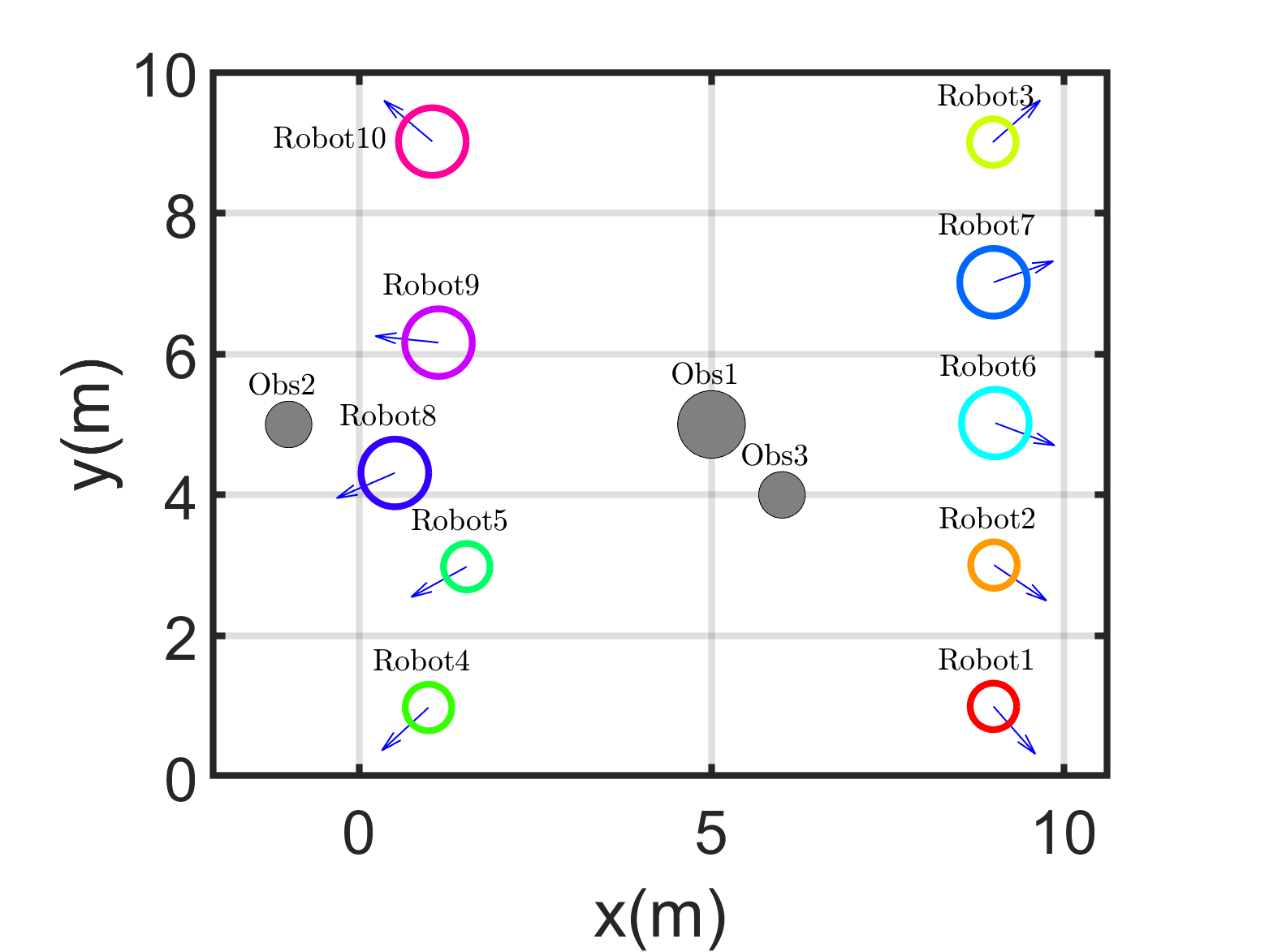}}
\end{minipage}
 \caption{\justifying Example III: SOATT snapshot between ten robots and multi-obstacle obtained by the proposed method at different times. (a) and (f) $t=0\mathrm{s}$. (b) and (g) $t=10\mathrm{s}$. (c) and (h) $t=40/3\mathrm{s}$. (d) and (i) $t=50/3\mathrm{s}$. (e) and (j) $t=20\mathrm{s}$ where top: $\mathbf{v}_{\text{obs}3}=[0.12,0.12]^\text{T} \mathrm{m/s}$, bottom: $\mathbf{v}_{\text{obs}3}=[0.15,0.1]^\text{T} \mathrm{m/s}$.}\label{figs6}
 \end{figure*}

In addition, we also can observed from Fig. \ref{h} that the avoidance directions of the robot avoiding obstacles \text{obs1} and \text{obs2} always are chosen as the side with the minimum attitude angle.
We further consider CA between ten robots and multiple obstacles in Example III, where $\text{obs}1$ is a static obstacle, the other two are dynamic obstacles with $\mathbf{v}_{\text{obs}2}=[-0.3,-0.2]^\text{T}\mathrm{m/s}$.  The initial state is shown in Fig. \ref{figs6}(a) and Fig. \ref{figs6}(f). Notice that Fig. \ref{figs6}(a) is same as Fig. \ref{figs6}(f), however, the velocity of the dynamic obstacle $\text{obs}3$ is different, where $\mathbf{v}_{\text{obs}3}=[0.12,0.12]^\text{T}\mathrm{m/s}$ in Fig. \ref{figs6}(a),
$\mathbf{v}_{\text{obs}3}=[0.15,0.1]^\text{T} \mathrm{m/s}$ in Fig. \ref{figs6}(f). This leads to both Robot$5$ and Robot$9$ show different CA behaviors, see Fig. \ref{figs6}(c)-(d) and Fig. \ref{figs6}(h)-(i), manifesting as Robot$5$ and Robot$9$ respectively passing above and below obstacle $\text{obs}1$. It is observed from Fig. \ref{figs6}(f), both Robot$5$ and Robot$9$ do not reach their respective goals accurately. The reason is due to physical constraint Eq. (\ref{equ.16c}), limitations on control inputs make two robots no enough velocity to return to the destination quickly after avoiding \text{obs}3. \textbf{A dynamic illustration is given in \href{https://youtu.be/xQv-UaGLbqM}{https://youtu.be/xQv-UaGLbqM}.}

\begin{figure}
  \centering
  \vspace{-6pt}
\setlength{\abovecaptionskip}{-0.1cm}   
\begin{minipage}[]{5.5in}
\subfigcapskip=-4pt
\subfigure[]{\includegraphics[width=1.83in]{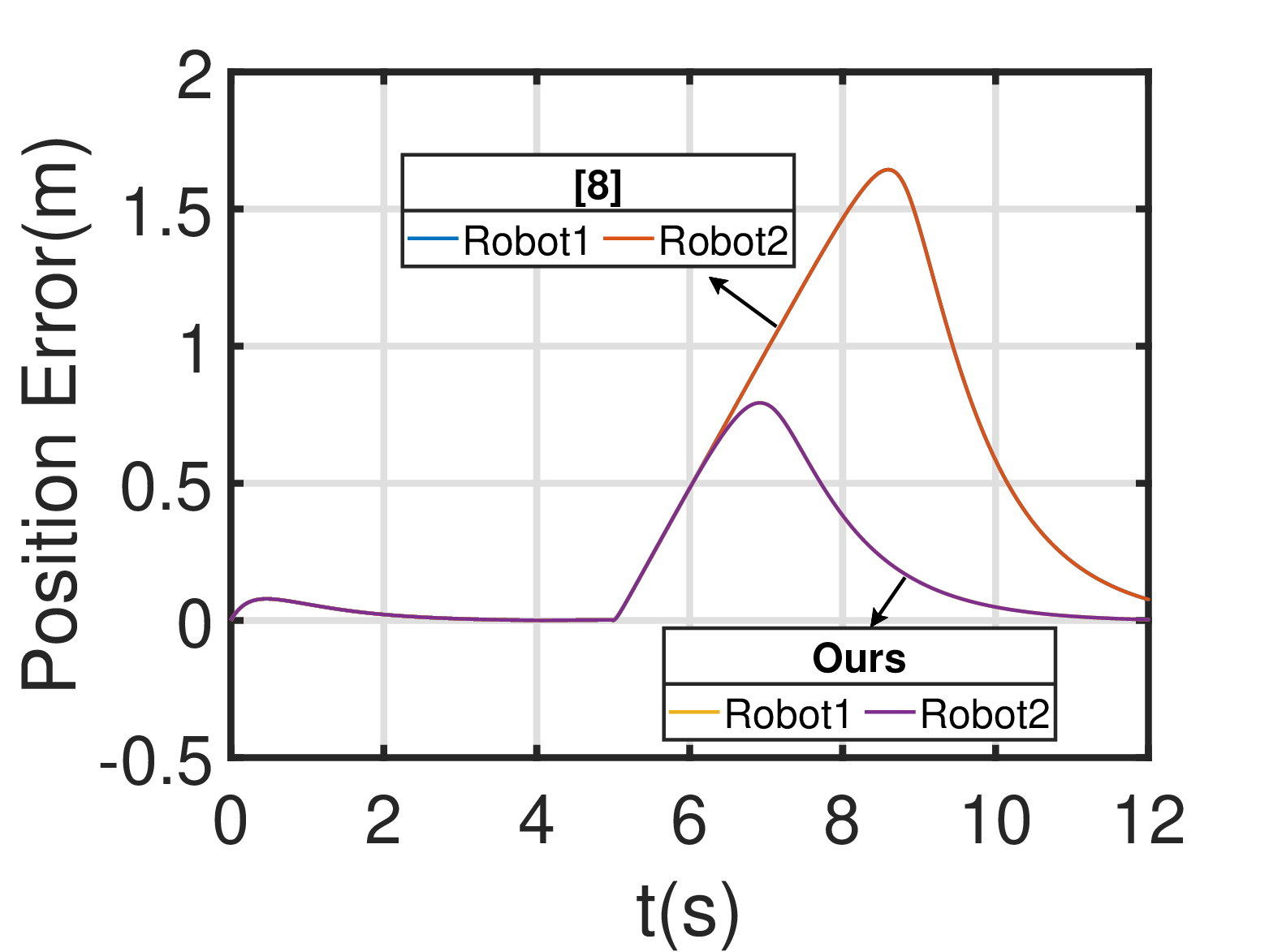}}
\hspace{-12pt}
\subfigure[]{\includegraphics[width=1.83in]{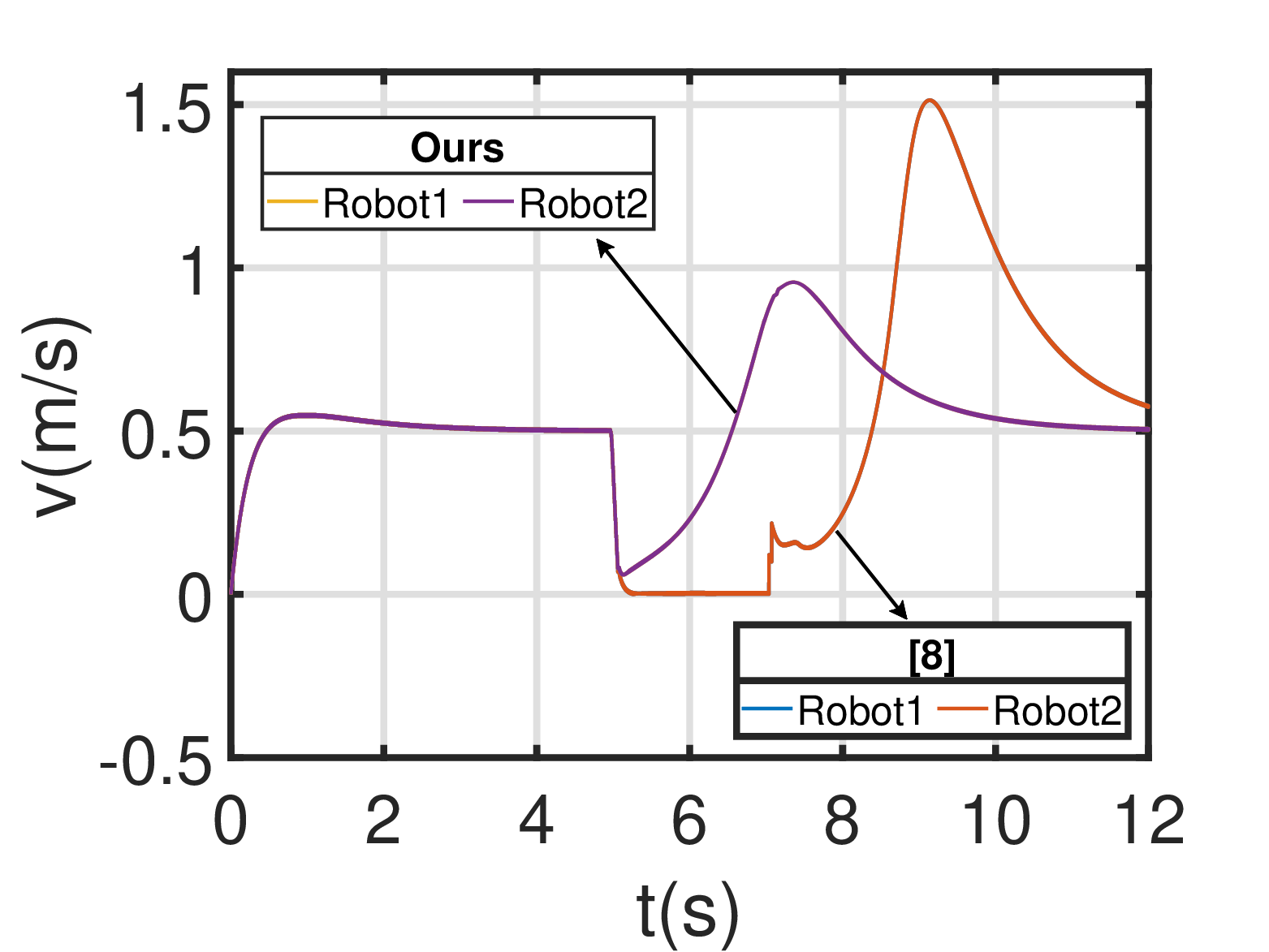}}
\end{minipage}
 \caption{\justifying Position errors and velocity profiles synthesised by our deconfliction strategy and one proposed by \cite{IJRR2023}. (a): Position errors. (b): Velocity profiles.}\label{figijrr}
 \end{figure}

 \begin{figure}[htb]
  \centering
\begin{minipage}[]{4.5in}
\hspace{-26pt}
\subfigure[]{\includegraphics[width=2.3in]{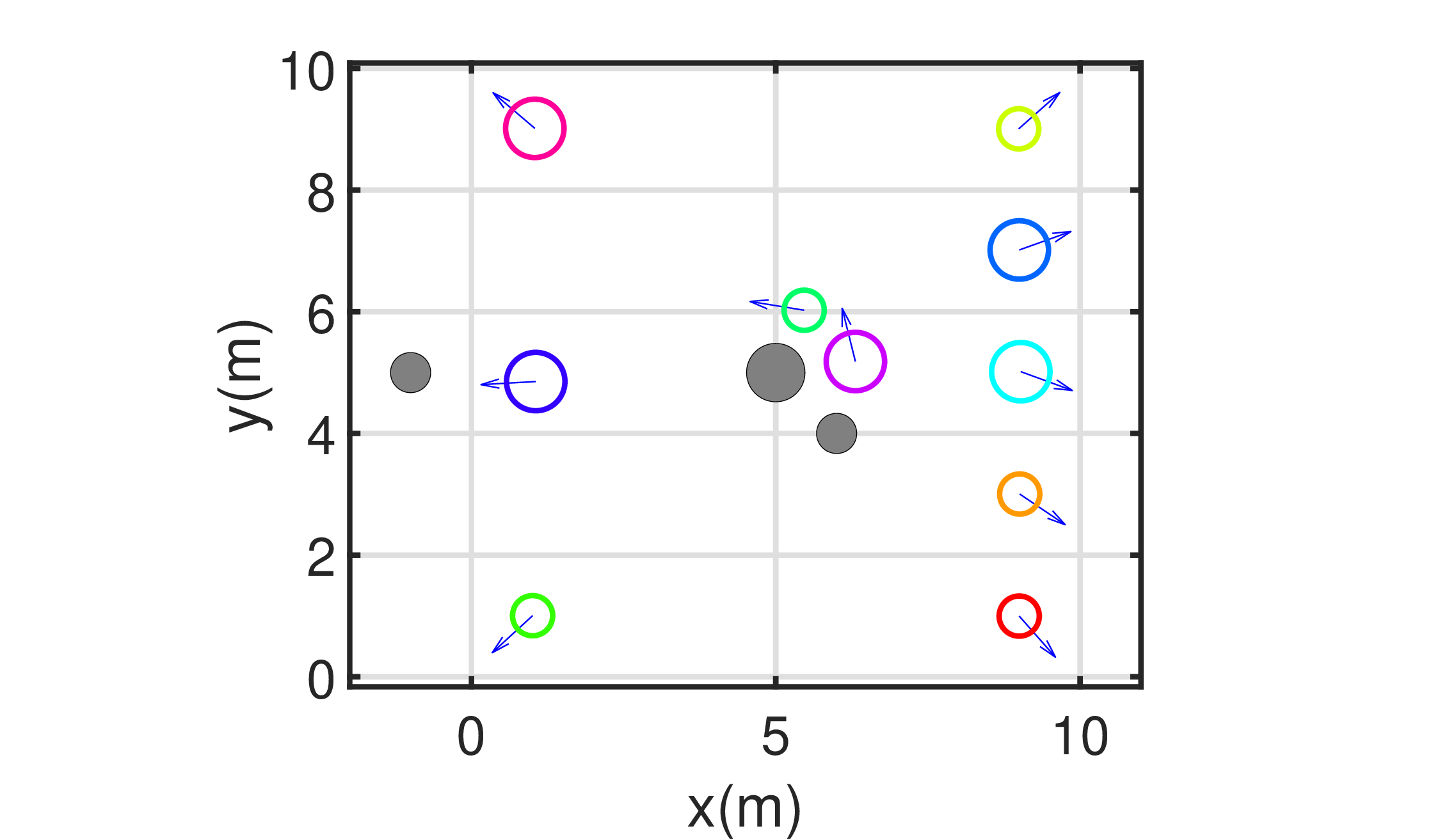}}
\hspace{-40pt}
\subfigure[]{\includegraphics[width=2.3in]{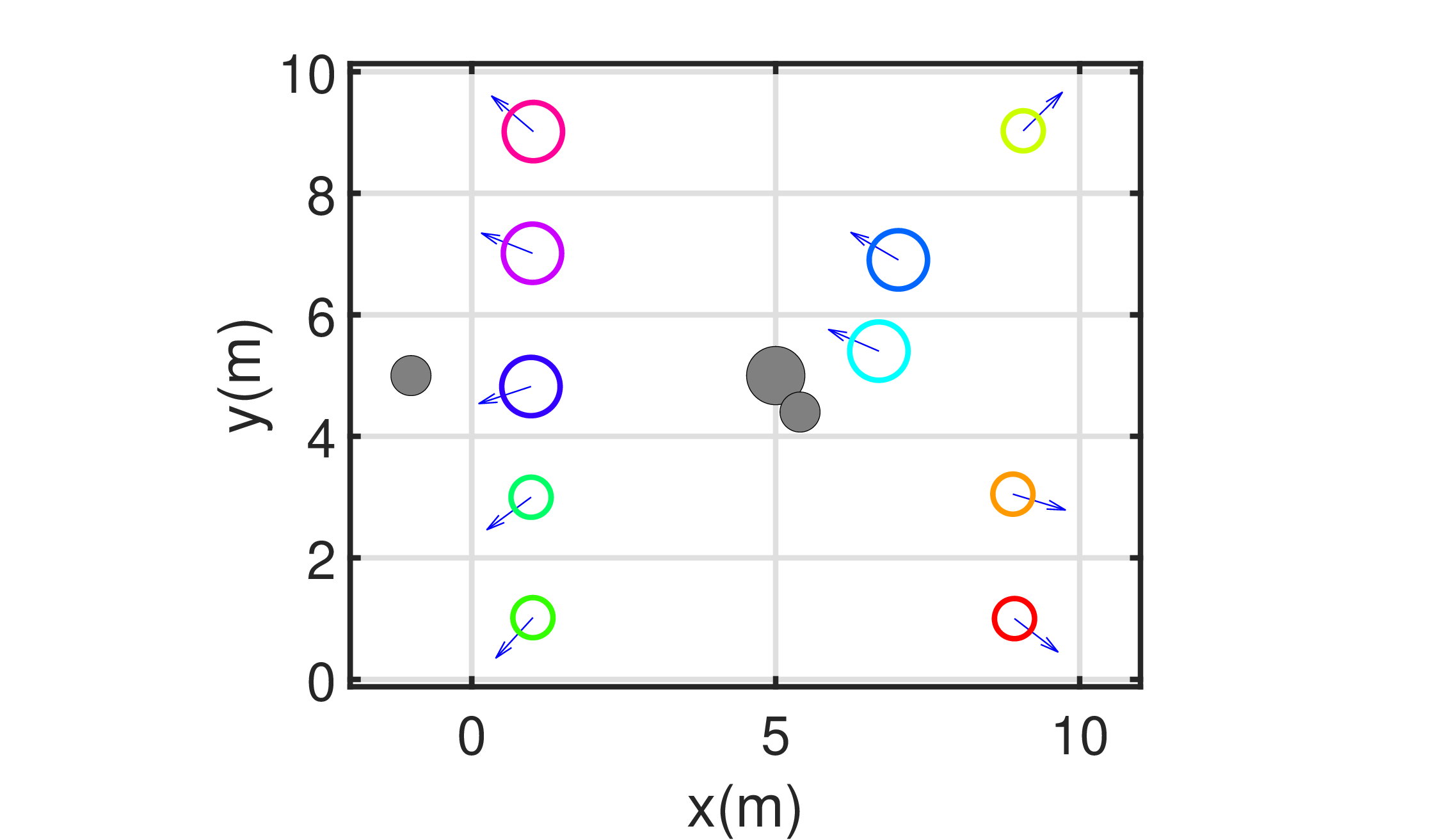}}
\end{minipage}
 \caption{\justifying SOATT results obtained by the fixed-side avoidance rule corresponding to Fig. \ref{figs6}(f) and Fig. \ref{figs6}(a), respectively. (a): Left-hand rule. (b): Right-hand rule.}\label{figs7}
 \end{figure}
 \vspace{-10pt}
\subsection{Comparison on deconfliction strategy}
In this part, we compare our deconfliction method with \cite{IJRR2023}. The principle behind both is to escape deadlock by deviating from the preferred trajectory. Therefore, intervention time and proximity of the actual trajectory on the desired trajectory are used to evaluate the performance. Intervention time is the duration from the start to the end of a robot’s CA action. Three metrics (root mean square error (RMSE), mean average error (MAE) and standard deviation) are used to quantify proximity of the actual trajectory on the desired trajectory, and lower values are desirable. Considering a pair of robots that gradually approach each other from a distance of $6\mathrm{m}$, with the expectation that they maintain a safe distance of $0.96\mathrm{m}$. Performance indices (Intervention Time, RMSE, MAE, Standard Deviation) achieved by our method and \cite{IJRR2023} are $[2.1980,0.2893,0.1667,0.2364]$ and $[3.853,0.6962,0.4465,0.5341]$ respectively.
Fig. \ref{figijrr} also gives position errors $||\mathbf{p}-\mathbf{p}_d||$ and velocity $||\mathbf{v}||$ profiles synthesised by our deconfliction strategy and one proposed by \cite{IJRR2023}.
It is obvious from Fig. \ref{figijrr} that when deadlock is detected (around $t=5\mathrm{s}$), both our method and \cite{IJRR2023}
start to deviate from the desired trajectory, \emph{i.e}, the position errors starts to be greater than zero. These two methods successfully escape the deadlock and reach the final destinations, however, robots in \cite{IJRR2023} inevitably stagnate for a period of time to resolve conflicts, and then rush towards their destination at a higher speed. In this example, the stagnation time is around $2\mathrm{s}$, which is not desirable for the real-time application. The performance indices is inferior to ours, which is also caused by the stagnation. \textbf{A dynamic illustration is given in \href{https://youtu.be/9ayrzR9qpuw}{https://youtu.be/9ayrzR9qpuw}.}
Moreover, the whole control process is divided into three phase: before deadlock, during deadlock and after deadlock, which are controller by three controllers respectively in \cite{IJRR2023}. In our method, the robot is controlled by a solitary CBF-QP controller. This is beneficial to reduce energy consumption caused by switching between different control modules to some extent.

\begin{figure}
  \centering
\setlength{\abovecaptionskip}{-0.1cm}   
\begin{minipage}[]{5.5in}
\subfigcapskip=-2pt
\subfigure[]{\includegraphics[scale=0.103]{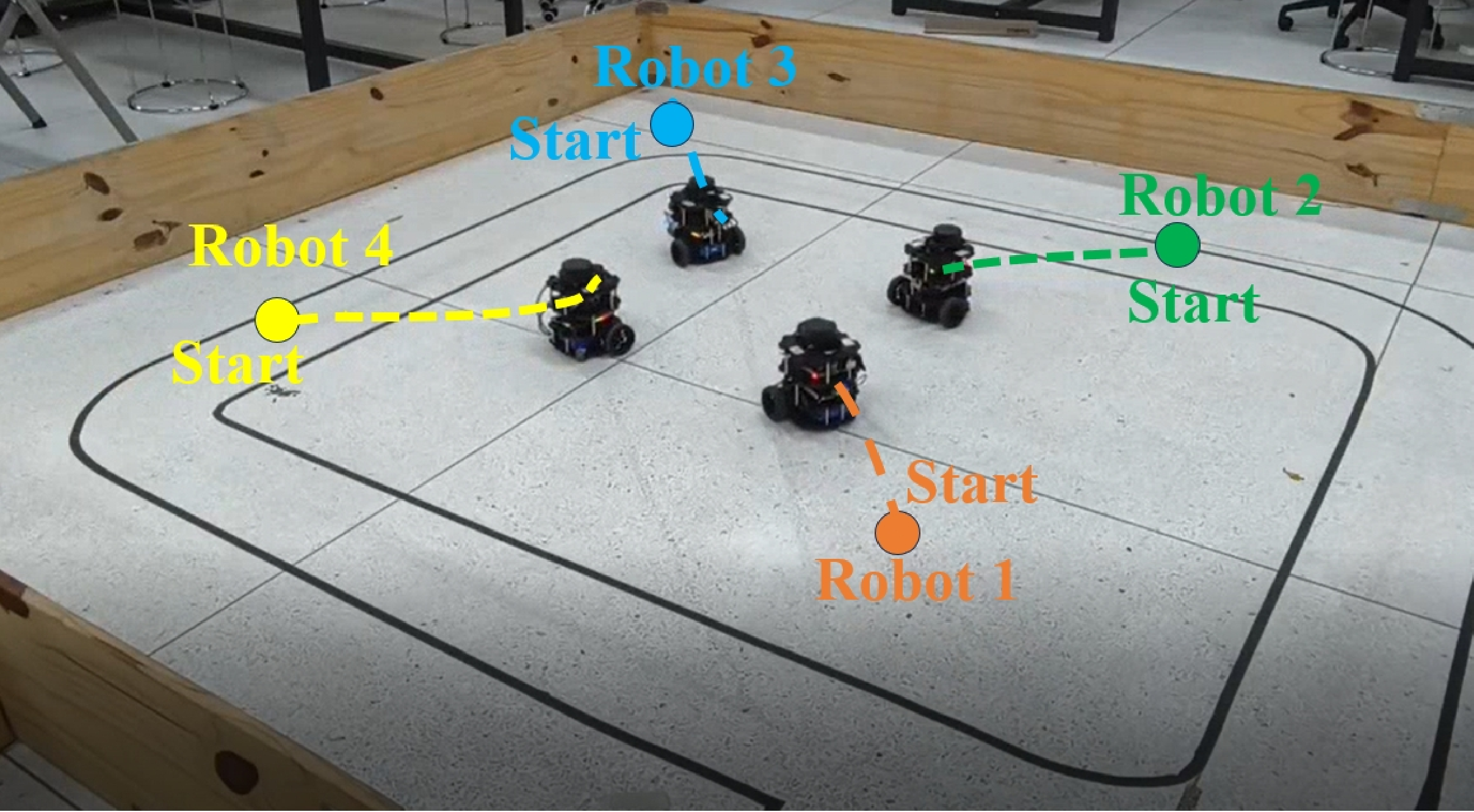}}
\hspace{1pt}
\subfigure[]{\includegraphics[scale=0.103]{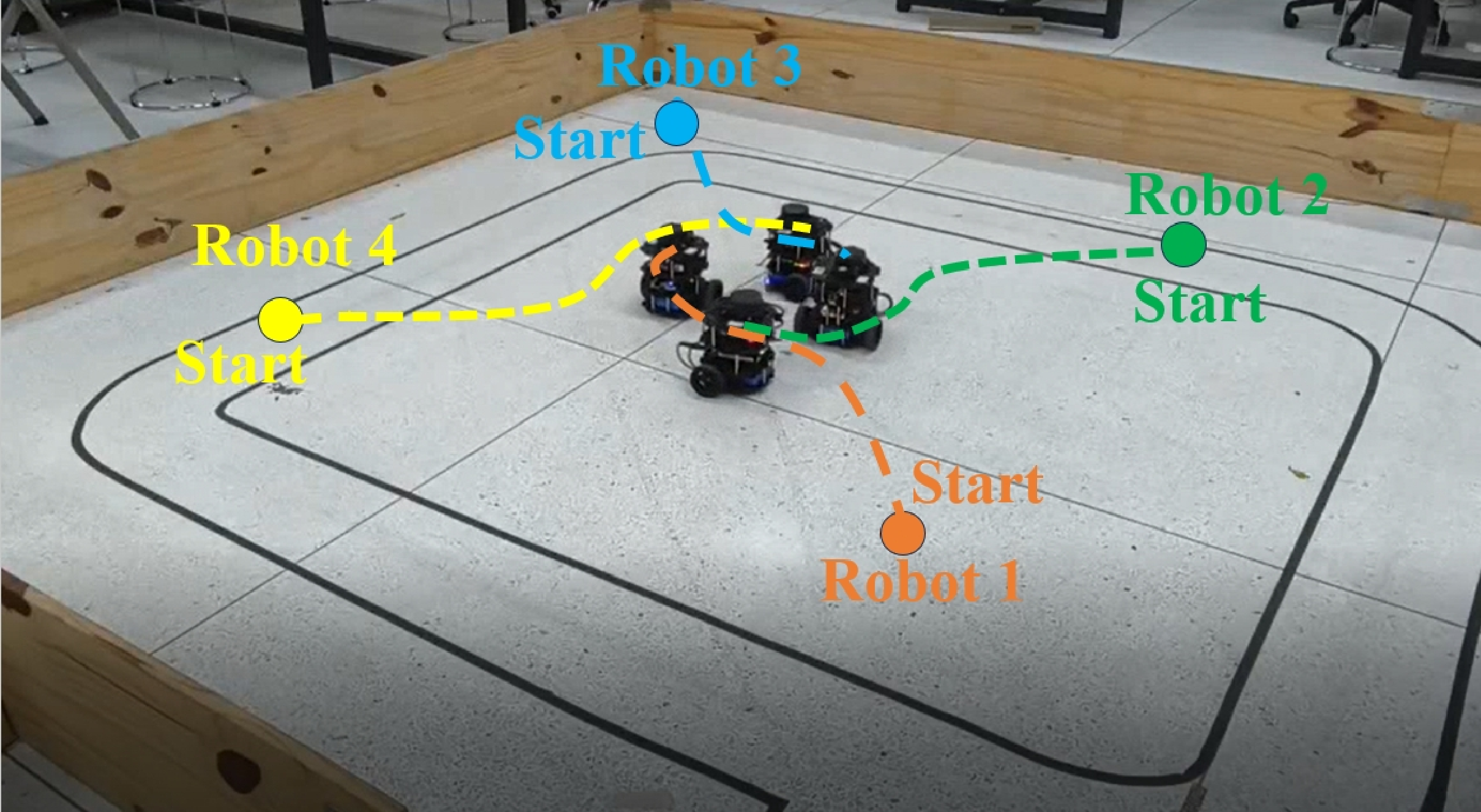}}
\end{minipage}
\begin{minipage}[]{5in}
\vspace{-3pt}
\subfigcapskip=-2pt
\subfigure[]{\includegraphics[scale=0.103]{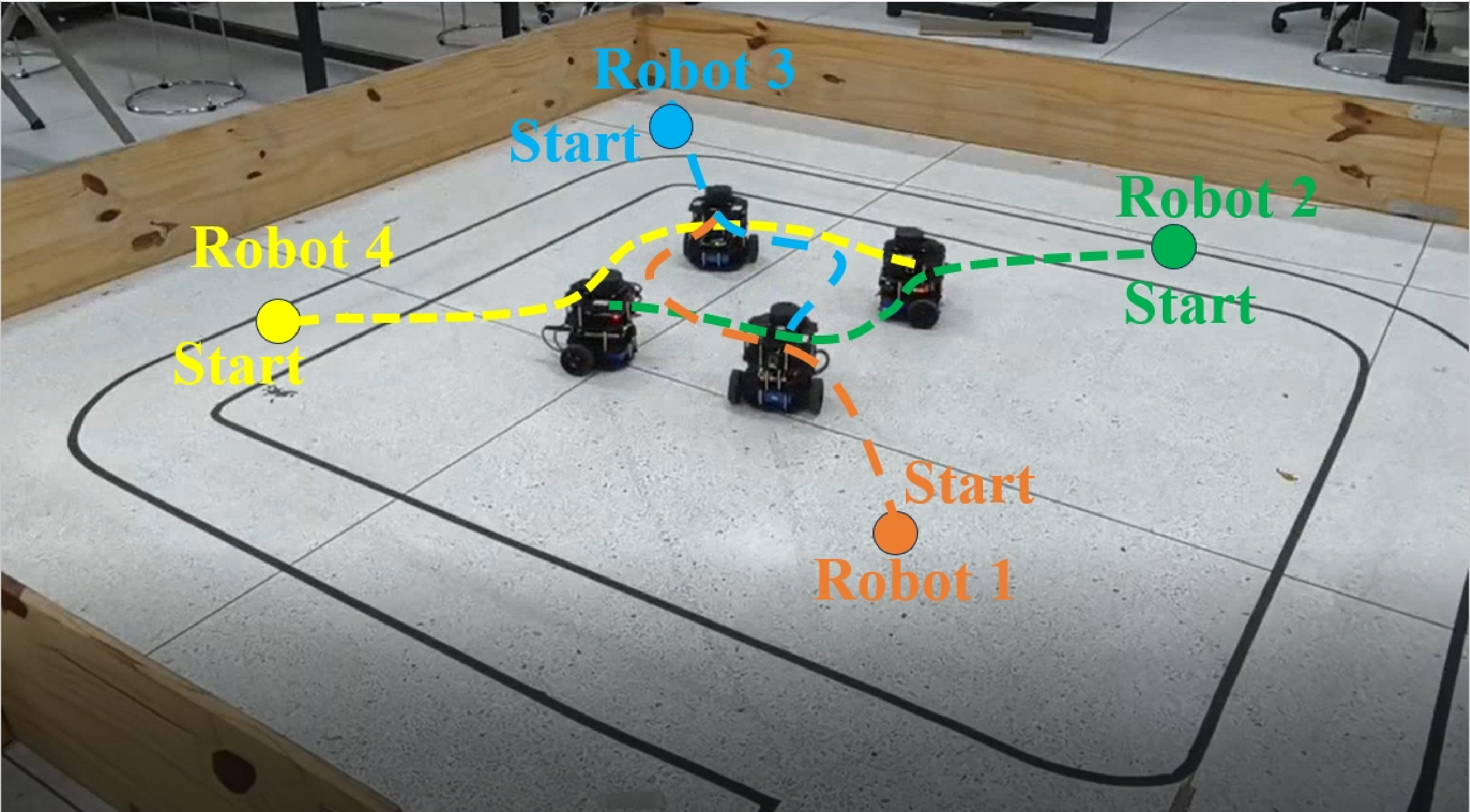}}
\hspace{1pt}
\subfigure[]{\includegraphics[scale=0.103]{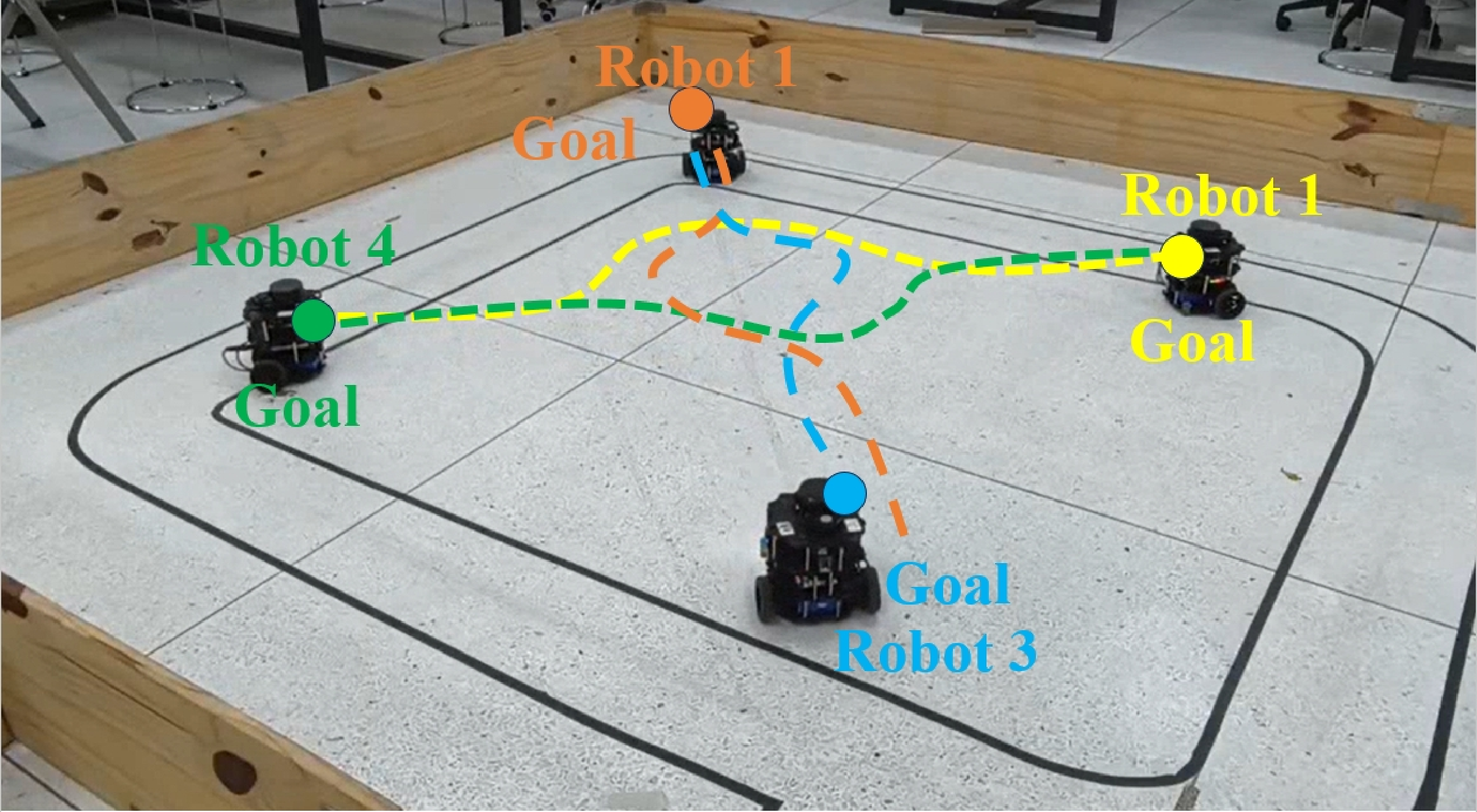}}
\end{minipage}
 \caption{\justifying SOATT snapshot between four robots. (a) Before collision avoidance. (b) During collision avoidance. (c) After collision avoidance. (d) Real trajectories.}\label{fig.e2}
 \end{figure}

\begin{figure}
  \centering
\setlength{\abovecaptionskip}{-0.1cm}   
\begin{minipage}[]{5.5in}
\subfigcapskip=-2pt
\subfigure[]{\includegraphics[scale=0.104]{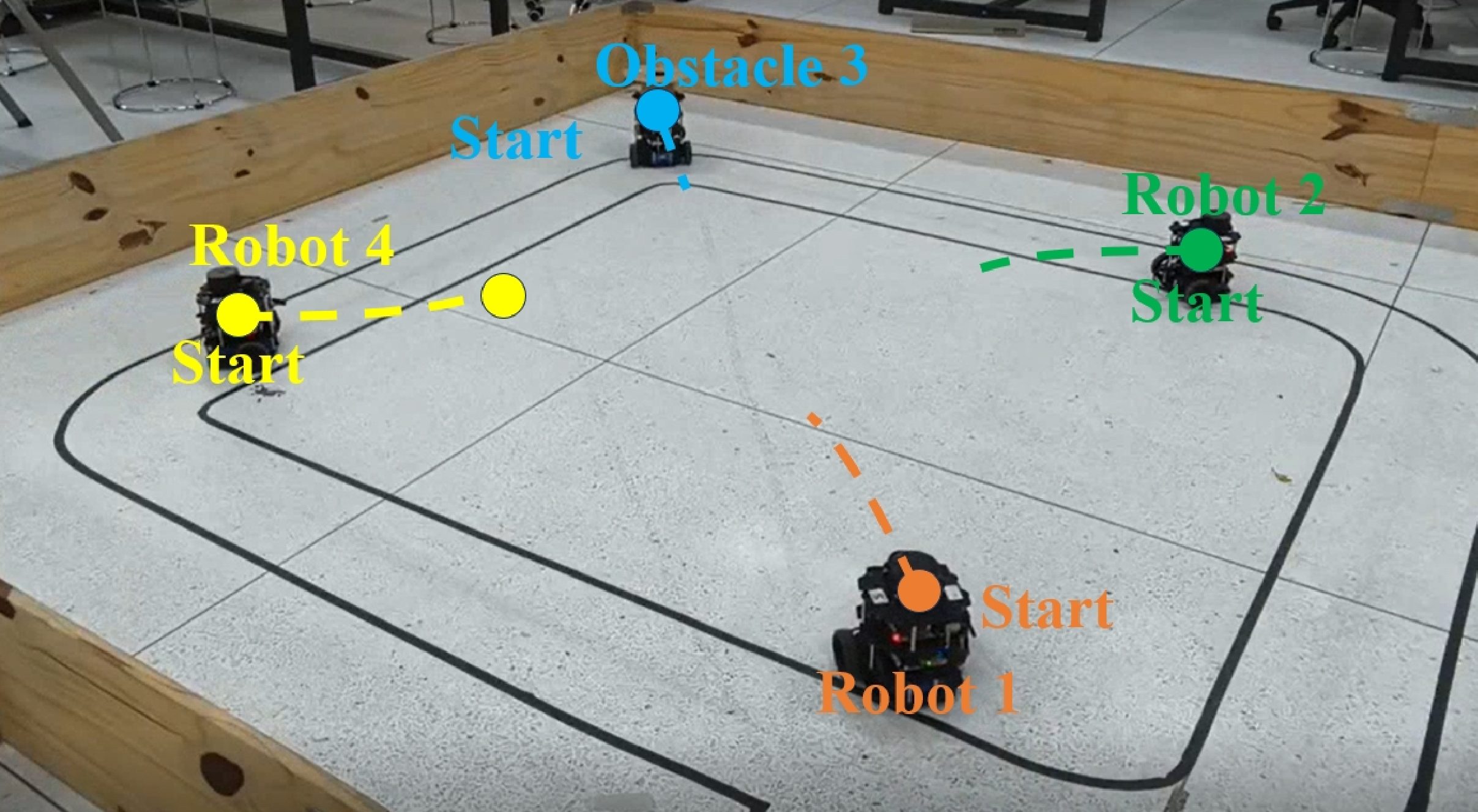}}
\hspace{1pt}
\subfigure[]{\includegraphics[scale=0.104]{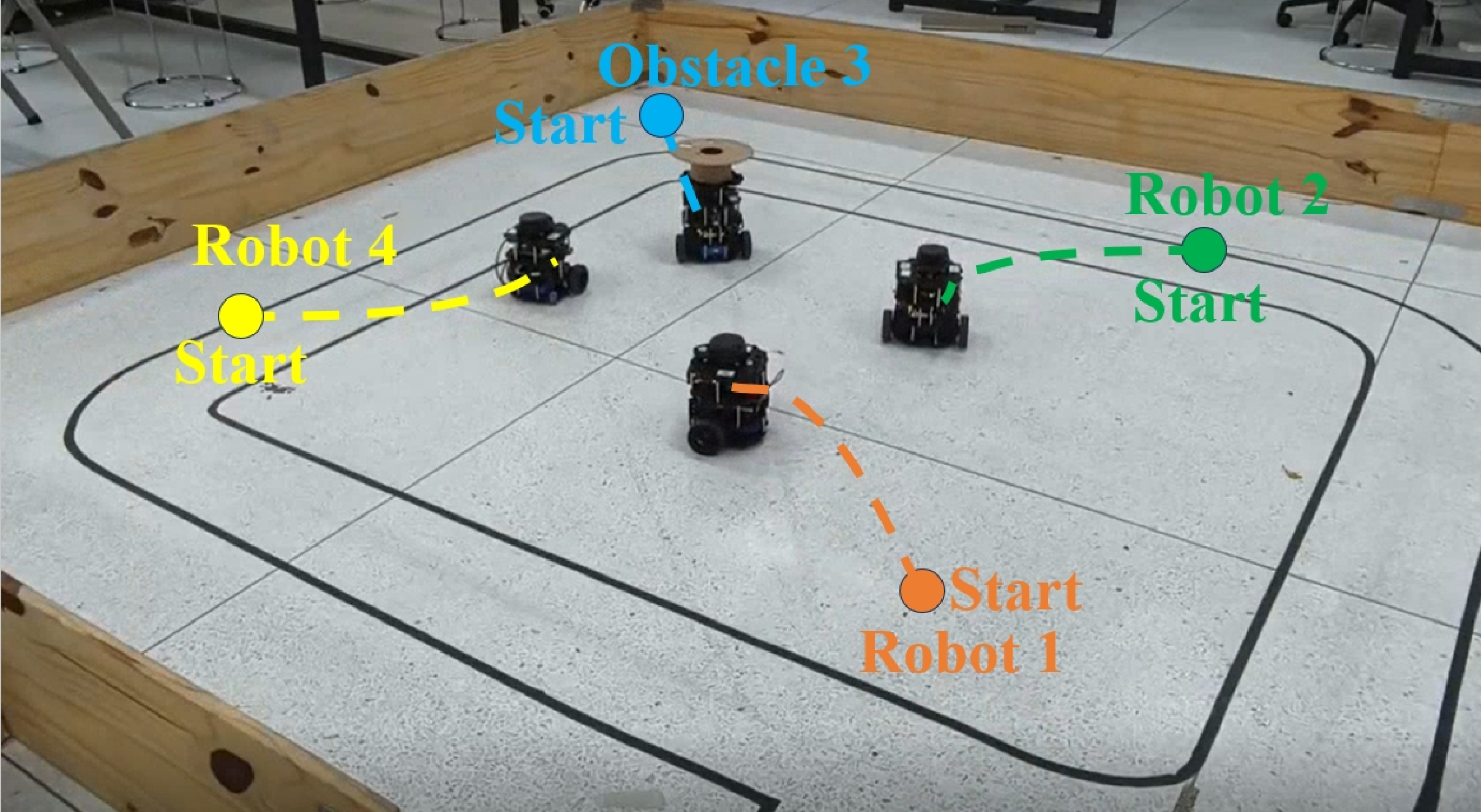}}
\end{minipage}
\begin{minipage}[]{5in}
\vspace{-3pt}
\subfigcapskip=-2pt
\subfigure[]{\includegraphics[scale=0.104]{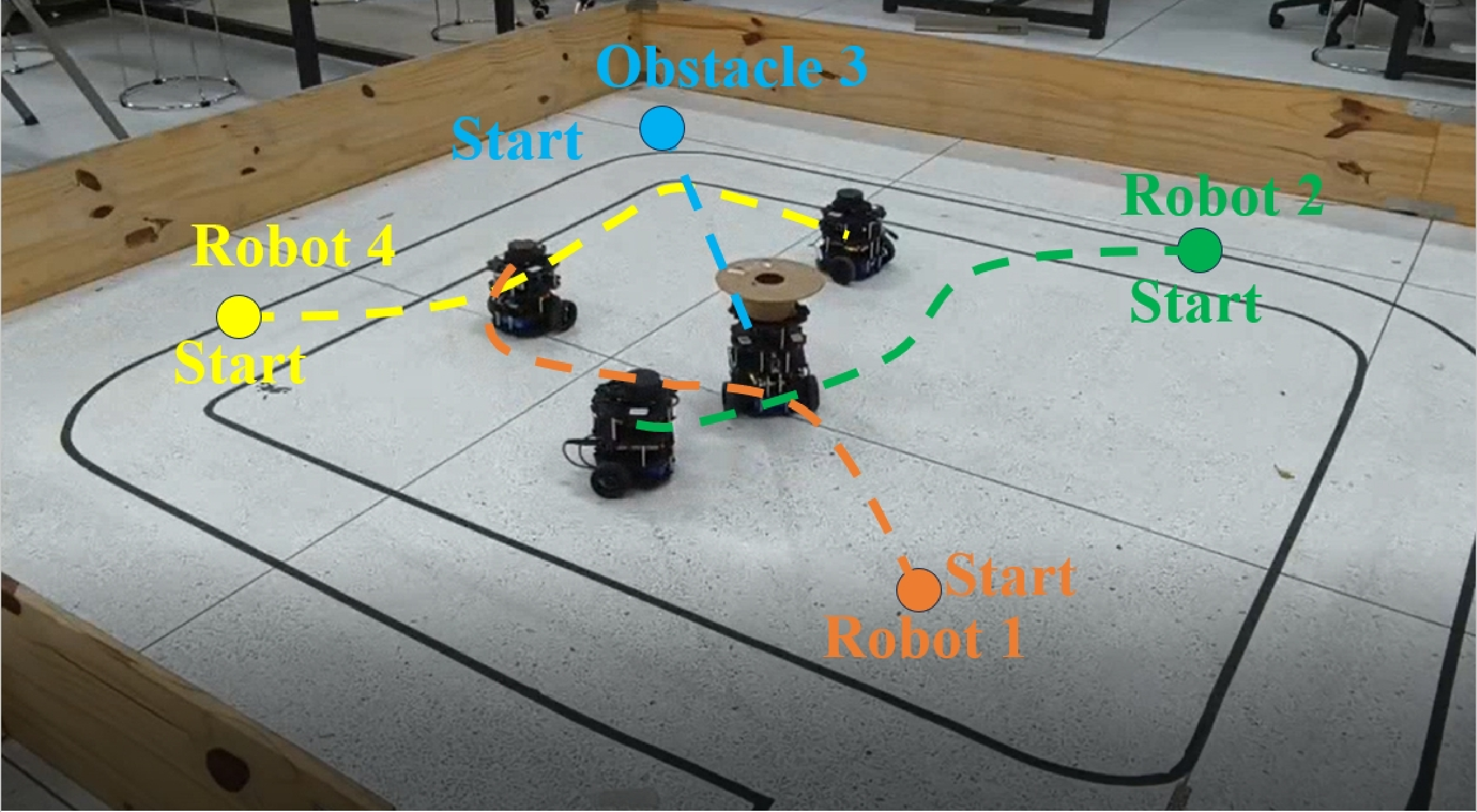}}
\hspace{1pt}
\subfigure[]{\includegraphics[scale=0.104]{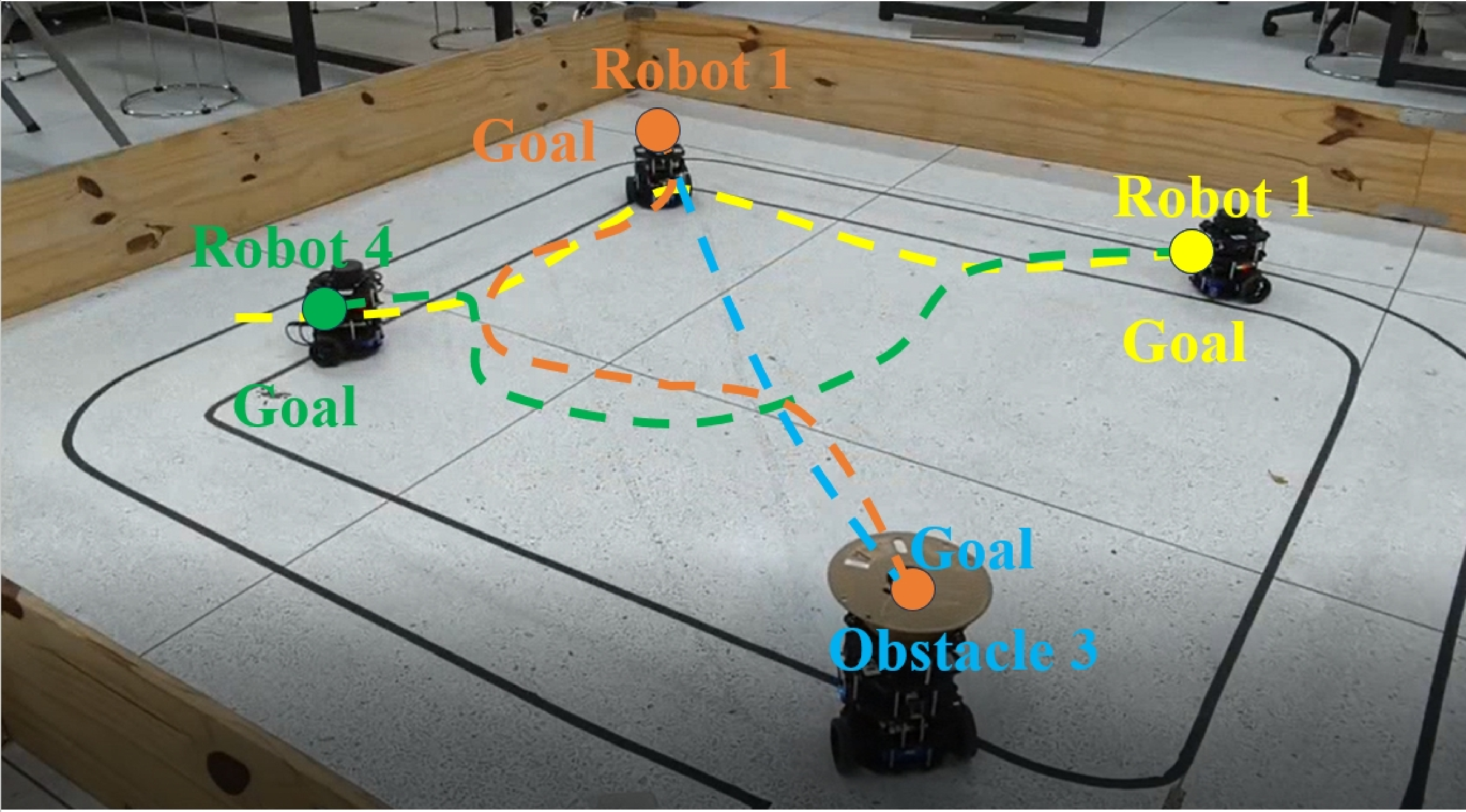}}
\end{minipage}
 \caption{\justifying SOATT snapshot between three robots and a dynamic obstacle. (a) Before collision avoidance. (b) During collision avoidance. (c) After collision avoidance. (d) Real trajectories.}\label{fig.e3}
 \end{figure}
 \subsection{Rationality of avoidance direction}
 SOATT results obtained by the left-hand avoidance rule and the right-hand rule are given in Fig. \ref{figs7}, based on Fig. \ref{figs6}(f) and Fig. \ref{figs6}(a), respectively.
 \textbf{A dynamic illustration is given in \href{https://youtu.be/XJqBnoqtHiU}{https://youtu.be/XJqBnoqtHiU},} of which Robots 5 and 9 have been trying to move to the left due to the left-hand rule setting, but obstacle Obs3 is moving to their left, repelling them from moving to the left, and as a result, they cannot reach their destination. These two examples demonstrate the irrationality of fixed-side avoidance. In contrast, the proposed method always make the robot to turn at the minimum attitude angle. This results can also be observed from Fig. \ref{h}.
\subsection{Physical experiments}
Physical experiment is implemented in TurtleBot 3 Burger. The obtained experimental results are shown in Fig. \ref{fig.e2} and Fig. \ref{fig.e3}. In these two experiments, we consider the position swapping scenario. Fig. \ref{fig.e2} shows SOATT snapshot between four robots. Fig. \ref{fig.e3} shows SOATT snapshot between three robots and a dynamic obstacle where Robot3 is set as a dynamic obstacle without the CA controller. We give trajectory of four phases: before CA, during CA, after CA and final real trajectory.
Following Fig. \ref{fig.e2} and Fig. \ref{fig.e3}, it is obvious that the collision is avoided and all robots reach their destinations.

\section{Conclusion}
A predictive cooperative CA method has been proposed, which provided an additional buffer zone for measurement uncertainty while relatively guaranteed smoothness of the CA trajectory. A deadlock avoidance strategy with the minimum attitude angle has been designed, avoiding a large detours. The proposed method could help robots quickly escape the deadlock state compared to existing methods, without obvious stagnation.

In this paper, we assume there are sufficient collision avoidance space for robots. In the future, we will consider multi-robot motion coordination problems with global optimal behavior under limited space constraints. In addition, combined optimal control and reinforcement learning based online multi-robot motion coordination is being considered deeply.

\bibliographystyle{IEEEtran}
\bibliography{reference}

\end{document}